\documentclass[11pt]{article}
\usepackage{amsmath}
\usepackage{graphicx}
\usepackage{amsfonts}
\usepackage{amsthm}
\usepackage{setspace}
\usepackage{tikz}
\usepackage{amssymb}
\usetikzlibrary{patterns}
\usepackage{sgame}
\usepackage{color}
\usepackage{hyperref}
\usepackage{times}
\usepackage{enumitem}
\usepackage{csquotes}
\usepackage{multirow,array}
\usepackage[none]{hyphenat}
\usepackage[top=0.95in, bottom=0.89in, left=0.8in, right=0.8in]{geometry}

\newtheorem{Proposition}{\hskip\parindent\bf{Proposition}}

\newtheorem{Theorem}{\hskip\parindent\bf{Theorem}}
\newtheorem*{Theorem1}{\hskip\parindent\bf{Theorem 1'}}
\newtheorem*{Theorem2}{\hskip\parindent\bf{Theorem 2'}}

\newtheorem{Lemma}{\hskip\parindent\bf{Lemma}}[section]

\newtheorem{Refinement}{\hskip\parindent\bf{Refinement}}

\begin{document}

\title{\vspace{-2cm}Crime Aggregation, Deterrence, and Witness Credibility\footnote{Preliminary versions of this project have been circulated under various titles. We thank S. Nageeb Ali, Bocar Ba, Sandeep Baliga, Arjada Bardhi, Laura Doval, Mehmet Ekmekci, Alex Frankel, Chishio Furukawa, George Georgiadis, Bob Gibbons,  Andrei Gomberg, Yingni Guo, Andreas Kleiner,
Anton Kolotinin, Frances Xu Lee, Annie Liang, Matt Notowidigdo, Wojciech Olszewski, Alessandro Pavan,
Joyce Sadka,
Larry Samuelson, Ron Siegel, Vasiliki Skreta, Juuso Toikka, Rakesh Vohra, Alex White, Alex Wolitzky, Boli Xu, and our seminar audiences for helpful comments. Strulovici acknowledges financial support from the National Science Foundation (NSF Grant No.1151410). Pei: Department of Economics, Northwestern University. harrydp@northwestern.edu. Strulovici:
Department of Economics, Northwestern University. b-strulovici@northwestern.edu.}}
\author{Harry Pei \and Bruno Strulovici}
\date{\today}

\maketitle

\noindent
\textbf{Abstract:} We present a model for the equilibrium frequency of offenses and the informativeness of witness reports when potential offenders can commit multiple offenses and witnesses are subject to retaliation risk and idiosyncratic reporting preferences. We compare two ways of handling multiple accusations discussed in legal scholarship: (i) When convictions are based on the probability that the defendant committed \textit{at least one, unspecified offense} and entail a severe punishment, potential offenders induce negative correlation in witnesses' private information, which leads to uninformative reports, information aggregation failures, and frequent offenses in equilibrium. Moreover, lowering the punishment in case of conviction can improve deterrence and the informativeness of witnesses' reports. (ii) When accusations are treated separately to adjudicate guilt and conviction entails a severe punishment, witness reports are highly informative and offenses are infrequent in equilibrium.\\

\vspace{.1cm}

\noindent \textbf{Keywords:} soft evidence, deterrence, negative correlation, coordination.\\
\noindent \textbf{JEL Codes:} D82, D83, K42.

\numberwithin{equation}{section}
\begin{spacing}{1.5}

\section{Introduction}
When a defendant faces multiple charges, the legal norm is to consider these charges separately and to convict the defendant if there are {\em specific} charges whose corresponding evidence meets the appropriate standard of proof.

While this separation of charges is standard, its desirability for deterrence and fairness is by no means obvious. Consider a defendant who may have committed two offenses with probability $0.8$ each, independent of each other. If the conviction threshold for each offense is $0.9$, the defendant is acquitted on both counts, even though the probability that he is guilty of \textit{at least one} offense is $1-0.2 \times 0.2 = 0.96$. By contrast, a defendant accused of a single offense may be convicted even if his probability of guilt is $0.91$, and thus lower than the first defendant's.

This issue is most salient when defendants face multiple accusations that are hard to prove beyond a reasonable doubt, such as abuses of power, extortions, and sexual assaults. In such cases, evidence often relies on witness testimonies and may thus be affected by witnesses' incentives to tell the truth.

Legal scholarship has explored the possibility of aggregating charges into an overall probability of guilt instead of treating charges separately (see Cohen 1977, Bar Hillel 1984, Robertson and Vignaux 1993). In particular, Harel and Porat (2009) define the \textit{Aggregate Probabilities Principle} (``APP'') as follows: a defendant is convicted if the probability that he has committed some \textit{unspecified} offense exceeds a given threshold. They compare APP to the \textit{Distinct Probabilities Principle} (``DPP''), which requires that the principal be convicted if there is at least one specific offense for which the probability that the principal committed this offense exceeds some exogenous threshold. Harel and Porat argue that APP can reduce adjudication errors, improve deterrence, and reduce the cost of enforcement, and advocate using APP to varying degrees in both civil and criminal instances of the law.

While the object of these studies is of clear importance, their arguments rely on the assumptions that the {\em distribution of the defendant's guilt across charges is exogenous}, the {\em quality of evidence across charges is exogenous}, and guilt is \textit{independently distributed across charges}. This approach ignores any strategic consideration in the behaviors of potential offenders and witnesses. In particular, it cannot speak to how the introduction of APP may affect the incentives of potential offenders and the informativeness of witness testimonies.

This paper compares the effects of APP and DPP from a strategic perspective. In our model, the probability of committing offenses and the informativeness of accusations (or absence thereof) are endogenous. Since potential offenders choose their actions strategically, distinct offenses need not be independently distributed. In fact, we show that APP can induce potential offenders to introduce \textit{negative correlation} in witnesses' private information and severely undermine the informativeness of witnesses' reports in equilibrium.
This negative correlation violates the independence assumption used in the legal literature to analyze APP and undermines its conclusions.

In our model, a potential offender, the \textit{principal}, has several opportunities to commit offenses, each of which is associated with a distinct witness, or {\em agent}, who observes whether the corresponding offense takes place. For example, the principal may be an employer with multiple opportunities of violating the law and agents may be employees in a position to witness these violations, as victims or potential whistleblowers. Our model encompasses various types of reportable misbehavior, from the most serious crimes to trivial violations of laws and social norms.

Agents simultaneously decide whether to accuse the principal on the basis of three considerations: (i) a preference for punishing offenses, (ii) a risk of facing retaliation or social stigma, which is higher when accusations fail to get the principal convicted, and (iii) some (possibly small) idiosyncratic private benefits or costs of getting the principal convicted, which are independent of whether offenses have taken place.\footnote{In our model, some abuses go unreported and some charges of abuse are not deemed credible enough to lead to a conviction. Both features are consistent with the empirical evidence on abuses and reports of abuse. These patterns are documented in a study of harassment in the U.S. military by the RAND corporation (2018) and studies of police brutality or inaction by Ba (2018) and Ba and Rivera (2019) using data from the city of Chicago. Similar partterns arise in a 2016 survey conducted by the USMSPB, which  concluded that 21\% of women and 8.7\% of men experienced at least one of 12 categorized behaviors of sexual harassment, of which only a small fraction was followed by charges. According to data released by USMSPB, among the harassment charges filed in 2017, only 16\% led to ``\textit{merit resolutions},'' i.e., to outcomes favorable to the charging parties.}

A Bayesian judge then observes agents' reports (accusations, or absence thereof) and decides whether to convict or acquit the principal. When contemplating the commission of offenses, the principal trades off the utility from committing offenses with the expected cost of punishment from conviction.

We compare  the principal's incentive to commit offenses and the informativeness of witness reports when either APP or DPP is used to adjudicate guilt.  APP and DPP are identical when there is only one offense under consideration, but differ when the principal may have committed two or more offenses. For example, a defendant who is surely guilty of exactly one of three accusations, uniformly across accusations,
will surely be convicted if APP is the criterion used for conviction, no matter what standard of proof is applied,
but will be acquitted under DPP and the conviction standard is even as weak as the 50\% threshold used for preponderance of evidence. Understanding the comparative benefits of these criteria is relevant not only for criminal law and civil law but also for corporate decisions such as whether to fire an employee facing multiple allegations of misconduct.

To highlight the main forces at play, we start in Section \ref{sec3} by comparing APP and DPP when (i) the principal can commit {\em at most two offenses} and (ii) conviction entails a large punishment relative to the principal's gain from the commission of these offenses. Theorem~\ref{Theorem1} shows that when APP is used as the adjudication criterion, the principal commits at most one offense in every equilibrium. This strategic restraint induces \textit{negative correlation} in agents' private information: when an agent observes an offense, he believes that the other agent is unlikely to observe one. This negative correlation exposes agents to the risk of contradicting each other and to retaliation. In equilibrium, this reduces the informativeness of witness reports and, despite the threat of a large punishment in case of conviction, prompts the principal to commit offenses with high probability.

When DPP is used, by contrast, the principal's  decisions to commit distinct offenses are independently distributed in equilibrium and, hence, so are witnesses' private observations (Theorem \ref{Theorem2}). As the punishment in case of conviction becomes arbitrarily large, agents' accusations become arbitrarily informative and the probability of offense converges to zero.\footnote{In the benchmark model with only one offense and one witness, APP and DPP coincide. Proposition \ref{Prop1} shows that agent's report becomes arbitrarily informative and the probability of offense vanishes as the punishment to the convicted principal increases.}

The logic of Theorem~\ref{Theorem1} can be described more explicitly in two steps. First, when conviction entails a large enough punishment, the principal is convicted in equilibrium only if {\em both} agents accuse him (Lemma \ref{L3.2}). Intuitively, suppose that one accusation sufficed to convict the principal with positive probability. Since each accusation \textit{strictly} increases the probability that the principal is guilty of at least one offense, a Bayesian judge would then surely convict the principal when both agents accuse him. If conviction entails a large punishment, this would give the principal a strict incentive not to commit any offense. This, in turn, would prompt a Bayesian judge never to convict the principal and lead to a contradiction (Lemma \ref{L3.1}).

This first step implies that the principal's decisions to commit offenses are \textit{strategic substitutes}, because he goes unpunished when he faces only one accusation but is convicted with positive probability when facing two accusations (Lemma \ref{L3.3}). It also implies that agents' decisions to accuse the principal are \textit{strategic complements}, because an agent who accuses the principal is more likely to convict him if the other agent also accuses. These observations lead to the second step of the argument, which unfolds as follows.  In equilibrium, the principal may commit zero or one offense, but never two, which induces \textit{negative correlation} in agents' private information. Witnessing an offense reduces the probability that the other agent also witnessed an offense and, due to agents' reporting complementarity, weakens the incentive to accuse the principal. Likewise, not witnessing an offense leads to a higher probability that the other agent witnessed an offense and raises the incentive to accuse the principal. Agents' incentives are distorted in way that reduces the informativeness of their reports. This, in turn, increases the equilibrium probability of offenses because reports are less responsive to the commission of offenses.

By contrast, DPP preserves the informativeness of witnesses' reports by disentangling a potential offender's incentives to commit distinct offenses and, hence, by removing any correlation in witnesses' private information. Unlike the probability that the principal is guilty of \textit{at least one} offense, the probability that he is guilty of a \textit{specific} offense need not increase, other things equal, when a larger set of agents accuse him of other offenses. With DPP,
the probability that the principal is convicted is \textit{linear} in the number of accusations and the principal's decisions to commit distinct offenses are neither complements nor substitutes. This implies that the principal's incentives to commit distinct offenses operate independently of one another and that agents' private observations  are uncorrelated. As a result, agents' coordination motive no longer undermines the informativeness of their reports. Finally, these linear conviction probabilities are consistent with the DPP criterion, because the judge's belief concerning the occurrence of each offense reaches the conviction threshold if and only if the agent who can observe this offense accuses the principal, and is unaffected by the reports of other agents.

In practice, one rationale for aggregating different charges against a defendant is that these charges may shed light on the defendant's propensity to commit offenses, i.e., on the defendant's type, or ``character.'' In Section \ref{sec4}, we account for this potential heterogeneity by allowing the principal to be either a \textit{virtuous type}, whose benefit from committing an offense is either zero or negative, or an \textit{opportunistic type} whose benefit from committing an offense is strictly positive. The principal's type is unobserved by other parties.

The insights of Theorems \ref{Theorem1} and \ref{Theorem2} extend to this setting. When conviction entails a sufficiently large punishment and the judge uses APP, two reports are required to convict the principal in every equilibrium. In contrast to the baseline model, an opportunistic principal may now commit \textit{two offenses} with strictly positive probability. However, agents' private observations remain \textit{negatively correlated}. As in the baseline model, this negative correlation undermines the informativeness of agents' reports and leads to a high probability of offense in equilibrium. By contrast, when DPP is used as the conviction criterion,
the offenses witnessed by the agents are uncorrelated and
the probability with which the principal is convicted is a linear function in the number of accusations.
As the punishment in case of conviction becomes arbitrarily large (relative to an opportunistic type's benefit from committing an offense), agents' reports become arbitrarily informative and offenses become arbitrarily infrequent.

In Section \ref{sec5}, we explore the effects of APP and DPP when the principal can commit more than two offenses and, correspondingly, face accusations by more than two witnesses. When APP is used and conviction entails a large enough punishment, the informativeness of each agent's report becomes so weak that conviction occurs only when \textit{all} agents accuse the principal. Theorem~\ref{Theorem3} shows that, as the number of agents increases, the aggregate informativeness of {\em all} reports pooled together \textit{decreases} even though there are more reports available and the frequency of offenses increases in equilibrium. However, the equilibrium probability that an agent accuses the principal is \textit{increasing} in the number of agents. This distinguishes our result from theories of public good provision, in which each agent contributes less
as the number of agents increases.

In summary, our results provide a rationale for using DPP rather than APP in judicial settings. Our analysis is also relevant for firms and organizations considering decisions such as whether to fire an employee or to sanction some of their members.
In these cases, DPP raises concerns for fairness and justice and APP can be justified ex post. Schauer and Zeckhauser (1996) observe that ``...\textit{although sound reasons for the criminal law's refusal to cumulate multiple low-probability accusations exist, the reasons for such refusal are often inapt in other settings. Taking adverse decisions based on cumulating multiple  low-probability charges is often justifiable both morally and mathematically}.''
However, we show that enshrining APP in corporate decisions can have an undesirable effect on ex ante incentives for both potential offenders and potential witnesses. Our findings echo some critiques of procedures that link accusations across potential victims.\footnote{For example, Keith Hiatt, director of the Technology Program at the Human Rights Center at UC Berkeley School of Law, noted concerning the multiple-accusations approach taken by the online platform Callisto that ``it may also codify an entrenched attitude that women need to have corroborating evidence to be believed.'' {\em New York Times, ``The War on Campus Sexual Assault Goes Digital''.}}

Motivated by applications in which the use of APP is unavoidable and the use of DPP is inconvenient, we show in Proposition 3 that mitigating punishment to the convicted principal can improve deterrence. More precisely, there exist intermediate punishment levels for which the probability of conviction is \textit{concave} in the number of accusations, and a single accusation suffices to convict the principal with positive probability. The
principal's decisions to commit offenses are now \textit{strategic complements}, which induces \textit{positive correlation} in witnesses' private information. By a reverse logic to the high-punishment case, this increases the informativeness of witness reports and lowers the frequency of offenses. When the principal does commit at least one offense, he may now commit multiple offenses. In equilibrium, the frequency of guilty principals is lower but the severity of their actions is higher than in the high-punishment case. Our result thus points to a tradeoff between reducing the fraction of offenders and reducing the severity of offender's actions.

While we did not specify any social welfare function, our analysis unveils several tradeoffs between improving the quality of judicial decisions and deterrence, between ex ante incentives and ex post fairness, and between the frequency and the severity of committed offenses. We summarize these lessons in Section \ref{secX}, which are useful for a social planner considering various adjudication rules to maximize an objective function that incorporates the various facets of these tradeoffs.

Section \ref{sec6} proposes several extensions that illustrate the robustness of our results, which include but not limited to the following considerations: (1) the punishment to the principal also depends on the probability he has committed multiple offenses; (2) the principal's marginal benefit from committing an offense is decreasing in the number of committed offenses; (3) false accusations may be exposed ex post with positive probability and results in false accuser(s) being punished; (4) agents have an intrinsic preference for reporting the truth, irrespective of how this affects the principal's conviction, (5) the set of offense opportunities is observed only
by the principal.

\paragraph{Related Literature:}  Models of information aggregation and strategic communication often assume that the fact of interest is exogenously generated. In applications to civil and criminal law, however, facts are generated by individuals whose incentives interact with how information is aggregated, communicated, and ultimately incorporated into judicial decisions. By endogenizing agents' private information and incentives through the strategic commission of offenses, this paper provides a new perspective on information aggregation and strategic communication, which emphasizes the interaction between the ``state of the world'' and the choices made at later stages to aggregate and
use information, and illustrates the significant, complex, and perhaps unintended, consequences of this interaction.

First, our results provide a novel explanation for information aggregation failures. In Scharfstein and Stein (1990), Banerjee (1992), Bikhchandani et al. (1992), Ottaviani and S{\o}rensen (2000), and Smith and S{\o}rensen (2000), agents fail to act on their private information because they can observe informative actions taken by their predecessors.\footnote{Information aggregation can also fail due to individual biases (Morgan and Stocken 2008) and voters using pivotal reasoning (Austen-Smith and Banks 1996, Bhattacharya 2013).} Here, by contrast, agents move simultaneously and information aggregation fails because of the negative correlation in  agents' private information, combined with agents' incentives to coordinate their reports.\footnote{Strulovici (2020) studies a sequential learning model in which an agent is less likely to have an informative signal, other things equal, if another agent has found such a signal. This creates negative correlation in the {\em informativeness} of agents' signals, rather than the {\em direction} of these signals, which hampers social learning.}

Our paper contributes to the literature on voting by studying a game in which both the voting rule and the correlation between agents' private information are endogenous. This stands in contrast to existing works in this literature in which at least one of these two ingredients are exogenous. This includes voting models with endogenous information acquisition (Persico 2004), voting with negatively correlated private information or payoffs (Schmitz and Tr\"{o}ger 2012 and Ali, Mihm, and Siga 2018), and dynamic voting models in which information acquisition may induce negative correlation in voters' continuation values (Strulovici 2010).


Second, our paper contributes to the literature on strategic information transmission with multiple senders (Battaglini 2002, 2017, Ambrus and Takahashi 2008, and Ekmekci and Lauermann 2019). In contrast to these works, senders in our model communicate information about the \textit{principal's actions} and the correlation between their private signals is endogenous. Our results shed light on how this endogenous correlation structure, combined with senders' endogenous coordination motive, affects communication informativeness.

Third, our paper contributes to the law and economics literature by (i) studying decision rules that aggregate the probabilities of offenses, (ii) endogenizing the informativeness of witness testimonies, and (iii) analyzing the interplay between an individual's incentive to commit offenses and witnesses' incentives to report the truth.\footnote{Silva (2019) studies a model with multiple suspects and constructs a mechanism that elicits truthful confessions among suspects. In Baliga, Bueno de Mesquita and  Wolitzky (2020), only one of the potential assailants has an opportunity to commit an offense. In both papers, the negative correlation in whether the suspects are guilty or innocent is exogenous.} Our results justify a key feature of criminal justice systems, which is to treat distinct accusations separately in conviction decisions. Our finding (Proposition~\ref{Prop3}) that a lower punishment can reduce the probability of offense stands in contrast to Becker's (1968) well-known observation that maximal punishments save on law-enforcement costs.\footnote{Stigler (1970) observes that several punishment levels should be used when criminals can choose between different levels of crime. The rationale there is to provide marginal incentives not to commit the worst crimes. In this scenario, applying the maximal punishment to the worst crimes remains optimal from the perspective of deterring crimes.}

Lee and Suen (2020) study the timing of reports by victims and libelers when a criminal commits offenses against two agents with exogenous probability. They provide an explanation for the well-documented fact that victims sometimes delay their accusations. Their analysis and ours consider complementary aspects of witnesses' reporting incentives. Cheng and Hsiaw (2020) adopt a global game perspective to study the reporting incentives of a continuum of agents who observe conditionally independent signals of the state of the world. Naess (2020) also considers reporting incentives and, among other results, finds that making reporting costly may improve social welfare. The principal's strategic restraint that emerges endogenously in our model and the negative correlation that it induces on the agents' private information are distinctive features of our analysis.

\section{Model}\label{sec2}
\paragraph{Overview:} We consider a game between a potential offender (the ``principal''), $n$ potential witnesses or victims (the ``agents''), and a Bayesian judge, which unfolds in three stages.

In the first stage, the principal privately observes his
benefit from committing offenses and then chooses which offenses to commit within a given set of opportunities. For tractability, our analysis considers two principal types: a ``virtuous'' type who has no benefit from committing the offense and an ``opportunistic'' type who has a strictly positive benefit per committed offense.\footnote{One could consider more types; for instance, some ethical principals could have a negative benefit from committing the offense. In our model they would behave identically to the virtuous type.} Our results are first presented when the principal is opportunistic with probability $1$ (Section \ref{sec3}). This allows us to describe as simply as possible the forces at play and to show that our results do not rely on the presence of type heterogeneity.\footnote{This also addresses an ideological and legal concern about assuming heterogeneous defendant propensities to commit offenses. With heterogeneous types, accusing someone of an offense affects the belief about the person's type or ``character'' and, consequently, the probability of guilt regarding other accusations. This may conflict with rules that forbid the use of ``character evidence'', such as the Federal Rule of Evidence 404 in the United States.} We then extend our analysis to allow for both principal types (Section \ref{sec4}). There is a fixed set of opportunities to commit offenses in the baseline model. We discuss the possibility that this set be stochastic and privately observed by the principal in Section \ref{sec6}, and explain why the negative correlation underlying our main result still applies to this more general case.

In the second stage, each agent observes whether the offense associated with him has taken place. Each offense opportunity is associated with exactly one agent. Each agent then independently sends a report about his observation, either accusing the principal or stating that he did not observe any offense, which is formally similar to staying silent. Agents can choose not to accuse the principal when they have observed an offense and vice versa. Agents' observations and reports are assumed to be binary throughout the paper.

In the third stage, a ``judge,'' who could represent a judge in a legal setting or a different kind of evaluator in a non-legal setting, observes the reports of all agents and adjudicates guilt. We compare two criteria to adjudicate guilt: the \textit{aggregate probabilities principle} (``APP'', defined in (\ref{2.4})) and the \textit{distinct probabilities principle} (``DPP'', defined in (\ref{2.5})). In the baseline model, we consider a binary punishment decision: either the principal is punished (which may be interpreted as losing his job or reputation, or receiving a harsh sentence), or he is not. In Section \ref{sec6}, we discuss an extension, in which the punishment to a convicted principal is higher if he is believed to have committed multiple offenses. As explained in that section, this possibility reinforces our main insights.

The principal trades off his (possibly, null) benefit from committing offenses with the risk of punishment. Agents choose their reports based on the following considerations: (i) a preference for convicting a guilty principal, (ii) a cost of retaliation or social stigma if they accuse the principal, which is strictly higher if the principal is acquitted than if he is convicted, (iii) an idiosyncratic taste shock (e.g., affinity or grudge) for seeing the principal acquitted or convicted, and, in an extension, (iv) a desire for telling the truth irrespective of the adjudication outcome. For tractability, the components of the principal's and the agents' payoffs are assumed to be additively separable. To guarantee that all report profiles may occur in equilibrium, we also include an infinitesimal fraction of behavioral agents, who with some probability always accuse the principal and with the complement probability always abstain from accusing him. We do not explicitly microfound the judge's preference. However, both adjudication rules maximize some quadratic payoff functions that we could stipulate. The conviction threshold is taken as exogenous and could also easily be microfounded as a tradeoff between type~I and type~II errors.


\subsection{Baseline Model}
We consider a three-stage game between a principal, $n$ agents, and a judge. In stage $1$, the principal privately observes his type $t \in \{t^v,t^o\}$, where $t = t^v$ means that the principal is \textit{virtuous} and $t = t^o$ means that he is \textit{opportunistic}. The principal then chooses an $n$-dimensional vector $\boldsymbol{\theta} \equiv (\theta_1,...,\theta_n) \in \{0,1\}^n$, where $\theta_i =1$ (respectively, $\theta_i = 0$) means that the principal commits (does not commit) an offense witnessed by agent $i$.

In stage $2$, each agent $i \in \{1,2,...,n\}$ privately observes $\theta_i$, the realization of a payoff shock $\omega_i \in \mathbb{R}$, and whether he is strategic or behavioral. Each agent $i$ then chooses between accusing the principal ($a_i=1$) or not ($a_i =0$). If agent $i$ is strategic, he chooses $a_i$ to maximize his expected payoff. If he is behavioral, with probability $\alpha \in (0,1)$ he always accuses the principal and with probability $1-\alpha$ he never accuses him.\footnote{The assumption that each agent may be behavioral, even with an infinitesimally small probability, guarantees that all reporting profiles are on path and refines away unreasonable equilibria in which the principal commits offenses against all agents with probability $1$ and is surely convicted even when no agent accuses him. Our results hold for more general specifications of behavioral agents' strategies, in which agents' reports depend on their observation of $\theta_i$.}

In stage $3$, the judge observes $\boldsymbol{a} \equiv (a_1,...,a_n) \in \{0,1\}^n$, and makes a decision $s \in \{0,1\}$, where $s=1$ stands for convicting the principal and $s=0$ stands for acquitting him.

\paragraph{Type Distributions:} The random variables $t$, $\{\omega_i\}_{i=1}^n$, and agents' types (strategic vs. behavioral) are independently distributed. We let $\pi^o \in (0,1]$ denote the probability that the principal is opportunistic. The variables $\omega_i$ for $i =1,\ldots, n$ are drawn from a normal distribution with mean $\mu$ and variance $\sigma^2$. We let $\Phi(\cdot)$ and $\phi(\cdot)$ denote their cdf and pdf. Each agent is strategic with probability $\delta \in (0,1)$. We are interested in asymptotic results as $\delta$ goes to $1$, which means that the fraction of behavioral agents is arbitrarily small.

\paragraph{Payoffs:} The principal's realized payoff is
\begin{equation}\label{2.1}
   \mathbf{1} \{ t=t^o \} \cdot \sum_{i=1}^n \theta_i - sL,
\end{equation}
which means that a virtuous principal does not benefit from committing offenses, and the opportunistic type trades off the benefit (normalized to $1$) from committing each offense with the cost of being convicted $L>0$.

If agent $i$ is strategic, his realized payoff is
\begin{equation}\label{2.2}
u_i(\omega_i,\theta_i,a_i)  \equiv \left\{ \begin{array}{ll}
0 & \textrm{ if} \quad s=1 \\
\omega_i - b
    \Big( (1-\gamma) \theta_i +\gamma f(\theta_i,\theta_{-i}) \Big) -ca_i & \textrm{ if} \quad s=0,
\end{array} \right.
\end{equation}
where $b>0$, $c>0$, and $\gamma \in [0,1]$ are parameters, and $f(\theta_i,\theta_{-i})$ is strictly increasing in both arguments.

This specification normalizes to 0 agents' utility when the principal is convicted. This is without loss of generality for the purpose of analyzing agents' incentives, since an agent's reporting decision depends only on the \textit{difference} between his utility when the principal is convicted and his utility when the principal is acquitted.\footnote{This formulation does rule out the possibility that agents have a preference for reporting the truth irrespective of the adjudication outcome. We study to this extension in Section \ref{sec6} and generalize our main results.}

The parameter $b>0$ captures each agent's preference for punishing offenders; $c>0$ captures the retaliation or social stigma cost that an agent incurs if he accuses the principal but fails to convict him;\footnote{In the baseline model, we normalize the social stigma (or loss from retaliation) to zero when the principal is convicted. Our results hold as long as the social stigma (or loss from retaliation) is strictly larger when the principal is acquitted rather than convicted.} and $\gamma \in [0,1]$ captures the intensity of an agent's \textit{social preference}, i.e.,
the extent to which his payoff internalizes
offenses committed against (or witnessed by) other agents. Finally, $\omega_i$ is an idiosyncratic shock that captures agent~$i$'s preference for acquitting the principal. A lower $\omega_i$ means that agent $i$ has a stronger incentive to accuse the principal.

We compare equilibrium outcomes when the judge uses either APP or DPP to adjudicate guilt. These rules differ only in terms of how to measure the probability of guilt when the principal is capable of committing \textit{multiple offenses}. Both rules may be microfounded by assigning some appropriate quadratic payoff function to the judge. We let $\pi^* \in (0,1)$ denote the standard of proof used by the judge to adjudicate guilt. We assume that $\pi^*$ is exogenous although it, too, could be microfounded by specifying costs for making type I and type II errors. Let
\begin{equation}\label{2.3}
\overline{\theta} \equiv \max_{i \in \{1,2,...,n\}} \theta_i.
\end{equation}
The principal has committed \textit{at least one offense} if $\overline{\theta}=1$, and no offense if $\overline{\theta}=0$.

The first adjudication rule, APP, requires that
\begin{equation}\label{2.4}
s  \left\{ \begin{array}{ll}
=1 & \textrm{ if} \quad \Pr\Big(\overline{\theta}=1\Big|\boldsymbol{a}\Big) > \pi^*\\
\in \{0,1\} & \textrm{ if} \quad \Pr\Big(\overline{\theta}=1\Big|\boldsymbol{a}\Big) = \pi^*\\
=0 & \textrm{ if} \quad \Pr\Big(\overline{\theta}=1\Big|\boldsymbol{a}\Big) < \pi^*.
\end{array} \right.
\end{equation}
This rule convicts the principal when the probability that he has committed \textit{at least one offense} exceeds the exogenous threshold $\pi^*$. This rule has been advocated when an individual is accused of wrongdoing outside the criminal process, for example when managers are charged with discriminating against minority workers, or when supervisors are charged with abusing their subordinates. Schauer and Zeckhauser (1996) argue that the aggregation of multiple low-probability accusations to discipline agents is often appropriate in corporate and other non-judicial settings. Harel and Porat (2009, p.~263) advocate the use of APP in some judicial settings, observing that {\em acquitting defendants who are almost surely guilty of some unspecified crime is neither just nor efficient}.

The second adjudication rule, DPP, requires that
\begin{equation}\label{2.5}
s  \left\{ \begin{array}{ll}
=1 & \textrm{ if} \quad \max_{i \in \{1,2,...,n\}} \Pr\big(\theta_i=1\big|\boldsymbol{a}\big) > \pi^*\\
\in \{0,1\} & \textrm{ if} \quad \max_{i \in \{1,2,...,n\}} \Pr\big(\theta_i=1\big|\boldsymbol{a}\big) = \pi^*\\
=0 & \textrm{ if} \quad \max_{i \in \{1,2,...,n\}} \Pr\big(\theta_i=1\big|\boldsymbol{a}\big) < \pi^*.
\end{array} \right.
\end{equation}
DPP is close to the criterion used to adjudicate guilt by most criminal justice systems when  a defendant is charged with multiple offenses: the court examines each charge individually and convicts the defendant if the probability that the defendant committed some {\em specific} offense exceeds some evidentiary threshold.\footnote{Actual sentencing procedures are of course more complex. In some cases, sentences are cumulated across offenses for which the defendant is found guilty; in others, they are substitutes and only the most severe punishment among the charges for which the defendant was found guilty is meted out. Our analysis can accommodate these variations, as explained in Section \ref{sec6}. The key feature that we, and legal scholars before us, emphasize is that DPP treats charges independently of one another to adjudicate guilt.}

When there is only one agent, (\ref{2.4}) and (\ref{2.5}) coincide. When there are two or more agents, (\ref{2.4}) and (\ref{2.5}) are distinct criteria. In the example presented in the introduction, using $\pi^*=0.9$, a defendant who commits two offenses with probability 0.8 each is convicted under decision rule (\ref{2.4}) but acquitted under decision rule (\ref{2.5}).

\paragraph{Solution Concept:} We focus on proper equilibria (Myerson 1978) that satisfy two refinements, described below, which will simply be referred to as \textit{equilibria}. Formally,
a proper equilibrium consists of a strategy profile $\big\{ \sigma^o, \sigma^v, (\sigma_i)_{i=1}^n, q \big\}$ where
\begin{itemize}
  \item $\sigma^o \in \Delta \big(\{0,1\}^n \big)$ and $\sigma^v  \in \Delta \big(\{0,1\}^n \big)$  are the strategies followed by an opportunistic type and a virtuous type of principal, respectively;
  \item $\sigma_i: \mathbb{R} \times \{0,1\} \rightarrow \Delta \{0,1\}$ is agent $i$'s strategy when he is strategic and maps each realization of $\omega_i$ and $\theta_i$ into a probability of accusing the principal;
  \item $q: \{0,1\}^n \rightarrow [0,1]$ is the judge's strategy and defines the probability that the judge convicts the principal after observing $\boldsymbol{a} \in \{0,1\}^n$: $q(\boldsymbol{a}) = \Pr(s = 1|\boldsymbol{a})$.
\end{itemize}
Our first refinement requires that the principal be acquitted if no agent accuses him.
\begin{Refinement}[No Conviction Unless Accused]
$q(0,0,...,0)=0$.
\end{Refinement}
Refinement 1 rules out equilibria in which an opportunistic principal commits offenses against all agents with probability one and is convicted with probability one regardless of agents' reports. In these equilibria, the principal is convicted on the sole basis of a judge's prior belief. These equilibria violate the legal principle of the presumption of innocence. Our second refinement confers their meaning to agents' messages.
\begin{Refinement}[Monotonicity]
For every $i \in \{1,2,...,n\}$, $q(1,a_{-i}) \geq q(0,a_{-i})$ for every $a_{-i} \in \{0,1\}^{n-1}$, and there exists $a_{-i} \in \{0,1\}^{n-1}$ such that $q(1,a_{-i}) > q(0,a_{-i})$.
\end{Refinement}
Refinement 2 requires, first, that the conviction probability be nondecreasing in the set of agents who accuse the principal and, second, that each agent's report have a \textit{nontrivial influence} on the adjudication outcome, for at least one reporting profile of other agents. This refinement implies that
each accusation is a move against the principal, and it is consistent with our interpretation of $c$ as a retaliation cost. Intuitively, a principal retaliates against messages that reduce his payoff and does not retaliate against other messages.\footnote{A microfoundation for this monotonicity refinement is provided by Chassang and Padr\'{o} i Miquel (2019), in which the principal optimally commits to a retaliation plan $\widetilde{c}_i : \{0,1\} \rightarrow [0,c]$ privately against agent $i$, which maps agent $i$'s report to his loss from the principal's retaliation. Retaliation can only be carried out when
the principal is acquitted, and $c$ is the maximal damage the principal can inflict on each agent. The principal's optimal strategy is to inflict maximal retaliation against a message that increases his probability of conviction and zero retaliation against the other message.}

Lemma \ref{L3.1} establishes the existence of a proper equilibrium that satisfies both refinements, and provides preliminary observations that apply to both APP or DPP and to \textit{any number of agents}.
\begin{Lemma}\label{L3.1}
For every punishment $L$ above some threshold $\overline{L}\in \mathbb{R}_+$, there exists a proper equilibrium that satisfies Refinements 1 and 2. In every such equilibrium:
\begin{enumerate}
\item For every $i \in \{1,2,...,n\}$, agent $i$'s strategy is characterized by cutoffs $\omega_i^* >\omega_i^{**}$ such that agent $i$ accuses the principal when either $\theta_i = 1$ and $\omega_i \leq \omega_i^*$ or $\theta_i=0$ and $\omega_i \leq \omega_i^{**}$.
  \item The prior probability $\Pr\big(\overline{\theta}= 1 \big)$ that the principal commits at least one offense is strictly between $0$ and $1$.
 \item A virtuous principal never commits any offense.
\end{enumerate}
\end{Lemma}
The proof appears in Appendix A, except for the proof of existence, which is in Supplementary Appendix J. The  existence of an equilibrium satisfying Refinement 1 requires $L$ to be sufficiently large. For example, if $L<1$ and $\pi^o > \pi^*$, the principal commits an offense against all agents with probability $1$ when he is opportunistic, in every proper equilibrium, and is convicted even when no agent accuses him. In this case, no Nash equilibrium satisfies Refinement 1. To get a sense of how large $L$ has to be for the existence of an equilibrium, we performed some numerical simulations. For example, when $\pi^*=0.95$, $\omega_i$ follows a normal distribution with mean $0$ and variance $1$, the fraction of strategic agents is $\delta=0.95$, $b=1$ and $c=10$, a proper equilibrium that satisfies both refinements exists when $L \geq 5$.

\subsection{Discussion of the Assumptions}

\paragraph{Hidden vs. Public Reports} For tractability, our analysis is conducted under the assumption that agents' reports are public. The issues exposed in our analysis are not easily dispelled by relaxing this assumption. To see this, consider the case of two agents and suppose that agents' reports are made public only if both agents accuse the principal. This rids agents from the risk of social stigma or retaliation but, by the same token, allows agents to accuse the principal with impunity and replaces false negatives with false positives. It is easy to see that hiding reports in this way can never lead to highly informative reports and low offense frequency (see Remarks 1 and 2). We show that the effects of retaliation (or social stigma) costs are subtle and depend critically on the correlation between agents' private information: a positive retaliation cost
undermines the informativeness of agents' reports when agents' private information is negatively correlated (Theorem~\ref{Theorem1}), and
improves the informativeness of reports when private information is positively correlated (Proposition~\ref{Prop3}).

\paragraph{Simultaneous-Move \textit{vs} Sequential-Move:} In our baseline model, agents simultaneously decide whether to file accusations against a potential offender. In practice, such decisions are often made \textit{sequentially} and agents can often choose not only whether to file an accusation but also {\em when} to file it. We explain in Section \ref{sec7} that the forces underlying our results also apply when reporting is sequential.

To illustrate this, suppose that there are two agents, who make their reports in a predetermined order. An agent who witnessed an offense and who is the first to report may be concerned that the second agent will not accuse the principal, which is more likely if offenses are negatively correlated. Moreover, the second-reporting agent will not accuse the principal, regardless of what he has witnessed, if the first-moving agent remains silent. Similar issues arise if agents' order of move is uncertain.

\paragraph{General Punishment Function:} In our baseline model, the conviction decision is binary. In principle, the punishment administered to a convicted defendant could depend on the entire probability distribution of the number of offenses that the defendant is guilty of committing. For example, a defendant almost surely guilty of {\em at least two offenses} could be given a higher punishment than a defendant almost surely guilty of just one offense.

Our main result is robust with respect to such concerns. Theorem 1 establishes that when the punishment in case of conviction is large enough, an opportunistic principal commits \textit{at most one} offense in equilibrium even when the punishment for committing multiple offenses is the same as the punishment for committing a single offense. The incentives to commit two or more offenses are {\em a fortiori} even weaker when the punishment for defendants likely of committing multiple offenses is higher, which reinforces our results.

\section{Main Results}\label{sec3}

We show that when there are \textit{multiple potential witnesses}, APP leads to uninformative reports and ineffective deterrence (Theorem \ref{Theorem1}), while DPP leads to informative reports and effective deterrence (Theorem \ref{Theorem2}).

To highlight the forces at play in their simplest form, this section assumes that the principal is surely opportunistic, i.e., $\pi^o=1$, agents do not directly care about offenses that they cannot observed, i.e., $\gamma=0$, and compares decision rules (\ref{2.4}) and (\ref{2.5}) when there are \textit{one or two potential witnesses}. The results are generalized to multiple principal types in Section \ref{sec4}, three or more agents in Section \ref{sec5}, and agents who directly care about all offenses in Section~\ref{sec6}.

\subsection{Benchmark: Single Potential Witness}\label{sec3.2}
When there is only one agent,
Refinement 1 implies that the principal may be convicted only if the agent accuses him, i.e., $q(0) = 0$ and $q(1) > 0$. If $\theta=1$, the agent accuses the principal when
\begin{equation}\label{3.1}
    \omega-b \leq (1-q(1)) (\omega-b-c) \quad \textrm{or, equivalently,} \quad \omega \leq \omega^* \equiv b-c\frac{1-q(1)}{q(1)}.
\end{equation}
If $\theta=0$, the agent accuses the principal when
\begin{equation}\label{3.2}
  \omega \leq (1-q(1)) (\omega-c) \quad \textrm{or, equivalently,} \quad  \omega \leq \omega^{**} \equiv -c\frac{1-q(1)}{q(1)}.
\end{equation}
Let $\Pr(\boldsymbol{\theta}=1)$ be the prior probability that the principal commits the offense and
$\Pr(\boldsymbol{\theta}=1|\boldsymbol{a}=1)$ be the judge's posterior belief when the agent accuses the principal. Bayes rule implies that
\begin{equation}\label{3.3}
 \underbrace{\frac{\Pr(\textrm{agent reports }| \textrm{ }\theta=1)}{\Pr(\textrm{agent reports }|\textrm{ }\theta=0)} }_{\equiv \mathcal{I}}   \cdot \frac{\Pr(\boldsymbol{\theta}=1)}{1-\Pr(\boldsymbol{\theta}=1)} = \frac{\Pr(\boldsymbol{\theta}=1|\boldsymbol{a}=1)}{1-\Pr(\boldsymbol{\theta}=1|\boldsymbol{a}=1)}.
\end{equation}
The ratio $\mathcal{I}$ measures the \textit{informativeness} of the agent's accusation about the principal's guilt.
\begin{Proposition}\label{Prop1}
Suppose that $n=1$.
\begin{enumerate}
  \item There exists $\overline{L} >0$ such that for any $L > \overline{L}$ and any equilibrium, the judge assigns probability $\pi^*$ to the principal being guilty when the agent accuses the principal.
  \item As $L$ goes to $+\infty$ and the fraction of behavioral agents $1-\delta$ goes to 0,\footnote{Throughout the paper, the limit is first taken first with respect to $\delta$ and then with respect to $L$. For example, Proposition~\ref{Prop1} states that $\lim_{L\rightarrow +\infty} (\lim_{\delta \rightarrow 1} \Pr(\boldsymbol{\theta}=1)) = 0$.} The informativeness ratio $\mathcal{I}$ goes to $+\infty$ and the prior probability of offense $\Pr(\boldsymbol{\theta}=1)$ goes to $0$.
\end{enumerate}
\end{Proposition}
The proof is in Appendix B. Proposition~\ref{Prop1} shows that when the punishment in case of conviction is large and there can be at most one accusation against the principal, the outcome of any equilibrium has highly desirable properties: accusations are arbitrarily informative and the frequency of offenses becomes arbitrarily close to zero.

Proposition \ref{Prop1} may be understood as follows: In equilibrium, the ex ante probability of offense is interior (Lemma \ref{L3.1}), and the principal must be indifferent between committing an offense and abstaining from it. This indifference condition implies that $q(1)$ vanishes to $0$ as $L$ goes to infinity. From (\ref{3.1}) and (\ref{3.2}), the agent's reporting cutoffs are decreasing in $L$, but  their distance $\omega^*-\omega^{**}$ remains constant and equal to $b$.

As the fraction $1-\delta$ of behavioral agents goes to $0$, the informativeness ratio
$\mathcal{I} $ becomes approximately equal to $\Phi(\omega^*)/\Phi(\omega^*-b)$.
Since $\omega^*-\omega^{**}=b$, the ratio
$\Phi(\omega^*)/\Phi(\omega^*-b)$
goes to infinity as $\omega^* \rightarrow -\infty$.\footnote{This is a property of the Gaussian distribution and, more generally, of thin-tailed distributions.} Since $q(1)$ is strictly between $0$ and $1$ when $L$ is large enough, the judge's posterior belief equals $\pi^*$ after observing the agent's report. Equation (\ref{3.3}) then implies that the prior probability of offense vanishes to $0$.

\paragraph{Remark 1:} Proposition~\ref{Prop1} fails when the agent's retaliation cost $c$ is exactly equal to zero. In this case, it is straightforward to see that the agent's accusation thresholds are fixed, equal to $b$ and $0$, independently of $L$. This implies that the prior probability of committing an offense is also fixed in equilibrium, independently of $L$.\footnote{This must be true to maintain a judge's incentive to convict the principal with interior probability when the agent accuses him.} Introducing a slightly positive retaliation cost when $L$ is large leads to arbitrarily informative reports and reduces the probability of offense to arbitrarily low levels in equilibrium.

\subsection{Two Potential Witnesses: Aggregating the Probabilities of Distinct Offenses}\label{sec3.3}
Suppose now that there are two agents and that adjudication rule (\ref{2.4}) is used. Recalling that $\overline{\theta}$ describes whether at least one offense took place (from equation~(\ref{2.3})), Bayes rule implies that for every $\boldsymbol{a} \in \{0,1\}^2$,
\begin{equation}\label{3.4}
\frac{\Pr(\boldsymbol{a}|\overline{\theta}=1)}{\Pr(\boldsymbol{a}|\overline{\theta}=0)} \cdot \frac{\Pr(\overline{\theta}=1)}{1-\Pr(\overline{\theta}=1)}=
\frac{\Pr(\overline{\theta}=1|\boldsymbol{a})}{1-\Pr(\overline{\theta}=1|\boldsymbol{a})}.
\end{equation}
The ratio
\begin{equation}\label{3.5}
\mathcal{I} (\boldsymbol{a})  \equiv  \frac{\Pr(\boldsymbol{a}|\overline{\theta}=1)}{\Pr(\boldsymbol{a}|\overline{\theta}=0)},
\end{equation}
is a sufficient statistic for the judge's posterior belief after observing $\boldsymbol{a}$, and therefore, it measures the \textit{informativeness} of reporting profile $\boldsymbol{a}$.
\begin{Theorem}\label{Theorem1}
When $n=2$ and the judge uses decision rule (\ref{2.4}), there exists $\overline{L}$ such that when $L > \overline{L}$,
\begin{enumerate}
  \item \textbf{Endogenous Negative Correlation Between Offenses:} In every equilibrium
  \begin{equation}\label{3.6}
    \Pr(\theta_1=1|\theta_2=1)<\Pr(\theta_1=1|\theta_2=0) \textrm{ and } \Pr(\theta_2=1|\theta_1=1)<\Pr(\theta_2=1|\theta_1=0).
  \end{equation}
\end{enumerate}
For every $\varepsilon>0$, there exists $\overline{L}_{\varepsilon} \in \mathbb{R}_+$, such that
in every equilibrium when $L>\overline{L}_{\varepsilon}$ and $\delta \in (0,1)$,
\begin{enumerate}
  \item[2.] \textbf{Low Informativeness of Reports \& Ineffective Deterrence:} $\max_{\boldsymbol{a} \in \{0,1\}^2} \mathcal{I} (\boldsymbol{a})<1 +\varepsilon$.
\item[3.]  \textbf{Ineffective Deterrence:} $\Pr(\overline{\theta}=1)>\pi^*-\varepsilon$.
\end{enumerate}
\end{Theorem}
Theorem \ref{Theorem1} shows that when conviction entails a large punishment relative to the principal's benefit from committing an offense, aggregating the probabilities of distinct offenses to adjudicate guilt induces \textit{negative correlation} across the occurrences of distinct offenses. This negative correlation undermines the informativeness of agents' reports and results in a high probability of offenses.
As $L \rightarrow +\infty$,\footnote{Theorem \ref{Theorem1} is stated for the same order of limits as the one used in Proposition \ref{Prop1}, i.e., $\lim_{L \rightarrow \infty} \lim_{\delta \rightarrow 1}$.}
agents' reports become arbitrarily uninformative and the equilibrium probability of offense converges to $\pi^*$. These conclusions stand in sharp contrast to the single-agent benchmark, in which the probability of offense goes to $0$ and the informativeness ratio goes to infinity as $L$ becomes arbitrarily large.

\paragraph{Remark 2:} Theorem 1 requires that the cost of social stigma or retaliation $c$ be strictly positive but does not impose any lower bound on $c$. In practice, one may devise mechanisms that shield to some extent agents from stigma and retaliation. For example, agents reports may be kept private except if multiple agents the principal. Even in this case, the expected cost of retaliation is still strictly positive as long as there is a strictly positive probability that information is leaked to the principal.

Theorem \ref{Theorem1} suggests that deterrence can be improved by lowering the conviction threshold $\pi^*$, which was taken as exogenously given. However, lowering $\pi^*$ increases the risk of convicting an innocent defendant. In our model, the probability that a convicted defendant is innocent equals $1-\pi^*$ in equilibrium. For example, if any judge were to use a conviction threshold $\pi^*$ of $10\%$, then in every equilibrium, each convicted individual has a $90\%$ chance of being innocent. In general, the conviction threshold may be chosen so as to strike a compromise between deterrence and wrongful convictions, and different values of $\pi^*$ may be justified by  different welfare costs of punishing the innocent.

One may conjecture that the lower witness credibility that arises with two agents is driven by agents' incentives to free-ride on others' reports, which is the intuition behind inefficient public good provision (e.g., Chamberlin 1974). The following comparative static result refutes this conjecture by showing that, other things equal, increasing the number of agents results in a {\em strictly higher probability} that agents accuse the principal in every equilibrium. This comparative static is generalized to three or more agents in Theorem \ref{Theorem3}.
\begin{Proposition}\label{Prop2}
When the judge uses decision rule (\ref{2.4}), there exists $\overline{L}>0$ such that the following holds for every $L > \overline{L}$: Let $\omega_{i,2}^*(L)$ and $\omega_{i,2}^{**}(L)$ denote agent $i$'s accusation cutoffs in some equilibrium of the two-agent setting and $\omega^*(L)$ and $\omega^{**}(L)$ are the accusation cutoffs in some equilibrium of the single-agent setting. Then, $\omega_{i,2}^* (L) > \omega^*(L)$ and $\omega_{i,2}^{**} (L) > \omega^{**}(L)$ for every $i \in \{1,2\}$.
\end{Proposition}

The remaining of this section explains in more detail why agents' private observations are negatively correlated and how this negative correlation interacts with agents' coordination motive.

\paragraph{Step 1: Equilibrium Conviction Probability} We first explain why, when $L$ is large enough, the principal is convicted with positive probability \textit{only when}
he is accused by both agents:
\begin{Lemma}\label{L3.2}
There exists $\overline{L} \in \mathbb{R}_+$ such that for any $L >\overline{L}$ and any corresponding equilibrium,
\begin{equation}\label{3.7}
    q(0,0)=q(1,0)=q(0,1)=0 \textrm{ and } q(1,1) \in (0,1).
\end{equation}
\end{Lemma}
This result is formally proved in Online Appendix F. The argument proceeds by contradiction: Suppose that a single accusation suffices to convict the principal with strictly positive probability. Since each agent is strictly more likely to accuse the principal when he has witnessed an offense, each additional accusation increases the probability that the principal has committed {\em at least one} offense. Since a Bayesian judge weakly prefers to convict the principal with only one accusation, she strictly prefers to convict him if both agents file accusations.

Building on this observation, we show in the appendix that the incremental probability that the principal is convicted when he commits an additional offense (whether starting from zero offense or from a single offense) is uniformly bounded below away from zero. As the punishment $L$ in case of conviction becomes arbitrarily large, this implies that the marginal cost from committing an additional offense must  exceed the marginal benefit from the offense, which gives the principal a strict incentive not to commit any offense. This contradicts Lemma \ref{L3.1}, which states that the equilibrium probability of offense is strictly positive, and, hence, shows that (\ref{3.7}) must hold.

\paragraph{Step 2: Principal's Incentives \& Endogenous Negative Correlation} We examine the principal's incentives to commit offenses and establish a \textit{necessary and sufficient condition} for his choices of $\theta_1$ and $\theta_2$ to be strategic substitutes or strategic complements.
\begin{Lemma}\label{L3.3}
When the principal is opportunistic, offenses $\theta_1$ and $\theta_2$ are strategic substitutes if and only if
\begin{equation}\label{3.8}
    q (1,1)+q (0,0)-q (1,0)-q (0,1)>0
\end{equation}
and strategic complements if and only if $q (1,1)+q (0,0)-q (1,0)-q (0,1)<0$.
\end{Lemma}
\begin{proof}[Proof of Lemma 3.2:]
For $i \in \{1,2\}$, let $\Psi_i^* \equiv \delta \Phi(\omega_i^*)+(1-\delta) \alpha$ and let
$\Psi_i^{**} \equiv \delta \Phi(\omega_i^{**})+(1-\delta) \alpha$.
The difference in the probability of conviction conditional on $(\theta_1,\theta_2)=(0,0)$ and on $(\theta_1,\theta_2)=(1,0)$ is
\begin{equation}\label{3.9}
(\Psi_1^*-\Psi_1^{**})
\Big(
(1-\Psi_2^{**}) \big(q (1,0)-q (0,0)\big)
+
\Psi_2^{**} \big(q (1,1)-q (0,1)\big)
\Big),
\end{equation}
while the difference in the probability of conviction conditional on
$(\theta_1,\theta_2)=(0,1)$ and on $(\theta_1,\theta_2)=(1,1)$ is
\begin{equation}\label{3.10}
(\Psi_1^*-\Psi_1^{**})
\Big(
(1-\Psi_2^{*}) \big(q (1,0)-q (0,0)\big)
+
\Psi_2^{*} \big(q (1,1)-q (0,1)\big)
\Big).
\end{equation}
Committing different offenses are strategic substitutes if and only if
(\ref{3.9}) is less than (\ref{3.10}) or, equivalently, if
\begin{equation*}
    (\Psi_1^*-\Psi_1^{**})(\Psi_2^*-\Psi_2^{**}) \Big(q (1,0)+q(0,1)-q(0,0)-q (1,1)\Big) < 0.
    \end{equation*}
Since $\omega_i^* > \omega_i^{**}$ for every $i \in \{1,2\}$, we have
$(\Psi_1^*-\Psi_1^{**})(\Psi_2^*-\Psi_2^{**})>0$, which implies
(\ref{3.8}).
\end{proof}
Lemmas~\ref{L3.2} and~\ref{L3.3} imply that in every equilibrium with $L$ large enough, conviction probabilities satisfy
(\ref{3.8}). This implies that the principal views offenses as strategic substitutes and, in particular, that he
\textit{cannot} be indifferent between committing no offense and committing two offenses. Since the equilibrium probability of offense is interior, by Lemma~\ref{L3.1}, the principal must be indifferent between committing no offense and committing one offense.\footnote{When the principal can be either opportunistic or virtuous, and the probability that the principal is virtuous exceeds $1-\pi^*$, we show in Section \ref{sec4} that the  opportunistic type must be indifferent between committing one and two offenses in equilibrium, and commits two offenses with strictly positive probability. Nevertheless, we show that agents' private observations remain \textit{negatively correlated} in this case.}
This shows that $\theta_1$ and $\theta_2$ are negatively correlated.

\paragraph{Step 3: Agents' Coordination Motive} From~\eqref{3.7}, the principal is convicted only if both agents accuse him. Therefore, an agent's incentive to accuse the principal is stronger when he expects the other agent to accuse the principal with higher probability. Formally, if $\theta_i=1$, then agent $i$ prefers to accuse the principal when:
\begin{equation}\label{3.11}
    \omega_i \leq \omega_i^* \equiv b -c \frac{1-q(1,1) Q_{1,j}}{q (1,1) Q_{1,j}}=b+c-\frac{c}{q(1,1) Q_{1,j}}
\end{equation}
where $Q_{1,j}$ is the probability that agent $j \neq i$ accuses the principal \textit{conditional on} $\theta_i=1$.
Similarly, if $\theta_i=0$, then agent $i$ prefers to accuse the principal when:
\begin{equation}\label{3.12}
    \omega_i \leq \omega_i^{**} \equiv -c \frac{1-q(1,1) Q_{0,j}}{q(1,1) Q_{0,j}}=c-\frac{c}{q(1,1) Q_{0,j}}
\end{equation}
where $Q_{0,j}$ is the probability that agent $j \neq i$ accuses the principal \textit{conditional on} $\theta_i=0$.
One can verify that $\omega_i^*$ is strictly increasing in $Q_{1,j}$, and $\omega_i^{**}$ is strictly increasing in $Q_{0,j}$.

Since each agent is more likely to accuse the principal when he has witnessed an offense (Lemma~\ref{L3.1}), the negative correlation between $\theta_1$ and $\theta_2$ implies that $Q_{1,j}<Q_{0,j}$ and, hence, that
$\omega_i^*-\omega_i^{**} \in (0,b)$. We show in Appendix C that $\omega_i^*-\omega_i^{**} \rightarrow 0$ as $L \rightarrow \infty$.
This stands in sharp contrast to the single-agent benchmark, in which the distance between the two reporting cutoffs is equal to $b$ regardless of the other parameters of the model. The decrease in the distance between reporting cutoffs
undermines the informativeness of agents' reports as measured by
$\max_{\boldsymbol{a} \in \{0,1\}^2} \mathcal{I} (\boldsymbol{a})$.
This informativeness measure converges to $1$, which corresponds to completely uninformative reports, as $L \rightarrow +\infty$. From~(\ref{3.7}), the judge's assigns a posterior probability of guilt equal to $\pi^*$ after observing $\boldsymbol{a}=(1,1)$. From~(\ref{3.4}),
\begin{equation*}
  \underbrace{\frac{\Pr(\boldsymbol{a}=(1,1)|\overline{\theta}=1)}{\Pr(\boldsymbol{a}=(1,1)|\overline{\theta}=0)}}_{\equiv \mathcal{I}(1,1)=\max_{\boldsymbol{a} \in \{0,1\}^2} \mathcal{I}(\boldsymbol{a})} \cdot \frac{\Pr(\overline{\theta}=1)}{1-\Pr(\overline{\theta}=1)}=\frac{\pi^*}{1-\pi^*},
\end{equation*}
which implies that the prior probability of  offense $\Pr(\overline{\theta}=1)$ must be close to $\pi^*$ when $L$ is large enough.


\subsection{Two Potential Witnesses: Separating the Probabilities of Distinct Offenses}\label{sec3.4}
We now turn to equilibrium outcomes when there are two agents and the judge uses DPP to adjudicate guilt, i.e., conviction is determined based on the maximal probability that the principal is guilty of some \textit{specific} offense.
\begin{Theorem}\label{Theorem2}
When $n=2$ and the judge uses decision rule (\ref{2.5}), there exists $\overline{L} \in \mathbb{R}_+$ such that for every $L > \overline{L}$ and every equilibrium,
\begin{enumerate}
  \item \textbf{Uncorrelated Offenses:} , $\Pr(\theta_i=1|\theta_j=1)=\Pr(\theta_i=1|\theta_j=0)$ for every $i \neq j$.
\item \textbf{Linear Conviction Probability:}  $\Pr(\theta_i=1|a_i=1)=\pi^*$ for every $i \in \{1,2\}$, and the conviction probability is linear in the number of accusations.
\end{enumerate}
As the fraction of behavioral agents goes to zero and $L\rightarrow +\infty$, the following asymptotic results hold:\footnote{As noted in Proposition \ref{Prop1}, the results are obtained by first taking the limit with respect to $\delta$ and then with respect to $L$.}
\begin{enumerate}
  \item[3.] \textbf{Effective Deterrence:} The equilibrium probability of offense
  $\Pr(\overline{\theta}=1)$
  converges to $0$.
  \item[4.] \textbf{Highly Informative Reports:} For every $i \in \{1,2\}$, the informativeness of agent $i$'s report about $\theta_i$, measured by $\frac{\Pr(a_i=1|\theta_i=1 )}{\Pr(a_i=1|\theta_i=0)}$,
goes to $+\infty$.\footnote{The same conclusion applies under other measures of informativeness, for example, the one
developed in Theorem \ref{Theorem1}, in which case one can show that
$\max_{\boldsymbol{a} \in \{0,1\}^2} \mathcal{I} (\boldsymbol{a})$ also goes to  infinity in the $L \rightarrow \infty$ limit.}
\end{enumerate}
\end{Theorem}
The proof is in Appendix D. Theorem \ref{Theorem2} shows that the offenses observed by distinct agents are \textit{uncorrelated}, and that the probability of convicting the principal is linear in the number of accusations. From Lemma \ref{L3.3}, this implies that the principal's incentives to commit distinct offenses are neither complements nor substitutes and, consequently, the principal's incentives to commit distinct offenses operate independently of each other. This feature preserves the credibility of the agents' reports and lowers the probability of offense in equilibrium.

We explain the intuition underlying Theorem \ref{Theorem2} in three steps, focusing primarily on why agents' private observations of offense are uncorrelated and why uncorrelated private information leads to linear conviction probabilities, informative reports, and effective deterrence in equilibrium.

\paragraph{Ruling Out Correlation:}
Suppose first, by way of contradiction, that $\theta_1$ and $\theta_2$ are \textit{negatively correlated}. In this case,
\begin{equation*}
    \Pr(\theta_1=1|\boldsymbol{a}=(1,1)) <  \Pr(\theta_1=1|\boldsymbol{a}=(1,0))
\end{equation*}
because an accusation by agent $2$ increases the probability that $\theta_2=1$ and, given the negative correlation between offenses, reduces the probability that $\theta_1=1$. A similar logic implies that $\Pr(\theta_2=1|\boldsymbol{a}=(1,1)) <  \Pr(\theta_2=1|\boldsymbol{a}=(0,1))$.
Under decision rule (\ref{2.5}), the above inequalities imply that $q(1,0) \geq q(1,1)$ and $q(0,1) \geq q(1,1)$. For this to satisfy Refinement 2, it has to be the case that
\begin{equation}\label{3.14}
q(1,1)=q(1,0)=q(0,1)=1 \textrm{ and } q(0,0)=0.
\end{equation}
When $L$ is large enough, one can show similarly to Lemma \ref{L3.2} that under such conviction probabilities, the principal has a strict incentive not to commit any offense, which contradicts Lemma~\ref{L3.1}'s claim that the equilibrium probability of offense is interior.

Next, suppose that $\theta_1$ and $\theta_2$ are \textit{positively correlated}, in which case
$\boldsymbol{a}=(1,1)$ is the unique maximizer of both
$\Pr\big(\theta_1=1\big|\boldsymbol{a}\big)$ and $\Pr\big(\theta_2=1\big|\boldsymbol{a}\big)$.
Under decision rule
(\ref{2.5}), we have
\begin{equation}\label{3.15}
q(1,1) \geq \max\{q(1,0),q(0,1)\} \geq q(0,0)=0
\end{equation}
where the first inequality is strict unless $\max\{q(1,0),q(0,1)\}=1$.
When $L$ is large enough, the same logic as in Lemma \ref{L3.2} shows that the principal has an incentive to commit an offense only if $q(1,1) \in (0,1)$ . The positive correlation between $\theta_1$ and $\theta_2$ then implies that $q(0,1)=q(1,0)=0$. This, together with Lemma \ref{L3.3}, shows that the principal's decisions to commit offenses are \textit{strategic substitutes} and contradicts the hypothesis that $\theta_1$ and $\theta_2$ are positively correlated.

\paragraph{Linear Conviction Probabilities:} Since $\theta_1$ and $\theta_2$ are \textit{uncorrelated}, the principal's decisions to commit distinct offenses
are neither substitutes or complements.
Lemma \ref{L3.3} and Refinement 1 then imply that $q(1,1)=q(1,0)+q(0,1)$.
Since the equilibrium probability of offenses is interior (Lemma \ref{L3.1}), the principal plays a completely mixed strategy in equilibrium. Taken together, these observations imply that conviction probabilities are symmetric, i.e., $q(1,0)=q(0,1)$. Otherwise, the principal would strictly prefer $\boldsymbol{\theta}=(1,0)$  to $\boldsymbol{\theta}=(0,1)$ or vice versa.

\paragraph{Informative Report \& Effective Deterrence:} The absence of correlation between $\theta_1$ and $\theta_2$ implies that an agent's belief about whether the other agent has witnessed an offense is independent of the first agent's observation.
From~(\ref{3.11}) and~(\ref{3.12}), the independence of $\theta_1$ and $\theta_2$ implies that $Q_{1,j}=Q_{0,j}$ for $j \in \{1,2\}$, which further implies that $\omega_j^*-\omega_j^{**}=b$.

As in the single-agent case,
agents' reporting cutoffs converge to $-\infty$ as $L$ becomes arbitrarily large, and the informativeness ratio of each agent's accusation, which is approximately equal to $\Phi(\omega^*)/\Phi(\omega^{*}-b)$, goes to $+\infty$. Since the judge's posterior about $\theta_i$ is equal to $\pi^*$ after observing $a_i=1$, and since the informativeness ratio goes to $+\infty$, the prior probability that $\theta_i=1$ must converge to $0$ as $L$ becomes arbitrarily large.

\paragraph{Remark:} Theorems~\ref{Theorem1} and~\ref{Theorem2} provide a justification for the use of DPP in criminal justice systems. They explain why aggregating probabilities of different offenses to adjudicate guilt can severely distort incentives of various actors and offer a rationale for prohibiting character evidence to establish the guilt of a defendant.

Implementing DPP in practice requires commitment power on the part of the adjudicator (judge, board of trustees of a firm, etc.). This aspect may be illustrated by the following thought experiment: Consider a setting with two defendants
and three plaintiffs. The offenses committed by Defendant~1 against plaintiffs 1 and 2 are perfectly negatively correlated, described in the following table:
{\small \begin{center}
\begin{tabular}{| c | c | c | c |}
\hline
   &  Pr(offense against plaintiff 1) & Pr(offense against plaintiff 2) & Pr(at least one offense ) \\
  \hline
  Defendant 1 & 49.5 \% & 49.5 \% & 99 \% \\ \hline
     &  Pr(offense against plaintiff 3) &  & Pr(at least one offense) \\
  \hline
  Defendant 2 & 51 \% & & 51 \% \\
  \hline
\end{tabular}
\end{center}}
\noindent Suppose that DPP is used together with the preponderance of evidence criterion ($\pi^* = 50\%$). Then, Defendant $1$ is acquitted and Defendant $2$ is convicted despite the fact that Defendant $1$ is almost certainly guilty.

Using DPP in this case seems particularly  problematic ex post, and it may be difficult to implement DPP in practice if the adjudicator lacks commitment  power. In non-legal settings, firms face more social pressure to fire a manager whose probabilities of abusing subordinates are given by the first row, political parties have incentives to ostracize party members with bad reputations (e.g., individuals who are believed to have committed at least some offenses with high probability) in order to restore their popularity.


Motivated by this commitment problem, we explore alternative remedies to reduce the probability of offense when
conviction decisions are made according to APP. Proposition \ref{Prop3} suggests that reducing the magnitude of punishment can help deter offenses by inducing a \textit{positive correlation} in agents' private information. This positive correlation, together with agents' coordination motive, encourages  each agent to accuse the principal when he has witnessed an offense. This improves the informativeness of all agents' accusations.
\begin{Proposition}\label{Prop3}
When the judge uses (\ref{2.4}), there exists, for every $c>0$, a nonempty interval $(\underline{L}(c),\overline{L}(c)) \subset \mathbb{R}_+$ such that $q(1,1)+q(0,0)-q(1,0)-q(0,1)<0$ for every $L$ in this interval and corresponding equilibrium.
\\For every $\varepsilon>0$, there exists $\underline{c}>0$, such that when $c>\underline{c}$ and $L \in (\underline{L}(c),\overline{L}(c))$, the equilibrium probability of offense is less than $\varepsilon$.
\end{Proposition}
The proof is in Online Appendix G. Taken together, Proposition \ref{Prop3} and Lemma \ref{L3.3} imply that
the principal's decisions to commit different offenses are \textit{strategic complements}.
Since the equilibrium probability of guilt is interior, this implies that the principal must be indifferent between committing no offense and committing two offenses. This induces \textit{positive correlation} in  agents' private information. Since an agent is less likely to face retaliation when the other agent has witnessed an offense, this positive correlation encourages an agent to accuse the principal after witnessing an offense and discourages him from doing so when he has not witnessed any offense. Unlike the case of negatively correlated private information, here the reporting cost $c$
improves the informativeness of agents' reports and decreases the equilibrium probability and the expected number of offenses.

Proposition \ref{Prop3} suggests that when the judge aggregates the probabilities of different offenses,
the optimal punishment that minimizes the probability of offense $\Pr(\overline{\theta}=1)$ is
\textit{interior}. This finding stands in sharp contrast to Becker's (1968) seminal analysis of criminal justice and law enforcement, according to which increasing the magnitude of punishment is efficient in reducing offenses. Our finding suggests a new rationale for avoiding harsh punishments.  This finding applies to settings in which evidence primarily consists of witnesses testimonies that is hard to corroborate with other forms of evidence.

While reducing $L$ may improve reports' informativeness and deter offenses, it also increases the number of offenses conditional on an at least one offense being committed. When the principal is guilty, he systematically commits multiple offenses.
Viewed from this perspective, Proposition~\ref{Prop3} shows a tradeoff between reducing the proportion of guilty individuals and reducing the severity of actions committed by these individuals, as measured by the number of offenses that they commit.

In addition, implementing the solution proposed by Proposition \ref{Prop3} is challenging in practice since it requires a careful calibration of $L$. This issue is particularly salient when the benefit from committing offenses is uncertain, since $L$ represents the magnitude of punishment relative to the benefit from committing an offense.

\section{Heterogenous Propensity to Commit Offenses}\label{sec4}
While the analysis so far point to several weaknesses of APP, a potential advantage of aggregating probabilities across offenses is to extract information about a defendant's ``character,'' defined as the propensity to commit offenses. This propensity may run the gamut from virtuous individuals who incur a disutility from the commission of offenses to serial offenders who experience a high utility from committing offenses.

This section considers this possibility, focusing for tractability on the case of the two principal types introduced in Section~\ref{sec2}: a virtuous one and an opportunistic one. Theorems \ref{Theorem1} and \ref{Theorem2} are extended to this setting.

We show that, under both decision rules, the opportunistic type may \textit{commit multiple offenses with positive probability}. Nevertheless, agents' private observations are still \textit{negatively correlated} under decision rule (\ref{2.4}) and \textit{uncorrelated} under decision rule (\ref{2.5}). Our predictions concerning the effect of these rules on the informativeness of witness testimonies and deterrence remain unchanged. Recall that $\pi^o \in (0,1]$ is the probability that the principal is opportunistic. Theorem 1' generalizes Theorem \ref{Theorem1}:
\begin{Theorem1}
When $n=2$ and the judge uses decision rule (\ref{2.4}), there exists $\overline{L}$ such that when $L > \overline{L}$,
\begin{enumerate}
  \item \textbf{Endogenous Negative Correlation Between Offenses:} In every equilibrium
  \begin{equation*}
    \Pr(\theta_1=1|\theta_2=1)<\Pr(\theta_1=1|\theta_2=0) \textrm{ and } \Pr(\theta_2=1|\theta_1=1)<\Pr(\theta_2=1|\theta_1=0).
  \end{equation*}
\end{enumerate}
For every $\varepsilon>0$, there exists $\overline{L}_{\varepsilon} \in \mathbb{R}_+$, such that in every equilibrium when $L>\overline{L}_{\varepsilon}$,
\begin{enumerate}
  \item[2.] \textbf{Low Informativeness of Reports:}
    \begin{equation}\label{4.1}
\max_{\boldsymbol{a} \in \{0,1\}^2} \mathcal{I} (\boldsymbol{a})
<
 \Big\{ \frac{\pi^*}{1-\pi^*} \Big/ \frac{\min\{\pi^*,\pi^o\}}{1-\min\{\pi^*,\pi^o\}} \Big\} +\varepsilon.
\end{equation}
\item[3.]  \textbf{Ineffective Deterrence:}
\begin{equation}\label{4.2}
\Pr(\overline{\theta}=1) > \min\{\pi^*,\pi^o\}-\varepsilon.
\end{equation}
\end{enumerate}
\end{Theorem1}
Theorem 1' shows that the negative correlation in agents' private information, the lack of informativeness of agents' reports, and the high frequency of offenses still arise when there is some unobserved heterogeneity in the principal's propensity to commit offenses. Theorem \ref{Theorem1} and Theorem 1' provide different asymptotic upper bounds on the informativeness of agents' reports and lower bounds on the probability of offenses:
\begin{itemize}
\item When the fraction $\pi^o$ of opportunistic types exceeds $\pi^*$, the equilibrium probability of offenses is close to the conviction cutoff $\pi^*$.
The informativeness of agents' accusations, measured by $\max_{\boldsymbol{a} \in \{0,1\}^2} \mathcal{I} (\boldsymbol{a})$
is arbitrarily close to $1$ as the punishment from conviction increases.
\item When the probability $\pi^o$ that the principal is opportunistic is less than $\pi^*$, the prior probability of offense equals $\pi^o$. This is the highest possible ex ante probability of guilt given that the virtuous type never commits any offense (Lemma \ref{L3.1}) and, in this sense, the worst possible outcome.

Given that in every equilibrium, there exists $\boldsymbol{a} \in \{0,1\}^2$ such that the principal is convicted with positive probability
when the judge observes $\boldsymbol{a}$,
the judge's posterior belief
that at least one offense has taken place is no less than $\pi^*$
after observing $\boldsymbol{a}$. This suggests a lower bound on the informativeness of agents' accusations:
    \begin{equation}\label{4.3}
\max_{\boldsymbol{a} \in \{0,1\}^2} \mathcal{I} (\boldsymbol{a}) \geq   \mathcal{I}_{min} \equiv \frac{\pi^*}{1-\pi^*}\Big/ \frac{\pi^o}{1-\pi^o}.
    \end{equation}
Theorem 1' shows that this lower bound $\mathcal{I}_{min}$ is attained in \textit{all equilibria}.
\end{itemize}

Theorem 1' is proved similarly to Theorem \ref{Theorem1}. Lemma \ref{L3.2} and Lemma \ref{L3.3} generalize to heterogeneous principal types and imply that the opportunistic type is never indifferent between committing zero and two offenses. The opportunistic type is either indifferent between committing zero and one offense or indifferent between committing one and two offenses. This leads to two disjoint cases in Theorem~1', whose separation depends on the prior probability of the virtuous type:
\begin{enumerate}
  \item When $\pi^o \geq \pi^*$, the opportunistic type is indifferent between committing zero and one offense, and equilibria have the same features as those predicted by Theorem~\ref{Theorem1}.
  \item When $\pi^o<\pi^*$ and $L$ is large enough, the opportunistic type always  commits at least one offense. Otherwise, agents' reports would be arbitrarily uninformative as $L$ becomes arbitrarily large, by the same logic as in Theorem~1. The posterior probability would have to be strictly lower than $\pi^*$ even when both agents accuse the principal. This would imply that the principal is never convicted and lead to a contradiction. Hence, the opportunistic type must be indifferent between committing one and two offenses.

Nevertheless, $\theta_1$ and $\theta_2$ remain \textit{negatively correlated}. To understand why, suppose by way of contradiction that $\theta_1$ and $\theta_2$ were independent or positively correlated. We would then have $Q_{0,j} \leq Q_{1,j}$, and the expressions for the reporting cutoffs (\ref{3.11}) and (\ref{3.12}) would imply that $\omega_j^*-\omega_j^{**} \geq b$.
From Lemma \ref{L3.2}, $q(1,1)$ must converge to $0$ as $L \rightarrow +\infty$ and
agents' reporting cutoffs both go to $-\infty$. As in the single-agent benchmark, the informativeness of agents' report would then go to $+\infty$.
Since an opportunistic type commits at least one offense, the prior probability that the defendant commits at least one offense is equal to $\pi^o$. Therefore, the posterior probability that the defendant committed an offense when both agents accuse the principal would have to exceed $\pi^*$. This contradicts Lemma \ref{L3.2}, which requires that the judge be indifferent between convicting and acquitting the defendant when both agents accuse him (since $q(1,1) \in (0,1)$) and, hence, that the judge's posterior be equal to the conviction threshold $\pi^*$.
\end{enumerate}
The next result generalizes Theorem \ref{Theorem2}, which studies equilibrium outcomes under DPP. The proof is similar to the proof of Theorem \ref{Theorem2} and omitted.
\begin{Theorem2}
When $n=2$ and the decision rule is (\ref{2.5}), there exists $\overline{L} \in \mathbb{R}_+$ such that when $L > \overline{L}$,
\begin{enumerate}
  \item \textbf{Uncorrelated Offenses:} , $\Pr(\theta_i=1|\theta_j=1)=\Pr(\theta_i=1|\theta_j=0)$ for every $i \neq j$.
\item \textbf{Linear Conviction Probability:}  $\Pr(\theta_i=1|a_i=1)=\pi^*$ for every $i \in \{1,2\}$, and the conviction probability is linear in the number of accusations.
\end{enumerate}
As the fraction of behavioral agents goes to zero and $L\rightarrow +\infty$, the following asymptotic results hold:\footnote{As before, the results are obtained by first taking the limit with respect to $\delta$ and then with respect to $L$.}\begin{enumerate}
  \item[3.] \textbf{Effective Deterrence:} The equilibrium probability that offense taking place
  $\Pr(\overline{\theta}=1)$
  converges to $0$.
  \item[4.] \textbf{Highly Informative Reports:} For every $i \in \{1,2\}$, the informativeness of agent $i$'s report about $\theta_i$, measured by $\frac{\Pr(a_i=1|\theta_i=1 )}{\Pr(a_i=1|\theta_i=0)}$,
goes to $+\infty$.
\end{enumerate}
\end{Theorem2}

\section{Three or More Agents}\label{sec5}

This section extends Theorems \ref{Theorem1} and \ref{Theorem2} to an arbitrary number of agents, reverting to the case of a single principal type, and provides a comparative static result (Theorem~\ref{Theorem3}) on the number of agents that generalizes Proposition \ref{Prop2}.

We start by generalizing the measure of informativeness used in Theorem \ref{Theorem1} to three or more agents. Bayes rule implies that
\begin{equation}\label{5.1}
\frac{\Pr(\boldsymbol{a}|\overline{\theta}=1)}{\Pr(\boldsymbol{a}|\overline{\theta}=0)} \cdot \frac{\Pr(\overline{\theta}=1)}{1-\Pr(\overline{\theta}=1)}=
\frac{\Pr(\overline{\theta}=1|\boldsymbol{a})}{1-\Pr(\overline{\theta}=1|\boldsymbol{a})}
\textrm{ for every } \boldsymbol{a} \in \{0,1\}^n.
\end{equation}
Therefore, the ratio
\begin{equation}\label{5.2}
    \mathcal{I}(\boldsymbol{a}) \equiv \frac{\Pr(\boldsymbol{a}|\overline{\theta}=1)}{\Pr(\boldsymbol{a}|\overline{\theta}=0)}
\end{equation}
measures the change in a judge's belief after observing report profile $\boldsymbol{a}$, which we use to measure the informativeness of $\boldsymbol{a}$. For tractability, we focus our analysis on \textit{symmetric} equilibria, i.e., equilibria in which the principal treats all agents symmetrically and all agents' equilibrium strategies are the same.
Proposition \ref{Prop4} generalizes the insights of Theorem \ref{Theorem1} to the case of three or more agents.
\begin{Proposition}\label{Prop4}
Suppose $n \geq 2$ and the judge uses decision rule (\ref{2.4}). For every $\varepsilon > 0$, there exists $\overline{L}_{\varepsilon}>0$
such that for every $L > \overline{L}_{\varepsilon}$ and $\delta \in (0,1)$,
\begin{itemize}
  \item[1.] There exists a symmetric equilibrium that satisfies Refinement 1.\footnote{We show in Online Appendix H.4 that every symmetric Bayes Nash equilibrium that satisfies Refinement 1 is a proper equilibrium that satisfies Refinement 2.}
\end{itemize}
\begin{itemize}
  \item[2.] In every symmetric equilibrium that satisfies Refinement 1,
  \item[(a)] $\Pr(\max_{j \neq i} \theta_j =1 |\theta_i=1)< \Pr(\max_{j \neq i} \theta_j =1 |\theta_i=0)$ for every $i \in \{1,2,...,n\}$;
  \item[(b)] $\max_{\boldsymbol{a} \in \{0,1\}^n} \mathcal{I} (\boldsymbol{a}) <1+\varepsilon$ and $\Pr(\overline{\theta}=1)>\pi^*-\varepsilon$.
\end{itemize}
\end{Proposition}
Statement 2(a) means that an agent who witnessed an offense assigns a lower probability to other agents having witnessed offenses than if did not witness any offense. We show that in every symmetric equilibrium, the principal commits either no offense or only one offense, which implies that agents' private observations remain \textit{negatively correlated}.
Statement 2(b) shows that agents' reports become arbitrarily uninformative and the  probability that the principal commits at least one offense converges to $\pi^*$ as the punishment $L$ becomes large relative to the benefit from committing offenses. This is similar to the prediction in Theorem \ref{Theorem1}.

The following result provides comparative statics with respect to the number of agents, showing in particular that each agent is {\em more} likely of accusing the principal as the number of agents increases.
\begin{Theorem}\label{Theorem3}
Suppose the judge uses decision rule (\ref{2.4}).
For every $k,n \in \mathbb{N}$ with $k>n$, there exists $\overline{L}>0$ such that for every $L > \overline{L}$, and
compare any symmetric equilibrium corresponding to $k$ agents to any symmetric equilibrium corresponding to $n$ agents:
\begin{itemize}
  \item[1.] \textbf{Lower Informativeness:} $\max_{\boldsymbol{a} \in \{0,1\}^k} \mathcal{I} (\boldsymbol{a}) < \max_{\boldsymbol{a} \in \{0,1\}^n} \mathcal{I} (\boldsymbol{a})$.
  \item[2.] \textbf{Higher Probability of Offense:} The equilibrium probability of offense $\Pr(\overline{\theta}=1)$ is strictly higher
  with $k$ agents than with $n$ agents.
  \item[3.] \textbf{Higher Probability of Filing Accusations:} Each agent's reporting cutoffs $(\omega^*,\omega^{**})$ are both strictly higher with $k$ agents than with $n$ agents.
\end{itemize}
\end{Theorem}
The proof is in Appendix E. Theorem \ref{Theorem3} shows that as the number of potential offenses increases,
the probability of offense increases,
the informativeness of reports decreases, and each agent is \textit{more likely} to accuse the principal.

This  last feature distinguishes our result from those on public good provision  (e.g., Chamberlin 1974), in which inefficiencies arise because agents free ride on one another's contributions. In our model, the lower informativeness of agents' reports as $n$ increases is due to an increased number of false accusations, not from agents abstaining at a higher rate from revealing observed offenses.

Next, we study the game's equilibrium outcomes when the judge uses DPP to adjudicate guilt.
Proposition \ref{Prop5} establishes the existence of equilibria in which agents' private information is uncorrelated and the conviction probability is a linear function of the number of accusations against the principal. These equilibria feature  arbitrarily informative reports and a vanishing probability of offense as the punishment $L$ becomes large.
\begin{Proposition}\label{Prop5}
When the judge uses decision rule (\ref{2.5}), there exists $\overline{L}>0$, such that for every
$L>\overline{L}$, there exist an equilibrium in which:
\begin{itemize}
  \item[1.] $\theta_i$ and $\theta_j$ are uncorrelated for every $i \neq j$,
  \item[2.] the probability that the principal is convicted is linear in the number of accusations.
\end{itemize}
In the limit as $\lim_{L \rightarrow \infty}$ and $\lim_{\delta \rightarrow 1}$,
\begin{enumerate}
  \item[3.] The probability $\Pr(\overline{\theta}=1)$ that the principal commits an offense converges to $0$.
  \item[4.] For every $i \in \{1,,\ldots, n\}$, the informativeness of agent $i$'s report, $\frac{\Pr(a_i=1|\theta_i=1 )}{\Pr(a_i=1|\theta_i=0)}$,
goes to $+\infty$.
\end{enumerate}
\end{Proposition}

\section{Welfare Implications of Conviction Rules}\label{secX}
While we did not specify any social welfare function, our analysis unveils several tradeoffs between improving the quality of judicial decisions and deterrence, between ex ante incentives and ex post fairness, and between the frequency and the severity of committed offenses. Our results may inform a social planner considering various adjudication rules (for example, choosing between APP and DPP, choosing $L$, $\pi^*$, and  $c$ under various constraints) to maximize an objective function that incorporates the various facets of these tradeoffs.

First, Theorems \ref{Theorem1} and \ref{Theorem2} show that when (i) a defendant necessarily incurs a large disutility from conviction relative to the benefit of committing the offense (a disparity that may be due, e.g., to the loss of one's job or reputation), and (ii) the conviction threshold $\pi^*$ is beyond the control of a social planner, APP induces a significantly higher offense rate than DPP.

Second, we obtain explicit formulas to describe a tradeoff between deterrence and fairness. Under both APP and DPP, a fraction $1-\pi^*$ of convicted defendants are innocent and a fraction $\pi^*$ of acquitted defendants are guilty. While offenses become less frequent when the standard of proof is lowered ($\pi^*$ decreases), this increases the fraction of innocent individuals among convicted defendants. The optimal cutoff $\pi^*$ could arise from \textit{maximizing a social welfare function} that trades off the effectiveness of deterrence with the fraction of false positive among cases that resulted in a conviction.\footnote{This tradeoff has been discussed in reduced-form by Harris (1970) and Miceli (1991), but it does not arise in Becker (1968) and Landes (1970), who ignore wrongful convictions. In Kaplow (2011), punishments are expressed in terms of fines, i.e., zero sum transfers that do not affect the social surplus.}

Similarly, suppose that a judge \textit{commits} to convict the principal whenever at least one report was filed. This would eliminate agents' need to coordinate their reports and, when the punishment $L$ is large, would give the principal a strict incentive not to commit any offense. However, in order to fulfill his commitment, the judge would have to convict the principal whenever some agent accuses the principal, even when the principal never commits any offense in equilibrium. This would lead to an undesirable outcome since \textit{all convicted individuals would be innocent} and the probability of convicting innocent individuals would be significant.\footnote{This paradox commonly arises in plea bargaining models, in which agents who reject pleas but are convicted at trial are known to be innocent (Grossman and Katz 1983, Reinganum 1988, and Siegel and Strulovici 2020).}

Third, our example in Section 3.3 illustrates how DPP may lead to ex post unfair outcomes: the defendant who is less likely to be guilty is convicted while the defendant who is almost surely guilty is acquitted. When ex post fairness has a strong influence on adjudication decisions, the use of APP may be viewed a constraint on the set of decision rules available to the social planner.

Fourth, we clarify that while APP is defined as a {\em rule} to adjudicate guilt, it is justified as a way to capture an ex post {\em social objective function} and show, paradoxically, that it may be socially undesirable to use APP as a rule even when the social objective is the one that motivates APP in the first place. Intuitively, we have that APP may be self-defeating once incentives are taken into account.

Fifth, deterrence may, counter-intuitively, be stronger if the punishment to a convicted defendant takes on intermediate values than extreme ones, as it induces positive correlation in witnesses' private observations and reduces their exposure to retaliation or social stigma. Proposition 3 shows that using a more lenient sentence in case of conviction can be effective in deterrence, without increasing the probability of convicting the innocent. These intermediate punishment levels result in an increase the number of offenses committed by guilty defendants, which points to a tradeoff between the frequency and the severity of committed offenses.

Lastly, the cost of reporting $c$ plays a subtle role on social welfare. Beyond the ethical concerns of protecting whistleblowers, Theorem~1 suggests that under APP, reducing $c$ can also improve the credibility of witness testimonies and reduce crime. By contrast, having a strictly positive reporting cost under DPP improves deterrence by lowering the fraction of agents submitting false positive reports.


\section{Extensions \& Robustness}\label{sec6}
This section presents several extensions of the baseline model, which convey the robustness our results. The details of the analysis can be found in Supplementary Appendix K.

\paragraph{Decreasing Marginal Benefits from Committing Offenses:} Our results continue to hold when the principal faces decreasing marginal returns
from committing offenses or receives a punishment larger than $L$ when he is believed to have committed multiple offenses. These changes motivate the principal to commit fewer offenses and, as in the baseline model, induce negative correlation in the agents' private information. Agents' coordination motive undermines the informativeness of their reports and increases the probability of offenses. In this extension, formalized in Supplementary Appendix K.3, the principal receives punishment $L$ if the probability that he committed at least one offense exceeds $\pi^*$ and receives punishment $L' (>L)$ if the probability that he committed two offenses exceeds some other cutoff $\pi^{**} \in (0,1)$. When $L$ is large,
the principal commits at most one offense in every symmetric equilibrium that satisfies
Refinement 1.

\paragraph{Accusation Costs:} An agent who accuses the principal may be subjected to retaliation or social stigma even when the principal is convicted. As long as the cost of retaliation or social stigma is strictly higher when the principal is acquitted than when he is convicted, this does not affect our results. More generally, the results generalize when each agent's loss from retaliation is decreasing in the number of accusations leveled against the principal, as long as this loss is strictly positive when the principal is acquitted. In fact, these variations strengthen agents' coordination motive and leaves unchanged the negative correlation affecting their private information.

\paragraph{Interdependent Preferences:} As anticipated in Section \ref{sec2}, an agent may directly care about other offenses committed by the principal, in addition to the offense that he may observe. This situation is modeled as follows: when agent $i$ is strategic, his payoff is normalized to $0$ when the principal is convicted and is equal to:
\begin{equation}\label{6.2}
   \omega_i - b
    \Big( (1-\gamma) \theta_i +\gamma f(\theta_i,\theta_{-i}) \Big) -ca_i
\end{equation}
when the principal is acquitted, where $f(\theta_i,\theta_{-i})$ is increasing in all arguments. The term $\gamma f(\theta_i,\theta_{-i})$ captures how $i$'s payoff depends on all the offenses committed by the principal.

We show in Supplementary Appendix K.1 that agents' reports become arbitrarily uninformative and the probability that the principal commits some offense converges to $\pi^*$ as $L$ becomes arbitrarily large. In fact, these social preferences undermine even further than in the baseline model the informativeness of agents' reports. Intuitively, when agent $i$'s payoff depends directly on $\theta_j$, agent $i$'s report $a_i$ becomes more responsive to his belief about $\theta_j$. This, together with the negative correlation between
$\theta_i$ and $\theta_j$, causes $i$'s report to become even less responsive to $\theta_i$ and less informative than in the baseline model.

\paragraph{Preference for Truth-telling:}

The baseline model took a consequentialist approach to model agents' preferences for telling the truth: an agent who observed an offense had a higher payoff when the principal was convicted, but not from accusing the principal per se. Suppose now that an agent who observed an offense receives a benefit $d$ $(>0)$ from accusing the principal, regardless of the adjudication outcome. Letting $l^* = \frac{\pi^*}{1-\pi^*}$, we show in Supplementary Appendix K.1 when there are two agents and $d< \frac{l^*}{l^*+2}c$, the informativeness $\frac{(1-\delta)\alpha + \delta \Phi(\omega^*)}{(1-\delta)\alpha +\delta \Phi(\omega^{**})}$ of agents reports is bounded above by
\begin{equation}\label{6.3}
    \frac{cl^*}{cl^*-(l^*+2)d}.
\end{equation}
This upper bound collapses to 1 (i.e., reports become completely uninformative) when the preference for truthtelling $d$ converges to 0, or when the retaliation cost $c$ becomes arbitrarily large. In particular, our results generalize when agents experience a small benefit from telling the truth.

\paragraph{Ex Post Evidence \& Punishing False Accusations:} Suppose that if the principal is wrongfully convicted, exonerating evidence arrives exogenously with probability $p^*$ and causes every accuser to be penalized by some constant $\ell \geq 0$. These punishments are equivalent to increasing the benefit $b$ from convicting the principal after witnessing an offense, and this extension is formally equivalent to the baseline model, as explained in Supplementary Appendix K.1.

\paragraph{Uncertain Number of Offense Opportunities:} In applications such as workplace bullying, physical assaults and discrimination, the number of potential offenses that the principal may commit is unobserved by the judge and by the victims. Suppose that nature randomly selects a subset $\widetilde{N} $ of $\{1,2,...,n\}$ of offense opportunities for the principal. We assume that (i) only agents with index in $\widetilde{N}$ can credibly accuse the principal and (ii) that agents who do not accuse the principal do not file any report. Assumption (i) is justified for instance if agents outside of $\widetilde N$ are inactive (equivalently, the set of agents in the model is stochastic, equal to $\widetilde N$, and only observed by the principal) or if they are active but the principal could easily refute any accusation by such agents, e.g., by providing an alibi. The principal privately observes the set $\widetilde{N}$. Agents in $\widetilde{N}$ observe the same information as in the baseline model and cannot observe the realization of~$\widetilde N$.

We informally argue that the logic behind our results is even stronger in this case than in the baseline model. Since the judge does not observe the cardinality of $\widetilde{N}$, his verdict depends only on the number of accusations (assuming for simplicity a symmetric treatment of accusers), not on the number of potential victims. Since the probability of convicting the principal is increasing in the number of accusations, the principal has a stronger incentive to commit offenses when there are fewer agents overall who can accuse him (i.e., when $|\widetilde{N}|$ is smaller). An agent who observes an offense therefore infers that $|\widetilde{N}|$ is more likely to be small, other things equal, and, hence, that the number of accusations by other agents is also likely to be small. This effect dampens an agent's incentive to accuse the principal when he has witnessed an offense and lowers the informativeness of agents' reports in equilibrium, by the same logic as in the baseline model.

\paragraph{Behavioral Agents and Retaliation Cost:} There is a close relationship between behavioral agents whose reporting strategy depends on their observation and strategic agents who are immune to retaliation. Suppose that a behavioral agent accuses the principal with probability $\overline{\alpha}$ when he has observed an offense and with probability $\underline{\alpha}$ when he has not, with $1> \overline{\alpha} \geq \underline{\alpha} >0$.
This formulation is equivalent to one in which these agents are strategic but immune to retaliation, so that their realized payoff is given by $(\omega_i - b\theta_i )(1-s)$. In equilibrium, such an agent uses reporting  cutoffs $b$ and $0$, depending on whether the agent has witnessed an offense or not, which corresponds to accusation probabilities $\overline{\alpha}=\Phi(b)$ and $\underline{\alpha}=\Phi(0)$, respectively. Therefore, this strategic agent behaves as though he were behavioral and playing  an informative cutoff-strategy.
As noted in Section~\ref{sec2}, all our results hold for such behavioral agents.
\section{Concluding Remarks}\label{sec7}

\paragraph{Relation to the Legal Scholarship:} Treating each charge made against an individual in isolation of other charges may seem a priori arbitrary, unfair, and ineffective. This observation has been formalized and explored in the legal scholarship and takes on a particular importance for individuals who face multiple accusations, each of which is hard to establish at the sufficient level of certainty.

The premise, common in the legal literature, that offenses are exogenously and independently distributed is particularly problematic when analyzing the aggregation of offense probabilities. This aggregation creates an incentive for defendants to strategically restrict the number of offenses, which introduces \textit{negative correlation} in the occurrence of offenses and violates the premise that offenses are independently distributed. Indeed, our analysis shows that aggregating offense probabilities has severe drawbacks once the incentives of potential offenders and accusers are taken into account.

Underlying this difficulty, the \textit{Aggregate Probabilities Principle} (APP) describes a {\em rule of punishment} rather than a {\em social objective function}. While it may be socially desirable to punish a defendant deemed sufficiently likely of committing some offense, even an unspecified one, the principle may be self-defeating and suboptimal from a social welfare perspective once incentives are taken into account.

\paragraph{Equilibrium Analysis {\em vs.} Nonequilibrium Adjustments:} Our results are derived from an equilibrium analysis, which presumes that players have correct expectations about the consequences of their actions and other players' strategies. When social rules change, as in case of a sudden crackdown on a specific type of offense, the introduction of new regulation, a drastic shift in social norms, or the emergence of new social media that change the social consequences of one's actions, equilibrium analysis may be viewed as a potential harbinger of issues that will emerge as economic and social actors learn to interact under these new rules or norms.\footnote{Foundations of Nash equilibrium based on players learning one another's strategies have a long history in economics. See, e.g., Fudenberg and Levine (1995).} This distinction seems particularly relevant in the context of the recent \textit{me too} movement, since abusers before the emergence of the movement likely underestimated the legal and professional consequences of their abusive behavior.

\paragraph{Shielding Accusers from Stigma through Secret Accusations:} To address the potential pressure that is sometimes experienced by lone accusers, institutions  have been developed under which reports are submitted to a third party and are only released when enough of them have been filed.\footnote{In particular, the nonprofit organization Callisto has a ``match'' feature, whereby a report is made official only if at least two victims name the same perpetrator. See www.projectcallisto.org.}

Such provisions increase the risk of wrongful accusations. Indeed, an agent holding a grudge against the principal has an opportunity to secretly accuse him in the hope that other agents, rightfully or not, may also accuse the principal. While these institutions are clearly well intentioned and worth considering, it is important to evaluate their consequences once all agents understand them.

Moreover, if accusations run the risk of being leaked, even with a small probability, this creates a strictly positive expected cost of retaliation, and brings us back to the baseline model.

\paragraph{Simultaneous vs Sequential Reporting:} The forces that underlie our results are also present in dynamic versions of our model, in which reports may be filed sequentially. First, the negative correlation between the agents' private information ($\theta_i$) continues to arise endogenously whenever a strategic principal is concerned about having too many reports made  against him. Second, an individual agent has an incentive to coordinate with other agents whenever he is unsure about whether his report is pivotal or not. In a dynamic setting, this incentive can materialize after a \textit{cold start} (i.e., where very few people have reported before and no agent wants to be the first accuser). It can also occur when an agent has observed many reports and is unsure of the number of reports needed to convict the principal (for example, if he faces uncertainty about the conviction standard $\pi^*$ used by the judge). The inefficiencies and lack of credibility caused by the agents'  coordination motive thus still arise in a dynamic environment.\footnote{See Lee and Suen (2020) for a model of strategic accusation in which the timing of accusation plays a major role.}

\newpage
\appendix
\section{Proof of Lemma 2.1}
\paragraph{Statement 1:} When $i$ is strategic and observes $(\omega_i, \theta_i)$, he chooses $a_i=1$ if and only if
\begin{equation*}
    \omega_i \sum_{\boldsymbol{a}_{-i} \in \{0,1\}^{n-1}}  \sigma_{-i}(\boldsymbol{a}_{-i}) \Big(q(1,\boldsymbol{a}_{-i})-q(0,\boldsymbol{a}_{-i})\Big)
    \leq
    \end{equation*}
    \begin{equation}\label{A.1}
    b \theta_i \sum_{\boldsymbol{a}_{-i} \in \{0,1\}^{n-1}}  \sigma_{-i}(\boldsymbol{a}_{-i}) \Big(q(1,\boldsymbol{a}_{-i})-q(0,\boldsymbol{a}_{-i})\Big)
    -c\sum_{\boldsymbol{a}_{-i} \in \{0,1\}^{n-1}}  \sigma_{-i}(\boldsymbol{a}_{-i}) \Big(1-q(1,\boldsymbol{a}_{-i})\Big).
\end{equation}
Refinement 2 implies that $q(1,\boldsymbol{a}_{-i})-q(0,\boldsymbol{a}_{-i}) \geq 0$ for every $\boldsymbol{a}_{-i} \in \{0,1\}^{n-1}$ with a strict inequality for some $\boldsymbol{a}_{-i}$.
The existence of behavioral agents implies that every $\boldsymbol{a}_{-i} \in \{0,1\}^{n-1}$ occurs with strictly positive probability. Therefore,
\begin{equation*}
\sum_{\boldsymbol{a}_{-i} \in \{0,1\}^{n-1}}  \sigma_{-i}(\boldsymbol{a}_{-i}) \Big(q(1,\boldsymbol{a}_{-i})-q(0,\boldsymbol{a}_{-i})\Big) > 0.
\end{equation*}
This, together with inequality~(\ref{A.1}), implies that agent $i$ accuses the principal if and only if $\omega_i$ lies below some cutoff that is strictly higher when $\theta_i=1$ than when $\theta_i=0$.

\paragraph{Statement 2:} Suppose toward a contradiction that $\Pr(\overline{\theta}=1)=0$ or, equivalently $\theta_1=...=\theta_n=0$ with probability $1$. Since every $\boldsymbol{a}$ occurs with strictly positive probability, this implies that $\Pr(\overline{\theta}=1|\boldsymbol{a})=0$ for every $\boldsymbol{a} \in \{0,1\}^n$.
Since $\pi^* \in (0,1)$, the principal is convicted with probability $0$ regardless of $\boldsymbol{a}$. Therefore, an opportunistic principal has a strict incentive to commit offense, which contradicts $\Pr(\overline{\theta}=1)=0$.

\paragraph{Statement 3:} Since $\omega_i^* > \omega_i^{**}$ for every $i \in \{1,2,...,n\}$ and since every report vector $\boldsymbol{a} \in \{0,1\}^n$ occurs with positive probability, Refinement 2 implies that for every $\boldsymbol{\theta} \succ \boldsymbol{\theta'}$, the principal is convicted with strictly higher when $\boldsymbol{\theta}$ than when $\boldsymbol{\theta'}$. Since a  virtuous principal does not benefit from committing offenses, he strictly prefers to commit no offense in any proper equilibrium that satisfies Refinements 1 and 2.

\section{Proof of Proposition 1}\label{secB}
Let $q = q(1)$ denote the probability of conviction when the principal is accused. The principal's expected cost of committing an offense is
\begin{equation}\label{B.1}
\delta L q \left(\Phi\left(b-c\frac{1-q}{q}\right)-
\Phi\left(-c\frac{1-q}{q}\right)\right),
\end{equation}
which is a continuous and strictly positive function of $q$ that converges to $0$ as $q \rightarrow 0$. Since the equilibrium probability of offense is strictly positive (Statement 2 of Lemma \ref{L3.1}),  the value of (\ref{B.1}) is less than or equal to $1$.

Suppose toward a contradiction that for every $\overline{L}>0$, there exists $L > \overline{L}$ under which in some proper equilibrium that satisfies Refinements 1 and 2, the value of (\ref{B.1}) is strictly less than $1$. An opportunistic principal then has a strict incentive to commit offense, i.e., $\Pr(\theta=1)=\pi^o$. Moreover, $q$ goes to $0$ and $\Phi(\omega^*)/\Phi(\omega^{**})$ goes to $+\infty$ as $\overline{L} \rightarrow \infty$. This implies the existence of $\overline{L}>0$ such that for every $L > \overline{L}$, and for every $q$ such that (\ref{B.1}) is strictly less than $1$, $\Pr(\theta=1|a=1)$ is strictly greater than $\pi^*$. This implies the judge surely convicts the principal and contradicts the fact that $q$ goes to $0$. Therefore, (\ref{B.1}) must be equal to $1$ when $L$ is large enough, $q$ lies in $(0,1)$, and$\Pr(\theta=1|a=1)=\pi^*$.

When $L \rightarrow +\infty$, $1/\delta L$ converges to $0$. Suppose toward a contradiction that $q_s$ converges to some limit $\underline{q} > 0$ along some sequence $\{L_n\}_{n=1}^{\infty}$ with $\lim_{n \rightarrow \infty} L_n=\infty$. Then, $\omega^*$ and $\omega^{**}$ respectively converge to $b-c(1-\underline{q})/\underline{q}$ and $-c(1-\underline{q})/\underline{q}$. The LHS of (\ref{3.5}) converges to
\begin{equation}\label{B.2}
    \delta \underline{q} \left(\Phi\left(b-c\frac{1-\underline{q}}{\underline{q}}\right)- \Phi\left(-c\frac{1-\underline{q}}{\underline{q}}\right) \right)
\end{equation}
which is strictly positive. This leads to a contradiction and shows that $q \rightarrow 0$.
The expressions for $\omega^*$ and $\omega^{**}$ in (\ref{3.1}) and (\ref{3.2}) imply that both cutoffs go to $-\infty$. This shows that
\begin{equation}\label{B.3}
\lim_{\omega^* \rightarrow -\infty}    \lim_{\delta \rightarrow 1} \frac{\delta \Phi(\omega^*)+(1-\delta)\alpha}{\delta \Phi(\omega^{*}-b)+(1-\delta)\alpha}=+\infty,
\end{equation}
where we use the observation that
$\lim_{\omega \rightarrow -\infty} \Phi(\omega)/\Phi(\omega-b) \rightarrow +\infty$  for every  $b>0$. From~(\ref{3.3}) and the fact that $\Pr(\boldsymbol{\theta}=1|\boldsymbol{a}=1)=\pi^*$, we conclude that $\Pr(\boldsymbol{\theta}=1)$ converges to $0$.

\section{Proof of Theorems 1 and 1'}\label{secC}
The following lemma establishes that all equilibria are symmetric.
\begin{Lemma}\label{LC.1}
When $n=2$ and the judge uses conviction rule (\ref{2.4}), there exists $\overline{L}>0$ such that when $L > \overline{L}$, the events  $(\theta_1,\theta_2)=(1,0)$ and $\theta_1,\theta_2)=(0,1)$ have the same probability and $(\omega_1^*,\omega_1^{**})=(\omega_2^*,\omega_2^{**})$.
\end{Lemma}
The proof of
Lemma \ref{LC.1} is in Supplementary Appendix I and the proof of Lemma \ref{L3.2}, which is used to prove Theorems 1 and 1', is in Online Appendix F. We now prove Theorems 1 and 1' taking Lemmas \ref{L3.2} and \ref{LC.1} as given.
We consider two cases separately, depending on the order of $\pi^o$ and $\pi^*$.

\subsection{Case 1: $\pi^o \geq \pi^*$}\label{secC.1}
Lemma \ref{L3.2} implies that $q(1,1)+q(0,0)-q(1,0)-q(0,1) >0$ and $\Pr(\overline{\theta}=1|\boldsymbol{a}) \leq \pi^*$ for every $\boldsymbol{a} \in \{0,1\}^2$.
Therefore, $\Pr(\overline{\theta}=1) <\pi^* \leq \pi^o$, which implies that an opportunistic principal chooses $\boldsymbol{\theta}=(0,0)$
with positive probability. From Lemma \ref{L3.3},
the principal's actions are strategic substitutes. Therefore, the principal he chooses $\boldsymbol{\theta}=(1,1)$ with zero probability, which implies statement 1.

Let $\pi$ be the ex ante probability of offense taking place.  Lemma \ref{LC.1} implies that $(\theta_1,\theta_2)$ equals $(1,0)$ and $(0,1)$ with equal probabilities.
Conditional on $\theta_i=0$, the probability that  $\boldsymbol{\theta}=(0,0)$ is
\begin{equation}\label{C.1}
\beta \equiv \frac{1-\pi}{1-\pi/2}.
\end{equation}
Let $Q_1 \equiv Q_{1,1}=Q_{1,2}$ and $Q_0 \equiv Q_{0,1}=Q_{0,2}$, where the equalities are guaranteed to hold by Lemma \ref{LC.1}.
Since $\boldsymbol{\theta}=(1,1)$ occurs with zero probability,
\begin{equation}\label{C.2}
    Q_{1} = \delta \Phi(\omega^{**}) +(1-\delta)\alpha
\end{equation}
and
\begin{equation}\label{C.3}
    Q_{0} = \delta \Big( \beta \Phi (\omega^{**}) +(1-\beta) \Phi (\omega^*) \Big)+(1-\delta) \alpha.
\end{equation}
Subtracting (\ref{3.12}) from (\ref{3.11}) yields
\begin{equation}\label{C.4}
\omega^*-\omega^{**}=    b - \frac{c}{q(1,1)} \cdot \frac{\displaystyle -1 + Q_0/Q_1}{Q_{0}}.
\end{equation}
\begin{Lemma}\label{LC.2}
$\omega^*-\omega^{**} \in (0,b)$.
\end{Lemma}
\begin{proof}[Proof of Lemma C.2:]
According to (\ref{C.2}) and (\ref{C.3}),
$\omega^*-\omega^{**}>0$ is equivalent to $Q_0 > Q_1$. To see this, suppose by way of contradiction that
$Q_{0}\leq Q_{1}$. Equation (\ref{C.4}) implies that $\omega^* \geq \omega^{**} + b > \omega^{**}$. The comparison between (\ref{3.11}) and (\ref{3.12}) then yields $Q_{0} > Q_{1}$, the desired contradiction. Since $Q_0>Q_1$, the term $ \frac{-1 + Q_0/Q_1}{Q_{0}}$ is strictly positive, which shows that $\omega^*-\omega^{**}<b$.
\end{proof}
Let
\begin{equation*}
\mathcal{I} \equiv  \frac{\Pr(a_1=a_2=1|\textrm{ } \overline{\theta}=1 )}{\Pr(a_1=a_2=1|\textrm{ } \overline{\theta}=0)}.
\end{equation*}
Since an opportunistic principal mixes between $(0,1)$, $(1,0)$, and $(0,0)$, we have
  \begin{equation}\label{C.5}
\mathcal{I} \equiv  \frac{\displaystyle \big(\delta \Phi(\omega^*) +(1-\delta)\alpha\big)\big(\delta \Phi(\omega^{**})+(1-\delta)\alpha\big)}{\displaystyle \big(\delta \Phi(\omega^{**})+(1-\delta)\alpha\big)^2}=\frac{\delta \Phi(\omega^*) +(1-\delta)\alpha}{\delta \Phi(\omega^{**})+(1-\delta)\alpha}.
\end{equation}
Since $q(1,1) \in (0,1)$, the judge assigns probability $\pi^*$ to $\overline{\theta}=1$ after observing $(a_1,a_2)=(1,1)$. This implies that
\begin{equation}\label{C.6}
   \frac{\pi}{1-\pi}=\frac{l^*}{\mathcal{I}} \quad \textrm{where} \quad l^* \equiv \frac{\pi^*}{1-\pi^*}.
\end{equation}
Plugging (\ref{C.6}) into (\ref{C.1}), we obtain the following expressions for
$\beta$ and $1-\beta$:
\begin{equation}\label{C.7}
    \beta = \frac{2 \mathcal{I}}{l^*+2\mathcal{I}} \textrm{ and } 1-\beta= \frac{l^*}{l^*+2\mathcal{I}}.
\end{equation}
Plugging (\ref{C.7}) into (\ref{C.2}) and (\ref{C.3}) then yields
\begin{equation}\label{C.8}
    \frac{Q_0}{Q_1} =\beta +(1-\beta)\mathcal{I} =\frac{(l^*+2) \mathcal{I}}{l^*+2\mathcal{I}}.
\end{equation}
Plugging
 (\ref{3.11}) and (\ref{3.12}) into (\ref{C.8}), we obtain
\begin{equation}\label{C.9}
    \frac{|\omega^*-c-b|}{|\omega^{**}-c|} =
     \frac{(l^*+2) \mathcal{I}}{l^*+2\mathcal{I}}.
\end{equation}
This leads to the following lemma.
\begin{Lemma}\label{LC.3}
If $\omega^* \rightarrow -\infty$, then $\mathcal{I} \rightarrow 1$ and $\pi \rightarrow \pi^*$.
\end{Lemma}
\begin{proof}[Proof of Lemma C.3:]
Since $\omega^*-\omega^{**} \in (0,b)$, the difference between
$|\omega^*-c-b|$ and $|\omega^{**}-c|$ is at most $b$. The LHS of (\ref{C.9}) converges to $1$ as $\omega^* \rightarrow -\infty$. Since the RHS of (\ref{C.9}) is strictly increasing in $\mathcal{I}$ and is equal to 1 when $\mathcal{I} = 1$, $\mathcal{I}$ must converge to 1 as $\omega^* \rightarrow -\infty$. Equation~(\ref{C.6}) then shows that $\pi$ converges to $\pi^*$.
\end{proof}
We now show that $\omega^* \rightarrow -\infty$
as $L \rightarrow + \infty$. Recall an opportunistic principal is indifference between committing zero and one offense if
\begin{equation}\label{C.10}
   (\delta L)^{-1}
    =q(1,1) \Big(
    \delta \Phi(\omega^{**})+(1-\delta)\alpha
    \Big)\Big(
    \Phi(\omega^*)-\Phi(\omega^{**})
    \Big).
\end{equation}
Suppose that there exists a sequence $\{L(n),\omega^* (n),\omega^{**}(n), q(n), \pi(n) \big\}_{n=1}^{\infty}$
such that
\begin{enumerate}
  \item $L(n) \geq \overline{L}$ for every $n \in \mathbb{N}$, and
  $\lim_{n \rightarrow \infty} L(n)=\infty$;
  \item for each $n \in \mathbb{N}$, $(\omega^* (n),\omega^{**}(n), q(n), \pi(n))$ is an equilibrium when $L=L(n)$;
  \item $\lim_{n \rightarrow \infty} \omega^{**} (n) = \omega^{**}$ for some finite $\omega^{**} \in \mathbb{R}$.
\end{enumerate}
Since $\delta \Phi(\omega^{**}(n))+(1-\delta)\alpha$ is bounded below away from $0$,
(\ref{C.10}) implies that
\begin{itemize}
  \item \textit{either} there exists a subsequence $\{ k_n \}_{n=1}^{\infty} \subset \mathbb{N}$ such that:
$\lim_{n \rightarrow \infty} q(k_n) = 0$.
  \item \textit{or} there exists a subsequence $\{ k_n \}_{n=1}^{\infty} \subset \mathbb{N}$ such that
    $\lim_{n \rightarrow \infty} \big( \Phi(\omega^*(k_n))-\Phi(\omega^{**}(k_n)) \big) = 0$. The finite limit of $\omega^{**}$ imposed by Condition 3 above then implies that $\lim_{n \rightarrow \infty} \big( \omega^*(k_n)-\omega^{**}(k_n) \big) = 0$.
\end{itemize}
First, suppose that $\lim_{n \rightarrow \infty} q(k_n) = 0$ for some subsequence $\{ k_n \}_{n=1}^{\infty}$. Then (\ref{3.11}) and (\ref{3.12}) imply that $\omega^*(k_n)$ and $\omega^{**}(k_n)$ both go to $-\infty$, which contradicts the condition that $\omega^{**} (n)$ converges to some finite number.
Next, suppose that $\lim_{n \rightarrow \infty}( \omega^*(k_n)-\omega^{**}(k_n)) = 0$
for some subsequence $\{ k_n \}_{n=1}^{\infty}$.
Since $\omega^{**}(k_n)$ converges to a finite limit, both $Q_0(k_n)$ and $Q_1(k_n)$ are bounded below away from $0$. Given the expressions for $Q_0$ and $Q_1$, this implies that $Q_0(k_n)/Q_1(k_n)$ converges to $1$ as $n \rightarrow \infty$. From the previous step, we know that there does not exist any
subsequence of $\{ k_n \}_{n=1}^{\infty}$ such that $q(k_n)$ converges to $0$. Equivalently, there exists $\eta >0$ such that $q(k_n) \geq \eta$ for every $n \in \mathbb{N}$.
Expression (\ref{C.4}) then implies that
$\omega^*(k_n)-\omega^{**}(k_n)$ converges to $b$, which contradicts the hypothesis that $\lim_{n \rightarrow \infty} \omega^*(k_n)-\omega^{**}(k_n) = 0$.

\subsection{Case 2: $\pi^o < \pi^*$}\label{secC.2}
First, we show that an opportunistic principal \textit{commits two offenses} with positive probability in every equilibrium.
Suppose toward a contradiction that he never commits two offenses. From Lemma \ref{L3.1}, a virtuous principal never commits any offense. Therefore, the equilibrium probability of that an offense occurs cannot exceed $\pi^o$, which is strictly less than $\pi^*$. As a result,
\begin{equation}\label{C.11}
    \mathcal{I} \equiv \frac{\Pr(a_1=a_2=1|\overline{\theta}=1)}{\Pr(a_1=a_2=1|\overline{\theta}=0)} \geq \frac{\pi^o}{1-\pi^o} \Big/ \frac{\pi^*}{1-\pi^*} > 1.
\end{equation}
The expressions for $Q_0$ and $Q_1$ in (\ref{C.2}) and (\ref{C.3}) also apply to this setting. The derivations contained in Appendix C.1 imply that for every $\varepsilon>0$, there exists $\overline{L}_{\varepsilon}>0$ such that when $L \geq L_{\varepsilon}$, the informativeness ratio is less than $1+\varepsilon$. This contradicts (\ref{C.11}), which requires that the informativeness ratio be strictly bounded away from $1$.

Since $q(1,1) \in (0,1)$ and $q(1,1)+q(0,0)-q(1,0)-q(0,1)>0$, Lemma \ref{L3.3} implies that an opportunistic principal
cannot be indifferent between committing zero and two offenses. Lemma \ref{LC.1} implies that he chooses $\boldsymbol{\theta} = (0,1), (1,0)$ and $(1,1)$ with positive probability, and chooses  $\boldsymbol{\theta} = (0,0)$ with zero probability. Therefore, $\Pr(\overline{\theta}=1)=\pi^o$.
The informativeness ratio is pinned down by Bayes rule:
\begin{equation}\label{C.12}
    \mathcal{I} \frac{\Pr(\overline{\theta}=1)}{1-\Pr(\overline{\theta}=1)} =\frac{\Pr(\overline{\theta}=1|a_1=a_2=1)}{1-\Pr(\overline{\theta}=1|a_1=a_2=1)}.
\end{equation}

We now show that $\theta_1$ and $\theta_2$ are negative correlated, as claimed in the first statement of both theorems. Suppose by way of contradiction that
$\theta_1$ and $\theta_2$ are independent or positively correlated,  Lemma \ref{LC.1} implies that $Q_1 \geq Q_0$. The reporting cutoff equations~(\ref{3.11}) and~(\ref{3.12}) then show that $\omega^*-\omega^{**}  \geq b$. When $L$ is large enough, we have $q(1,1) \rightarrow 0$ and $\omega^* \rightarrow -\infty$, and the analysis of the single-agent case shows that $\mathcal{I} \rightarrow \infty$. This contradicts (\ref{C.12})
and the conclusion $\Pr(\overline{\theta}=1|a_1=a_2=1)=\pi^*$ of Lemma \ref{L3.2}.

\section{Proofs of Theorem 2 and 2'}\label{secD}
Our proof uses the following result, which is proved in Online Appendix F.
\begin{Lemma}\label{LD.1}
There exists $\overline{L}>0$ such that $q(1,1)<1$ for every $L >\overline{L}$ and corresponding equilibrium.
\end{Lemma}
\paragraph{Uncorrelated Offenses:} First, suppose by way of contradiction that $\Pr(\theta_1=1|\theta_2=1)< \Pr(\theta_1=1|\theta_2=0)$.
Since $\theta_1$ and $\theta_2$ are both binary, $\Pr(\theta_2=1|\theta_1=1)< \Pr(\theta_2=1|\theta_1=0)$.
Since $\omega_i^*>\omega_i^{**}$ for $i \in \{1,2\}$, $\Pr(\theta_1=1|\boldsymbol{a}=(1,1))< \Pr(\theta_1=1|\boldsymbol{a}=(1,0))$ and
$\Pr(\theta_2=1|\boldsymbol{a}=(1,1))< \Pr(\theta_2=1|\boldsymbol{a}=(0,1))$.
Therefore,
\begin{equation}\label{D.1}
    \max_{i \in \{1,2\}} \Pr(\theta_i=1|\boldsymbol{a}=(1,1)) < \max \Big\{
    \max_{i \in \{1,2\}} \Pr(\theta_i=1|\boldsymbol{a}=(1,0)),
        \max_{i \in \{1,2\}} \Pr(\theta_i=1|\boldsymbol{a}=(0,1))
    \Big\}.
\end{equation}
Decision rule (\ref{2.5}) implies that $\max \{q(0,1),q(1,0)\} \geq q(1,1)$ and Refinement 2 requires that $q(1,1) \geq \max \{q(0,1),q(1,0)\}$. These inequalities imply that $q(1,1)=\max \{q(0,1),q(1,0)\}$. Refinement 1 requires that $q(0,0)=0$ and Lemma \ref{L3.1} implies that $\max \{q(1,1),q(1,0),q(0,1),q(0,0)\}>0$. Therefore, $q(1,1)=\max \{q(0,1),q(1,0)\}>0$. Inequality (\ref{D.1}) and decision rule (\ref{2.5}) rule out the possibility that $q(1,1)=\max \{q(0,1),q(1,0)\} \in (0,1)$. Therefore,
$q(1,1)=\max\{q(0,1),q(1,0)\}=1$, which contradicts Lemma \ref{LD.1}.

Next, suppose by way of contradiction that $\Pr(\theta_1=1|\theta_2=1)> \Pr(\theta_1=1|\theta_2=0)$.
Since $\theta_1$ and $\theta_2$ are both binary, $\Pr(\theta_2=1|\theta_1=1)>\Pr(\theta_2=1|\theta_1=0)$.
Since $\omega_i^*>\omega_i^{**}$ for $i \in \{1,2\}$, $\Pr(\theta_1=1|\boldsymbol{a}=(1,1))> \Pr(\theta_1=1|\boldsymbol{a}=(1,0))$ and
$\Pr(\theta_2=1|\boldsymbol{a}=(1,1))> \Pr(\theta_2=1|\boldsymbol{a}=(0,1))$.
Therefore,
\begin{equation}\label{D.2}
    \max_{i \in \{1,2\}} \Pr(\theta_i=1|\boldsymbol{a}=(1,1)) > \max \Big\{
    \max_{i \in \{1,2\}} \Pr(\theta_i=1|\boldsymbol{a}=(1,0)),
        \max_{i \in \{1,2\}} \Pr(\theta_i=1|\boldsymbol{a}=(0,1))
    \Big\}.
\end{equation}
From Lemma \ref{LD.1} and monotonicity, we have $q(1,1) \in (0,1)$. Decision rule (\ref{2.5}) then requires that
\begin{equation}
    \max_{i \in \{1,2\}} \Pr(\theta_i=1|\boldsymbol{a}=(1,1))=\pi^*.
\end{equation}
As a result, the RHS of (\ref{D.2}) is strictly less than $\pi^*$, which according to decision rule (\ref{2.5}) leads to $q(0,1)=q(1,0)=0$. Therefore, $q(1,1)+q(0,0)-q(1,0)-q(0,1)>0$. Under such conviction probabilities, the two offenses are strategic substitutes, which contradicts the hypothesis that $\Pr(\theta_1=1|\theta_2=1)> \Pr(\theta_1=1|\theta_2=0)$.

\paragraph{Linear Conviction Probabilities:} First, we show that choosing $\boldsymbol{\theta}=(0,0)$ is weakly optimal for an opportunistic principal. Suppose toward a contradiction that it is not. Lemma \ref{L3.1} then implies that $\Pr(\overline{\theta}=1)= \pi^o$. The conclusion in the first part suggests that $\Pr(\theta_1=1)=1-\sqrt{1-\pi^o}$,
$|\omega_i^*-\omega_i^{**}|=b$ for $i \in \{1,2\}$, and
\begin{equation*}
    \Pr(\theta_1=1|\boldsymbol{a}=(1,1))=\Pr(\theta_1=1|\boldsymbol{a}=(1,0))= \frac{\delta \Phi(\omega_1^*)+(1-\delta)\alpha}{\delta \Phi(\omega_1^{**})+(1-\delta)\alpha} \Pr(\theta_1=1).
\end{equation*}
For any $\pi^o>0$, the RHS is strictly greater than $\pi^*$ when $L$ is sufficiently large.
Under decision rule (\ref{2.5}), this implies that $q(1,1)=1$, which contradicts Lemma \ref{LD.1}.

Since $\Pr(\theta_1=1|\theta_2=1)= \Pr(\theta_1=1|\theta_2=0)$ and
$\boldsymbol{\theta}=(0,0)$ is weakly optimal for an opportunistic principal, for every $\pi^o \in (0,1)$, an opportunistic principal must be indifferent between all offense profiles $\boldsymbol{\theta} \in \{(0,0),(0,1),(1,0),(1,1)\}$: his actions are neither strict complements nor strict substitutes.
Lemma \ref{L3.3} then shows that $q(1,1)+q(0,0)=q(1,0)+q(0,1)$. Refinement 1 requires that $q(0,0)=0$. Suppose toward a contradiction that $q(1,0)>q(0,1)$. Then, an opportunistic principal strictly prefers $\boldsymbol{\theta}=(0,1)$ to $\boldsymbol{\theta}=(1,0)$, which contradicts the previous conclusion that he is indifferent between these profiles.

\paragraph{Limiting Properties:} Parts 1 and 2 of our proof imply that $|\omega_i^*-\omega_i^{**}|=b$ and that an opportunistic principal is indifferent between $\boldsymbol{\theta} \in \{(0,0),(0,1),(1,0),(1,1)\}$. The principal's indifference condition implies that $\omega_i^* \rightarrow -\infty$ as $\lim_{L \rightarrow +\infty} \lim_{\delta\rightarrow 1}$. As in Proposition \ref{Prop1}, the informativeness ratio of each agent's report converges to infinity and $\Pr(\theta_i=1)\rightarrow 0$. The latter implies that $\Pr(\overline{\theta}=1) \rightarrow 0$.

\section{Proofs of Theorem 3 and Proposition 2}\label{secE}
Our proof uses the following lemma, which is shown in Online Appendix H.
\begin{Lemma}\label{LE.1}
Fix any $n \geq 2$ and suppose that the judge uses decision rule (\ref{2.4}). There exists $\overline{L}>0$ such that for every $L > \overline{L}$ and every symmetric Bayes Nash equilibrium that satisfies Refinement 1
\begin{enumerate}
  \item the principal is convicted with positive probability only if $\boldsymbol{a}=(1,1,...,1)$;
  \item the principal commits at most one offense.
\end{enumerate}
\end{Lemma}
We derive formulas for agents' reporting cutoffs $(\omega_n^*,\omega_n^{**})$, the informativeness of reports $\mathcal{I}_n$ and the equilibrium probability of $\overline{\theta}=1$, denoted by $\pi_n$. Agent $i$'s reporting cutoff is
\begin{equation}\label{F.1}
    \omega_n^* =b+c-\frac{c}{q_n Q_{1,n}} \textrm{ when } \theta_i=1
\end{equation}
and
\begin{equation}\label{F.2}
    \omega_n^{**} =c-\frac{c}{q_n Q_{0,n}} \textrm{ when } \theta_i=0
\end{equation}
where
\begin{equation}\label{F.3}
    Q_{1,n} \equiv \Big(
    \delta \Phi(\omega_n^{**})+(1-\delta)\alpha
    \Big)^{n-1},
    \end{equation}
    \begin{equation}\label{F.4}
     Q_{0,n} \equiv \frac{n \mathcal{I}_n}{(n-1)l^*+n \mathcal{I}_n}\Big(
    \delta \Phi(\omega_n^{**})+(1-\delta) \alpha
    \Big)^{n-1}
    +
    \frac{(n-1)l^*}{(n-1)l^*+n \mathcal{I}_n}\Big(
    \delta \Phi(\omega_n^{**})+(1-\delta) \alpha
    \Big)^{n-2}\Big(
    \delta \Phi(\omega_n^{*})+(1-\delta) \alpha
    \Big).
\end{equation}
In any symmetric equilibrium, the aggregate informativeness of reports, defined in (\ref{5.2}), can be written as
\begin{equation*}
    \mathcal{I}_n =\frac{\delta \Phi(\omega_n^*)+(1-\delta)\alpha}{\delta \Phi(\omega_n^{**})+(1-\delta)\alpha}.
\end{equation*}
Since the judge is indifferent between convicting and acquitting the principal when there are $n$ accusations, we have
\begin{equation}\label{F.5}
    \mathcal{I}_n  =\frac{\pi^*}{1-\pi^*} \Big/ \frac{\pi_n}{1-\pi_n}.
\end{equation}
When $L$ is large enough, the principal is indifferent between committing an offense against a single agent and committing no offense, which leads to the indifference condition
\begin{equation}\label{F.6}
    \frac{1}{\delta L} =
q_n \Big(\Phi(\omega_n^*)-\Phi(\omega_n^{**})\Big)
    \Big(
    \delta \Phi(\omega_n^{**})+(1-\delta) \alpha
    \Big)^{n-1}.
\end{equation}

\paragraph{Reporting Cutoffs \& Distance Between Cutoffs:} In this part, we show that $\omega_k^* > \omega_n^*$. Suppose toward a contradiction that $\omega_k^* \leq \omega_n^*$. From (\ref{F.1}), we have
\begin{equation}\label{F.7}
    q_k \Big(
    \delta \Phi(\omega_k^{**})+(1-\delta)\alpha
    \Big)^{k-1} \leq q_n \Big(
    \delta \Phi(\omega_n^{**})+(1-\delta) \alpha
    \Big)^{n-1}.
\end{equation}
Therefore, $q_k Q_{1,k} \leq q_n Q_{1,n}$, which is equivalent to
\begin{equation*}
     q_k \Big(
    \delta \Phi(\omega_k^{**})+(1-\delta) \alpha
    \Big)^{k-1}
    \Big(\Phi(\omega_n^*)-\Phi(\omega_n^{**}) \Big)
    \leq q_n \Big(
    \delta \Phi(\omega_n^{**})+(1-\delta) \alpha
    \Big)^{n-1}\Big(\Phi(\omega_n^*)-\Phi(\omega_n^{**})\Big)
    \end{equation*}
    \begin{equation*}
    =q_k \Big(
    \delta \Phi(\omega_k^{**})+(1-\delta) \alpha
    \Big)^{k-1}\Big(\Phi(\omega_k^*)-\Phi(\omega_k^{**})\Big).
\end{equation*}
This implies that
\begin{equation}\label{F.8}
   \Phi(\omega_n^*)-\Phi(\omega_n^{**}) \leq \Phi(\omega_k^*)-\Phi(\omega_k^{**}).
\end{equation}
Since $\omega_k^* \leq \omega_n^*$, (\ref{F.8}) holds only if
\begin{equation}\label{F.9}
    \omega_n^*-\omega_n^{**}\leq \omega_k^*-\omega_k^{**},
\end{equation}
which in turn implies that $\omega_k^{**}\leq \omega_n^{**}$
and, therefore, that $q_k Q_{0,k} \leq q_n Q_{0,n}$.
Computing the two sides of (\ref{F.9}) by subtracting (\ref{F.2}) from (\ref{F.1}) for $n$ and $k$, we obtain
\begin{equation*}
    \omega_n^*-\omega_n^{**}=b-\frac{c}{q_n} \frac{Q_{0,n}-Q_{1,n}}{Q_{1,n}Q_{0,n}}
    \textrm{ and }
        \omega_k^*-\omega_k^{**}=b-\frac{c}{q_k} \frac{Q_{0,k}-Q_{1,k}}{Q_{1,k}Q_{0,k}}.
\end{equation*}
Since we have shown that
$q_k Q_{1,k} \leq q_n Q_{1,n}$ and $q_k Q_{0,k} \leq q_n Q_{0,n}$,
(\ref{F.9}) can hold only if
\begin{equation}\label{F.10}
    q_n (Q_{0,n}-Q_{1,n}) \geq q_k (Q_{0,k}-Q_{1,k}).
\end{equation}
We have
\begin{equation*}
    Q_{0,n}-Q_{1,n}
    = \frac{(n-1)l^*}{(n-1)l^*+n\mathcal{I}_n}
   \delta \Big(
    \Phi(\omega_n^*)-\Phi(\omega_n^{**})\Big)\Big(
    \delta \Phi(\omega_n^{**})+(1-\delta) \alpha
    \Big)^{n-2}
\end{equation*}
and
\begin{equation*}
    \delta \Big(
    \Phi(\omega_n^*)-\Phi(\omega_n^{**})\Big)\Big(
    \delta \Phi(\omega_n^{**})+(1-\delta) \alpha
    \Big)^{n-2}=L^{-1} q_n^{-1} \frac{1}{\delta \Phi(\omega_n^{**})+(1-\delta) \alpha}.
\end{equation*}
Combining this with  (\ref{F.6}) and (\ref{F.10}) yields
\begin{equation*}
    \frac{(n-1)l^*}{\displaystyle (n-1)l^*\Big(
    \delta \Phi(\omega_n^{**})+(1-\delta) \alpha
    \Big) + n \Big(
    \delta \Phi(\omega_n^{*})+(1-\delta) \alpha
    \Big)}
    \end{equation*}
    \begin{equation*}
    \geq
       \frac{(k-1)l^*}{\displaystyle (k-1)l^*\Big(
    \delta \Phi(\omega_k^{**})+(1-\delta) \alpha
    \Big) + k \Big(
    \delta \Phi(\omega_k^{*})+(1-\delta) \alpha
    \Big)},
\end{equation*}
which may be re-expressed as
\begin{equation*}
    (n-1)(k-1) l^* \Big(
    \delta \Phi(\omega_k^{**})+(1-\delta)\alpha
    \Big)+(n-1)k \Big(
    \delta \Phi(\omega_k^{*})+(1-\delta) \alpha
    \Big)
    \end{equation*}
    \begin{equation*}
    \geq   (n-1)(k-1) l^* \Big(
    \delta \Phi(\omega_n^{**})+(1-\delta) \alpha
    \Big)+(k-1)n \Big(
    \delta \Phi(\omega_n^{*})+(1-\delta) \alpha
    \Big).
\end{equation*}
This inequality cannot hold because $\delta \Phi(\omega_k^{**})+(1-\delta) \alpha < \delta \Phi(\omega_n^{**})+(1-\delta) \alpha$,
$\delta \Phi(\omega_k^{*})+(1-\delta) \alpha < \delta \Phi(\omega_n^{*})+(1-\delta) \alpha$ and $(n-1)k< (k-1)n$, where the last inequality comes from the assumption that $k>n$. This leads to a contradiction and shows that $\omega_k^* > \omega_n^*$ whenever $k>n$.

The relation $k>n$ was only used in the last step. Using the fact that $\omega_k^* > \omega_n^*$ and repeating the earlier argument up until (\ref{F.9}), we obtain
\begin{equation}\label{F.11}
    \omega_n^*-\omega_n^{**} > \omega_k^*-\omega_k^{**}.
\end{equation}
This, together with
$\omega_k^* > \omega_n^*$,
implies that $\omega_k^{**}> \omega_n^{**}$.

\paragraph{Report Informativeness and Probability of Offense:} We show that $\mathcal{I}_n > \mathcal{I}_k$.
Since $q(1,1,...,1)=\pi^*$, this together with (\ref{5.1}) proves that the  ex ante probability of offense is ranked as claimed by Theorem~\ref{Theorem3}.

Applying (\ref{F.1}) and (\ref{F.2}) to both $n$ and $k$, we get
\begin{equation}\label{F.12}
    \frac{\omega_n^*-b-c}{\omega_k^*-b-c} = \frac{q_k Q_{1,k}}{q_n Q_{1,n}} \quad \textrm{and} \quad
    \frac{\omega_n^{**}-c}{\omega_k^{**}-c} = \frac{q_k Q_{1,k} (\beta_k+(1-\beta_k)\mathcal{I}_k)}{q_n Q_{1,n}(\beta_n+(1-\beta_n)\mathcal{I}_n)}.
\end{equation}
First, we show that
\begin{equation}\label{F.13}
    \frac{\omega_n^*-b-c}{\omega_k^*-b-c} > \frac{\omega_n^{**}-c}{\omega_k^{**}-c}.
\end{equation}
Suppose toward a contradiction that the RHS of (\ref{F.13}) is at least as large as the LHS of (\ref{F.13}). Then,
\begin{equation}\label{F.14}
    \frac{\omega_n^{**}-c -(\omega_n^*-b-c)}{\omega_k^{**}-c-(\omega_k^*-b-c)} \geq \frac{\omega_n^*-b-c}{\omega_k^*-b-c}.
\end{equation}
The RHS of (\ref{F.14}) is strictly greater than $1$ since $0>\omega_k^*> \omega_n^*$ when $L$ is large enough. This implies that the LHS of (\ref{F.14}) is greater than $1$, which is equivalent to
\begin{equation*}
    b- (\omega_n^*-\omega_n^{**}) > b- (\omega_k^*-\omega_k^{**}).
\end{equation*}
This contradicts (\ref{F.11}), which was established earlier, and proves~(\ref{F.13}). This, together with (\ref{F.12}), implies that
\begin{equation*}
    \beta_k +(1-\beta_k) \mathcal{I}_k < \beta_n +(1-\beta_n) \mathcal{I}_n.
\end{equation*}
Plugging in the expressions of $\mathcal{I}_n$ and $\mathcal{I}_k$ obtain in (\ref{F.5}), we get
\begin{equation*}
    \mathcal{I}_k \big(k+(k-1)l^* \big) \big(n \mathcal{I}_n+(n-1)l^*\big)
    <
    \mathcal{I}_n \big(n+(n-1)l^* \big) \big(k \mathcal{I}_k+(k-1)l^* \big).
\end{equation*}
Letting $\Delta \equiv \mathcal{I}_k-\mathcal{I}_n$, the previous inequality reduces to
\begin{equation*}
(k-n) \mathcal{I}_n (\mathcal{I}_k -1)=    (k-n) \mathcal{I}_n (\mathcal{I}_n+\Delta -1) < k \Delta
    -\Big(
    l^* (k-1)(n-1)+nk
    \Big) \Delta.
\end{equation*}
Suppose toward a contradiction that $\Delta \geq 0$. Then,
the LHS is strictly positive since $\mathcal{I}_k>1$ and $k>n$. The RHS is negative since $l^* (k-1)(n-1)+nk>k$. This leads to the desired contradiction, and implies that $\Delta <0$ or, equivalently, that $\mathcal{I}_n > \mathcal{I}_k$. Equation~(\ref{5.1}) then implies that $\Pr(\overline{\theta}=1)$ increases when the number of agents increases from $n$ to $k$.
\end{spacing}


\newpage
\begin{thebibliography}{99}
\bibitem{pa} Ali, S. Nageeb, Maximilian Mihm and Lucas Siga (2018) ``Adverse Selection in Distributive Politics,'' Working Paper.
\bibitem{pa} Ambrus, Attila, and Satoru Takahashi (2008) ``Multi-sender Cheap Talk with Restricted State Spaces,'' \textit{Theoretical Economics}, 3, 1-27.
\bibitem{pa} Austen-Smith, David and Jeffrey Banks (1996) ``Information Aggregation, Rationality, and the Condorcet Jury Theorem,'' \textit{American Political Science Review}, 90(1), 34-45.
\bibitem{pa} Ba, Bocar (2018) ``Going the Extra Mile: The Cost of Complaint Filing, Accountability, and Law Enforcement Outcomes in Chicago,'' Working paper
\bibitem{pa} Ba, Bocar and Roman Rivera (2019) ``The Effect of Police Oversight on Crime and Allegations of Misconduct: Evidence from Chicago,'' Working paper.
\bibitem{pa} Baliga, Sandeep, Ethan Bueno de Mesquita and Alexander Wolitzky (2020) ``Deterrence with Imperfect Attribution,'' \textit{American Political Science Review}, forthcoming.
\bibitem{pa} Banerjee,  Abhijit (1992) ``A Simple Model of Herd Behavior,'' \textit{Quarterly Journal of Economics}, 107(3), 797-817.
\bibitem{pa} Bar-Hillel, Maya (1984) ``Probabilistic Analysis in Legal Factfinding,'' \textit{Acta Psychologica}, 56, 267-284.
\bibitem{pa} Battaglini, Marco (2002) ``Multiple Referrals and Multidimensional Cheap Talk,'' \textit{Econometrica}, 70(4), 1379-1401.
\bibitem{pa} Battaglini, Marco (2017) ``Public Protests and Policy Making,'' \textit{Quarterly Journal of Economics}, 132(1), 485-549.
\bibitem{pa} Becker, Gary (1968) ``Crime and Punishment: An Economic Approach,'' \textit{Journal of Political Economy}, 76(2), 169-217.
\bibitem{pa} Bhattacharya, Sourav (2013) ``Preference Monotonicity and Information Aggregation in Elections,'' \textit{Econometrica}, 81(3), 1229-1247.
\bibitem{pa}  Bikhchandani, Sushil,  David Hirshleifer, and Ivo Welch (1992) ``A Theory of Fads, Fashion, Custom, and Cultural Change as Information Cascades,'' \textit{Journal of Political Economy}, 100, 992-1026.
\bibitem{pa} Chamberlin, John (1974) ``Provision of Collective Goods As a Function of Group Size,'' \textit{The American Political Science Review}, 68(2), 707-716.
\bibitem{pa} Chassang, Sylvain and Gerard Padr\'{o} i Miquel (2019) ``Corruption, Intimidation and Whistle-Blowing: A Theory of Inference from Unverifiable Reports,'' \textit{Review of Economic Studies}, forthcoming.
\bibitem{pa} Cheng, Ing-Haw and Alice Hsiaw (2020) ``Reporting Sexual Misconduct in the MeToo Era,'' Working Paper.
\bibitem{pa} Cohen, Jonathan (1977) ``The Probable and The Provable,'' Oxford University Press.
\bibitem{pa} Dobbie, Will, Jacob Goldin, and Crystal S. Yang (2018) ``The Effects of Pretrial Detention on Conviction, Future Crime, and Employment: Evidence from Randomly Assigned Judges,'' \textit{American Economic Review}, 108(2), 201-240.
\bibitem{pa} Ekmekci, Mehmet and Stephan Lauermann (2019) ``Informal Elections with Dispersed Information,'' Working Paper.
\bibitem{pa} Fudenberg, Drew and David Levine (1995) ``The Theory of Learning in Games,'' MIT Press.
\bibitem{pa} Grossman, Gene and Michael Katz (1983) ``Plea Bargaining and Social Welfare,'' \textit{American Economic Review}, 73(4), 749-757.
\bibitem{pa} Harel, Alon and Ariel Porat (2009) ``Aggregating Probabilities Across Cases: Criminal Responsibility for Unspecified Offenses,'' \textit{Minnesota Law Review}, 482, 261-310.
\bibitem{pa} Harris, John (1970) ``On the Economics of Law and Order,'' \textit{Journal of Political Economy}, 78(1), 165-174.
\bibitem{pa} Kaplow, Louis (2011) ``On the Optimal Burden of Proof,'' \textit{Journal of Political Economy}, 119(6), 1104-1140.
\bibitem{pa} Landes, William (1971) ``An Economic Analysis of the Courts,'' \textit{Journal of Law and Economics}, 14(1), 61-107.
\bibitem{pa} Lee, Frances Xu and Wing Suen (2020) ``Credibility of Crime Allegations,'' \textit{American Economic Journal-Microeconomics}, 12, 220-259.
\bibitem{pa} Lynch, Gerard (1987) ``RICO: The Crime of Being a Criminal,'' \textit{Columbia Law Review}, 87(4), 661-764.
\bibitem{pa} Miceli, Thomas (1991) ``Optimal Criminal Procedure: Fairness and Deterrence,'' \textit{International Review of Law and Economics} 11(1), 3-10.
\bibitem{pa} Morgan, John and Phillip Stocken (2008) ``Information Aggregation in Polls,'' \textit{American Economic Review}, 98(3), 864-896.
\bibitem{pa} Morgan, Rachel and Grace Kena (2016) ``Criminal Victimization, 2016,'' Bulletin, Bureau of Justice Statistics.
\bibitem{pa} Myerson, Roger (1978) ``Refinements of the Nash Equilibrium Concept,'' \textit{International Journal of Game Theory}, 7(2), 73-80.
\bibitem{pa} Naess, Ole-Andreas Elvik (2020) ``Under-reporting of Crime,'' Working Paper.
\bibitem{pa} Ottaviani, Marco and Peter Norman S{\o}rensen (2000) ``Herd Behavior and Investment: Comment,'' \textit{American Economic Review}, 90(3), 695-704.
\bibitem{pa} Persico, Nicola (2004) ``Committee Design with Endogenous Information,'' \textit{Review of Economic Studies}, 70(1), 1-27.
\bibitem{pa} RAND Cooperation (2018) ``Sexual Assault and Sexual Harassment in the US Military,'' Technical Report.
\bibitem{pa} Reinganum, Jennifer (1988) ``Plea Bargaining and Prosecutorial Discretion,'' \textit{American Economic Review}, 78(4), 713-728.
\bibitem{pa} Scharfstein, David and Jeremy Stein (1990) ``Herd Behavior and Investment,'' \textit{Amercian Economic Review}, 80(3), 465-479.
\bibitem{pa} Robertson, Bernard and G. A. Vignaux (1993) ``Probabilit--The Logic of the Law,'' \textit{Oxford Journal of Legal Studies}, 13(4), 457-478.
\bibitem{pa} Schauer, Federick and Richard Zeckhauser (1996) ``On the Degree of Confidence for Adverse Decisions,'' \textit{Journal of Legal Studies}, 25(1), 27-52.
\bibitem{pa} Schmitz, Patrick and Thomas Tr\"{o}ger (2011) ``The Suboptimality of the Majority Rule,'' \textit{Games and Economic Behavior}, 651-665.
\bibitem{pa} Siegel, Ron and Bruno Strulovici (2020) ``Judicial Mechanism Design,'' Working Paper.
\bibitem{pa} Silva, Francesco (2019) ``If We Confess Our Sins,'' \textit{International Economic Review}, 60(3), 1389--1412.
\bibitem{pa} Smith, Lones and Peter Norman S{\o}rensen (2000) ``Pathological Outcomes of Observational Learning,'' \textit{Econometrica}, 68(2), 371-398.
\bibitem{pa} Stigler, George (1970) ``The Optimal Enforcement of Laws,'' \textit{Journal of Political Economy}, 78(3), 526-536.
\bibitem{pa} Strulovici, Bruno (2010) ``Learning while Voting: Determinants of Collective Experimentation,'' {\em Econometrica}, 78(3), 933--971.
\bibitem{pa} Strulovici, Bruno (2020) ``Can Society Function without Ethical Agents? An Informational Perspective,'' Working Paper, Northwestern University.
\bibitem{pa} U.S. Equal Employment Opportunity Commission (2017) ``Fiscal Year 2017 Enforcement And Litigation Data,'' Research Brief.
\bibitem{pa} USMSPB (2018) ``Update on Sexual Harassment in the Federal Workplace,'' Research Brief.
\end{thebibliography}
\end{document}


\title{Supplementary Appendix (Not for Publication) \\Crime Entanglement, Deterrence, and Witness Credibility}
\author{Harry Pei \and Bruno Strulovici}
\date{\today}

\maketitle

\appendix
\begin{spacing}{1.5}
\section*{I. $\quad$ Symmetry of Equilibrium}\label{secB}
From Lemma 3.1, $q(0,0)=q(1,0)=q(0,1)=0$ and $q(1,1) \in (0,1)$. Let $q  \equiv q(1,1)$, $\Psi_i^* \equiv \delta \Phi(\omega_i^*)+(1-\delta) \alpha$, and $\Psi_i^{**} \equiv \delta \Phi(\omega_i^{**})+(1-\delta) \alpha$.
Agent $i$'s reporting cutoffs are
\begin{equation}\label{B.1}
    \omega_i^*  =b -c \frac{1- qQ_{1,j}}{qQ_{1,j}}
\end{equation}
and
\begin{equation}\label{B.2}
    \omega_i^{**}  = -c \frac{1-qQ_{0,j}}{qQ_{0,j}}
\end{equation}
where $Q_{1,j}$ is the probability of $a_j=1$ conditional on $\theta_i=1$, given by
\begin{equation}\label{B.3}
    \Pr(\theta_j=1|\theta_i=1) \Psi_j^* + (1-\Pr(\theta_j=1|\theta_i=1))\Psi_j^{**},
\end{equation}
and $Q_{0,j}$ is the probability of $a_j=1$ conditional on $\theta_i=0$, given by
\begin{equation}\label{B.4}
      \Pr(\theta_j=1|\theta_i=0)  \Psi_j^* + (1-\Pr(\theta_j=1|\theta_i=0))  \Psi_j^{**}.
\end{equation}
When $\boldsymbol{\theta}$ changes from $(0,0)$ to $(1,0)$, the probability of conviction increases by
\begin{equation*}
\delta q \Phi(\omega_2^{**}) \Big(\Phi(\omega_1^*)-\Phi(\omega_1^{**})\Big).
\end{equation*}
When $\boldsymbol{\theta}$ changes from from $(0,1)$ to $(1,1)$, the probability of conviction increases by
\begin{equation*}
\delta q \Phi(\omega_2^{*}) \Big(\Phi(\omega_1^*)-\Phi(\omega_1^{**})\Big).
\end{equation*}
Therefore, $\boldsymbol{\theta}$ takes on the values $(1,0)$ and $(0,1)$ with the same, strictly positive probability, only if
\begin{equation}\label{B.5}
\frac{\Phi(\omega_1^*)}{\Phi(\omega_1^{**})}=\frac{\Phi(\omega_2^*)}{\Phi(\omega_2^{**})},
\end{equation}
and $\boldsymbol{\theta}$ takes on the value $(1,0)$ with a strictly higher probability than $(0,1)$ only if
\begin{equation}\label{B.6}
\frac{\Phi(\omega_1^*)}{\Phi(\omega_1^{**})}\leq \frac{\Phi(\omega_2^*)}{\Phi(\omega_2^{**})},
\end{equation}
with equality when $\boldsymbol{\theta}$ is equal to $(0,1)$ with strictly positive probability.

First, we show that the symmetry in the principal's equilibrium strategy implies the symmetry of the agents' reporting cutoffs.
Suppose toward a contradiction that $\omega_i^* > \omega_j^*$. Then $Q_{1,j} > Q_{1,i}$, which implies that $\omega_i^{**}< \omega_j^{**}$, and thatthat
\begin{equation}\label{B.7}
\frac{\Phi(\omega_i^*)}{\Phi(\omega_i^{**})} > \frac{\Phi(\omega_j^*)}{\Phi(\omega_j^{**})}.
\end{equation}
This contradicts the hypothesis that the principal chooses $\boldsymbol{\theta}$ being $(1,0)$ and $(0,1)$ with the same probability.

Next, we assume by way of contradiction that $\boldsymbol{\theta}$ equals $(1,0)$ with a strictly higher probability than$(0,1)$. This implies inequality (\ref{B.6}). The rest of the proof is divided in two cases: Section I.1 considers the case $\pi^o \geq \pi^*$ and Section I.2 considers the case $\pi^o < \pi^*$.

\subsection*{I.1 $\quad$ Case 1: $\pi^o \geq \pi^*$}
Lemma 2.1 implies that an opportunistic principal commits no offense with strictly positive probability. Lemmas 3.1 and 3.2 then imply that the principal commits two offenses with zero probability.

First, consider the case in which the principal chooses $\boldsymbol{\theta}=(0,1)$ with positive probability. As argued earlier, (\ref{B.6}) holds with equality, and for each $j \in \{1,2\}$ we have
\begin{equation}\label{B.8}
 Q_{1,j}=\Psi_j^{**} \textrm{ and }   Q_{0,j} = \Pr(\theta_j=1|\theta_i=0) \Psi_j^{*}
    + (1-\Pr(\theta_j=1|\theta_i=0)) \Psi_j^{**}.
\end{equation}
Therefore,
\begin{enumerate}
  \item If $\omega_1^* \geq \omega_2^*$, (\ref{B.1}) implies that $\omega_1^{**} < \omega_2^{**}$, which contradicts (\ref{B.6}).
  \item If $\omega_1^* < \omega_2^*$, we must have $\omega_1^{**} > \omega_2^{**}$, and $\Phi(\omega_1^{*})  \Phi(\omega_2^{**}) < \Phi(\omega_2^{*})  \Phi(\omega_1^{**})$, which leads to a contradiction.
\end{enumerate}
Second, consider the case in which the principal chooses $\boldsymbol{\theta}=(0,1)$ with zero probability. Upon observing $\theta_2=1$, agent $2$ attaches probability $1$ to $\boldsymbol{\theta}=(0,1)$ in every \textit{proper equilibrium}, because the principal's expected payoff is
strictly lower when he chooses $\boldsymbol{\theta}=(1,1)$ than when he chooses $\boldsymbol{\theta}=(0,1)$. Therefore, (\ref{B.8}) still applies, and $\omega_1^*=\omega_1^{**}+b$ and $\omega_2^*-\omega_2^{**} \in (0,b)$.

First, we show that $\omega_1^{**} < \omega_2^{**}$. Suppose toward a contradiction that $\omega_1^{**} \geq \omega_2^{**}$. Then, comparing the expressions for $\omega_1^{**}$ and $\omega_2^{**}$ yields
\begin{equation*}
    \Psi_2^{**} \geq \pi \Psi_1^* +(1-\pi) \Psi_1^{**} > \Psi_1^{**},
\end{equation*}
a contradiction. This shows that $\omega_1^{**} < \omega_2^{**}$.
Second, (\ref{B.6}) implies that $\omega_1^*< \omega_2^*$, and therefore, $Q_{1,2}<Q_{1,1}$. Since
$\Pr(\theta_2=1|\theta_1=1)=\Pr(\theta_1=1|\theta_2=1)=0$, we have
 $\omega_2^{**}< \omega_1^{**}$. This contradicts the previous conclusion that $\omega_1^{**} < \omega_2^{**}$.

\subsection*{I.2 $\quad$  $\pi^o < \pi^*$}\label{subB.2}
If an opportunistic principal chooses $\boldsymbol{\theta}=(1,1)$ with zero probability, then the proof follows the same steps as the previous case. If he chooses $\boldsymbol{\theta}=(1,1)$ with strictly positive probability, then Lemmas 3.1 and 3.2 imply that he chooses $\boldsymbol{\theta}=(0,0)$ with zero probability. First, consider the case in which $\boldsymbol{\theta}=(0,1)$ is chosen with strictly positive probability.
\begin{enumerate}
  \item If $\omega_1^* \geq \omega_2^*$, then (\ref{B.1}) implies that $Q_{1,2} \geq Q_{1,1}$ or, equivalently, that
  \begin{equation*}
    \Pr(\theta_2=1|\theta_1=1) \Psi_2^* +(1-\Pr(\theta_2=1|\theta_1=1)) \Psi_2^{**}
    \geq \Pr(\theta_1=1|\theta_2=1) \Psi_1^* +(1-\Pr(\theta_1=1|\theta_2=1)) \Psi_1^{**}.
  \end{equation*}
  The hypothesis that $\boldsymbol{\theta}=(1,0)$ has a strictly higher probability than $\boldsymbol{\theta}=(0,1)$ implies that
  $\Pr(\theta_2=1|\theta_1=1)<\Pr(\theta_1=1|\theta_2=1)$. The hypothesis that $\omega_1^* \geq \omega_2^*$ implies that $\Psi_1^*> \Psi_2^*$. From Lemma 2.1, we have $\Psi_i^*>\Psi_i^{**}$ for every $i \in \{1,2\}$. Therefore, the above inequality holds only if
$\omega_1^{**} < \omega_2^{**}$, which contradicts (\ref{B.6}).
  \item If $\omega_1^* < \omega_2^*$, then  $\omega_1^{**} > \omega_2^{**}$ and $\Phi(\omega_1^{*})  \Phi(\omega_2^{**}) < \Phi(\omega_2^{*})  \Phi(\omega_1^{**})$, which leads to a contradiction.
\end{enumerate}

\section*{J. $\quad$ Existence of Equilibrium}
Section J.1 proves existence of a proper equilibrium that satisfies Refinements 1 and 2 under the aggregate probabilities principle (or ``APP''). Section J.2 proves the existence of a proper equilibrium that satisfies Refinements 1 and 2 under the distinct probabilities principle (or ``DPP'').
\subsection*{J.1 $\quad$ Aggregate Probabilities Principle: Decision Rule (2.4)}\label{subA.1}
\paragraph{Case 1: $\boldsymbol{\pi^o \geq \pi^*}$} We show that when $L$ is large enough and $\pi^o \geq \pi^*$, there exists a \textit{symmetric} Bayes Nash equilibrium that satisfies Refinements 1 and 2 and possesses the following three properties:
\begin{enumerate}
  \item $q(\boldsymbol{a})>0$ if and only if $\boldsymbol{a}=(1,1,...,1)$.
  \item The opportunistic principal commits either no offense or exactly one offense.
  \item Each agent witnesses an offense with probability strictly between $0$ and $1$.
\end{enumerate}
This equilibrium is a proper equilibrium since all the information sets for all players occur with strictly positive probability. For every $\omega^*,\omega^{**} \in \mathbb{R}$, let $\Psi^* \equiv \delta \Phi(\omega^*) +(1-\delta) \alpha$ and
    $\Psi^{**} \equiv \delta \Phi(\omega^{**}) +(1-\delta) \alpha$.
\begin{PropositionJ1}
There exists $\overline{L} >0$ such that for every $L > \overline{L}$, there exists a triple $(\omega^*,\omega^{**},q) \in \mathbb{R}_- \times \mathbb{R}_- \times (0,1)$ that solves the following three equations:
\begin{equation}\label{A.1}
    \frac{q}{c} (\omega^*-c-b) =-\frac{1}{(\Psi^{**})^{n-1}}
\end{equation}
\begin{equation}\label{A.2}
        \frac{q}{c} (\omega^{**}-c) =-\frac{n}{n+(n-1)l^*} \frac{1}{(\Psi^{**})^{n-1}}
        -\frac{(n-1)l^*}{n+(n-1)l^*} \frac{1}{(\Psi^{**})^{n-2} \Psi^*}
\end{equation}
\begin{equation}\label{A.3}
    \frac{1}{\delta L}= q (\Psi^{**})^{n-1} (\Psi^*-\Psi^{**}).
\end{equation}
\end{PropositionJ1}
\begin{proof}
The proof consists of two steps. In \textbf{Step 1}, we show that the following expression is bounded below away from $0$:
\begin{equation}\label{A.4}
   A \equiv \inf_{(\omega^{*},\omega^{**}) \textrm{ that solves (\ref{A.1}) and (\ref{A.2}) when $q=1$}} \delta   \Big(
    \Phi(\omega^*)-\Phi(\omega^{**})
    \Big) (\Psi^{**})^{n-1}.
\end{equation}
This is shown by establishing lower bounds on $\Phi(\omega^{**})$ and $\Phi(\omega^*)-\Phi(\omega^{**})$, respectively.
Since $(\omega^*,\omega^{**})$ solves (\ref{A.1}) and (\ref{A.2}) when $q=1$, we have:
\begin{equation*}
    \omega^{**} \geq \omega^* -b \geq c -\frac{c}{(1-\delta)^{n-1} \alpha^{n-1}}
\end{equation*}
and therefore,
\begin{equation}\label{A.5}
\Phi(\omega^{**}) \geq \Phi\Big(
c -\frac{c}{(1-\delta)^{n-1} \alpha^{n-1}}
\Big).
\end{equation}
Next, we introduce the variable $\Delta \equiv \omega^*-\omega^{**}$, which is strictly between $0$ and $b$. Subtracting equation (\ref{A.2}) from (\ref{A.1}) and plugging in $q=1$, we have
\begin{equation}\label{A.6}
    \frac{b-\Delta}{c}
    =
    \frac{ l^*(n-1)}{l^* (n-1)+n} \big(\Psi^*\big)^{-1}
    \big(\Psi^{**}\big)^{-(n-1)}
    \big( \Psi^*-\Psi^{**} \big).
\end{equation}
We consider two cases separately,
\begin{enumerate}
\item When $\Delta  \geq b/2$,
\begin{equation}\label{A.7}
\Psi^*-\Psi^{**} \geq \frac{b\delta}{2} \phi(\omega^{**}) \geq \frac{b\delta}{2} \phi\Big(
c -\frac{c}{(1-\delta)^{n-1} \alpha^{n-1}}
\Big).
\end{equation}
\item When $\Delta < b/2$, (\ref{A.6}) implies that
  \begin{equation}\label{A.8}
    \Psi^*-\Psi^{**} \geq \frac{b((n-1)l^*+n)}{2(n-1)l^*c} \Psi^*
    \big(\Psi^{**}\big)^{n-1} \geq \frac{b((n-1)l^*+n)}{2(n-1)l^*c} \Big(
    \delta \Phi \big(c -\frac{c}{(1-\delta)^{n-1} \alpha^{n-1}} \big)+(1-\delta)\alpha
    \Big)^n.
  \end{equation}
\end{enumerate}
Taking the minimum of the right-hand sides of (\ref{A.7}) and (\ref{A.8}), we obtain a lower bound for $\Psi^*-\Psi^{**}$. This together with (\ref{A.5}) implies a strictly positive lower bound for (\ref{A.4}), which we denote by $\underline{A}$.

In \textbf{Step 2}, we show that when $L > \underline{A}^{-1}$, there exists a solution to (\ref{A.1}), (\ref{A.2}) and (\ref{A.3}). For every $(\Phi^*,\Phi^{**},q) \in [0,1]^2 \times [1/L,1]$, let $f \equiv (f_1,f_2,f_3):[0,1]^2 \times [1/L,1] \rightarrow [0,1]^2 \times [1/L,1]$
be the following mapping:
\begin{equation}\label{A.9}
    f_1 (\Phi^*,\Phi^{**},q) = \Phi \Big(b+
   c- \frac{c}{q (\Psi^{**})^{n-1}}
    \Big),
\end{equation}
\begin{equation}\label{A.10}
    f_2 (\Phi^*,\Phi^{**},q) = \Phi \Big(
  c-
\frac{c(n-1)l^*}{q((n-1)l^*+n)} \frac{1}{\Psi^* (\Psi^{**})^{n-2}}- \frac{cn}{q((n-1)l^*+n)} \frac{1}{(\Psi^{**})^{n-1}}
    \Big),
\end{equation}
\begin{equation}\label{A.11}
f_3 (\Phi^*,\Phi^{**},q) = \min \Big\{
    1, \frac{1}{\delta L} \frac{1}{\displaystyle  (\Psi^{**})^{n-1} \Big(\Phi^*-\Phi^{**}\Big)}
    \Big\}.
\end{equation}
where $\Psi^* \equiv \delta \Phi^* +(1-\delta)\alpha$ and $\Psi^{**} \equiv \delta \Phi^{**} +(1-\delta)\alpha$.
Since $f$ is continuous, Brouwer's fixed point theorem implies the existence of a fixed point.

Next, we show that if $(\Phi^*,\Phi^{**},q)$ is a fixed point, then $q <1$. This implies that every solution to the fixed point problem
solves the system of equations (\ref{A.1}), (\ref{A.2}) and (\ref{A.3}) as (\ref{A.11}) and (\ref{A.3}) are the same when $q <1$.
Suppose toward a contradiction that $q=1$. Then $\Phi^{-1}(\Phi^*)$ and $\Phi^{-1}(\Phi^{**})$ solve (\ref{A.1}) and (\ref{A.2}). From Part I of the proof, the assumption that $L >\underline{A}^{-1}$ implies that
\begin{equation*}
    \frac{1}{\delta L} \frac{1}{\displaystyle (\Psi^{**})^{n-1} \big(\Phi^*-\Phi^{**}\big)}<1.
\end{equation*}
 Therefore the RHS of (\ref{A.11}) is strictly less than  $1$. This contradicts the
claim that $(\Phi^*,\Phi^{**},1)$ is a fixed point of $f$. Therefore, the value of $q$ at the fixed point is strictly less than $1$.
\end{proof}
Given the tuple $(\omega^*,\omega^{**},q) \in \mathbb{R}_- \times \mathbb{R}_- \times (0,1)$, one can then uniquely pin down the equilibrium probability of an offense $\pi \equiv \Pr(\overline{\theta}=1)$ via the equation
\begin{equation}\label{A.12}
    \frac{\delta \Phi(\omega^*)+(1-\delta) \alpha}{\delta \Phi(\omega^{**})+(1-\delta)\alpha} = l^* \Big/ \frac{\pi}{1-\pi}.
\end{equation}
To see this, note that the principal is convicted with interior probability after the judge observes $\boldsymbol{a} = (1,1,...,1)$, and therefore, $\Pr(\overline{\theta}=1|\boldsymbol{a} = (1,1,...,1))=\pi^*$. Since equations (\ref{A.1}), (\ref{A.2}), (\ref{A.3}) and (\ref{A.12})
are sufficient conditions for a symmetric Bayes Nash equilibrium
that satisfies Refinement 1, the existence of a fixed point in (\ref{A.1}), (\ref{A.2}), and (\ref{A.3}) implies the existence of proper equilibrium that satisfies Refinements 1 and 2.

\paragraph{Case 2: $\boldsymbol{\pi^o < \pi^*}$} We show that when $L$ is large enough and $\pi^o < \pi^*$, there exists a \textit{symmetric} Bayes Nash equilibrium that satisfies Refinements 1 and 2, and possesses the following three properties:
\begin{enumerate}
  \item $q(\boldsymbol{a})>0$ if and only if $\boldsymbol{a}=(1,1,...,1)$.
  \item There exists $1 \leq k \leq n$ and $r \in [0,1]$ such that an opportunistic principal commits $k-1$ offenses with probability $r$
  and commits $k$ offenses with probability $1-r$.
  \item Each agent witnesses offense with probability strictly between $0$ and $1$.
\end{enumerate}
Such a Bayes Nash equilibrium is a proper equilibrium since all the information sets for all players occur with strictly positive probability. For every $\Psi^*$, $\Psi^{**}$, $k$ and $r$, let
\begin{equation}\label{A.12}
    Q_1(\Psi^*,\Psi^{**},k,r) \equiv \frac{r(k-1)\Psi^{**}+(1-r)k \Psi^*}{k-r} (\Psi^*)^{k-2} (\Psi^{**})^{n-k}
\end{equation}
and
\begin{equation}\label{A.13}
    Q_0(\Psi^*,\Psi^{**},k,r) \equiv \frac{\displaystyle r\pi^o (1-\frac{k-1}{n})(\Psi^*)^{k-1} (\Psi^{**})^{n-k} +(1-r) \pi^o (1-\frac{k}{n}) (\Psi^*)^{k} (\Psi^{**})^{n-k-1}+(1-\pi^o) (\Psi^{**})^{n-1}}
    {\displaystyle r\pi^o (1-\frac{k-1}{n})+ (1-r) \pi^o (1-\frac{k}{n}) + (1-\pi^o)}.
\end{equation}
Intuitively, $Q_1(\Psi^*,\Psi^{**},k,r)$ is an agent's expected probability that all other agents will accuse the principal conditional on $\theta_i=1$, and $Q_0(\Psi^*,\Psi^{**},k,r)$ is an agent's expected probability that all other agents will accuse the principal conditional on $\theta_i=0$. For every $k\in \{1,2,...,n\}$ and $r \in [0,1]$, let $\Lambda(k-r)$ be the set of values for $(\Psi^*,\Psi^{**},q) \in [0,1] \times [0,1] \times [0,1]$ such that
\begin{equation*}
    qL (\Psi^*)^{k-1} (\Psi^{**})^{n-k} (\Psi^*-\Psi^{**})=1 \textrm{ if } r \in (0,1),
\end{equation*}
\begin{equation*}
     qL (\Psi^*)^{k-1} (\Psi^{**})^{n-k} (\Psi^*-\Psi^{**}) \geq 1
     \geq qL (\Psi^*)^{k-2} (\Psi^{**})^{n-k+1} (\Psi^*-\Psi^{**})
     \textrm{ if } r=1
\end{equation*}
and
\begin{equation*}
     qL (\Psi^*)^{k} (\Psi^{**})^{n-k-1} (\Psi^*-\Psi^{**}) \geq 1
     \geq qL (\Psi^*)^{k-1} (\Psi^{**})^{n-k} (\Psi^*-\Psi^{**})
     \textrm{ if } r=0.
\end{equation*}
Letting
\begin{equation}\label{A.14}
    l^* \equiv \frac{\pi^* (1-\pi^o)}{(1-\pi^*)\pi^o},
\end{equation}
we have $l^* > 1$. Let $\Psi^* \equiv (1-\delta)\alpha +\delta \Phi(\omega^*)$ and $\Psi^{**} \equiv (1-\delta)\alpha +\delta \Phi(\omega^{**})$.
We show the following proposition.
\begin{PropositionJ2}
There exists $\overline{L} >0$ such that for every $L > \overline{L}$, there exists a tuple $(\omega^*,\omega^{**},q, k,r) \in \mathbb{R}_- \times \mathbb{R}_- \times (0,1] \times \{1,2,...,n\} \times [0,1]$ such that
\begin{equation}\label{A.15}
    (1-r) \Big(\frac{\Psi^*}{\Psi^{**}}\Big)^{k}
    +r \Big(\frac{\Psi^*}{\Psi^{**}}\Big)^{k-1}=l^*,
\end{equation}
\begin{equation}\label{A.16}
  (\Psi^*,\Psi^{**},q) \in \Lambda (k-r),
\end{equation}
\begin{equation}\label{A.17}
    \frac{q}{c} (\omega^*-c-b)=-\frac{1}{Q_1(\Psi^*,\Psi^{**},k,r)},
\end{equation}
and \begin{equation}\label{A.18}
    \frac{q}{c} (\omega^{**}-c)=-\frac{1}{Q_0(\Psi^*,\Psi^{**},k,r)}.
\end{equation}
\end{PropositionJ2}
\begin{proof} As in the proof of Proposition J.1, we show that
\begin{equation}\label{A.19}
   A \equiv \inf_{(\omega^{*},\omega^{**}) \textrm{ that solves (\ref{A.15}) and (\ref{A.16}) when $q=1$}}(\Psi^*)^{k-1} (\Psi^{**})^{n-k} (\Psi^*-\Psi^{**})
\end{equation}
is strictly bounded below away from $0$. This bound comes from the fact that
\begin{equation*}
    (\Psi^*)^{k-1} (\Psi^{**})^{n-k} \geq (\Psi^{**})^n,
\end{equation*}
and a similar argument as the one used in Proposition J.1 to establish strictly positive lower bounds for $\Psi^{**}$ and $\Psi^*-\Psi^{**}$. The resulting lower bound for $A$ is denoted by $\underline{A}$.

In what follows, we focus on $L \geq \underline{A}^{-1}$.
Let $\mathcal{I} \equiv \Psi^* /\Psi^{**}$ and $\beta \equiv k-r$. We introduce the following correspondence:
\begin{equation*}
    F: [0,1] \times [1,\frac{1}{(1-\delta)(1- \alpha)}] \times [1/L,1] \times [0,n] \rightrightarrows [0,1] \times [1,\frac{1}{(1-\delta)(1- \alpha)}] \times [1/L,1] \times [0,n],
\end{equation*}
where $F_2(\Psi^{**},\mathcal{I},q,k-r)$ is uniquely pinned down by
\begin{equation}\label{A.20}
     (1-r) F_2(\Psi^{**},\mathcal{I},q,k-r)^{k}
    +r F_2(\Psi^{**},\mathcal{I},q,k-r)^{k-1}=l^*,
\end{equation}
$F_3(\Psi^{**},\mathcal{I},q,k-r)$ is the set of $\min \{1,q'\}$ such that
\begin{equation}\label{A.21}
    \Big(\Psi^{**} F_2(\Psi^{**},\mathcal{I},q,k-r), \Psi^{**}, q' \Big) \in \Lambda (k-r),
\end{equation}
$F_1(\Psi^{**},\mathcal{I},q,k-r)$ is the set of $\Psi'$ such that there exists $q \in F_3(\Psi^{**},\mathcal{I},q,k-r)$ such that
\begin{equation}\label{A.22}
    \Psi'=\delta \Phi\Big(
    c-\frac{c}{q} \frac{1}{Q_0(F_2(\Psi^{**},\mathcal{I},q,k-r)\Psi^{**},\Psi^{**}, k,r)}
    \Big)+(1-\delta)\alpha,
\end{equation}
and
$F_4(\Psi^{**},\mathcal{I},q,k-r)$ is the set of integers $k-r$ such that there exists $q \in F_3(\Psi^{**},\mathcal{I},q,k-r)$ for which
\begin{equation}\label{A.23}
    \frac{qb}{c}=\frac{1}{Q_1}-\frac{1}{Q_0}.
\end{equation}
Since $F$ is defined on a non-empty, convex, and compact set and $F$ has a closed graph, Kakutani's fixed point theorem implies that $F$ has a fixed point. Proceeding as in the proof of Proposition J.1, one can show that
in every fixed point $(\Phi^{**},\mathcal{I},q, k-r)$, $q$ must be strictly less than $1$. Therefore, every fixed point of correspondence $F$
solves (\ref{A.15}), (\ref{A.16}), (\ref{A.17}) and (\ref{A.18}).
\end{proof}

\subsection*{J.2 $\quad$ Distinct Probabilities Principle: Decision Rule (2.5)}\label{subA.2}
We construct for $L$ large enough a symmetric proper equilibrium that satisfies Refinements 1 and 2. This also shows the existence part of Proposition 5.
\begin{enumerate}
  \item $q(\boldsymbol{\theta})$ is linear in the number of $1$s contained in $\boldsymbol{\theta}$,
  with $q(0,0,...,0)=0$.
  \item For every $i \neq j$, $\Pr(\theta_i=1|\theta_j=1)=\Pr(\theta_i=1|\theta_j=0)$ and
  $(\omega_i^*,\omega_i^{**})=(\omega_j^*,\omega_j^{**})$.
  \item An opportunistic principal is indifferent between every $\boldsymbol{\theta} \in \{0,1\}^n$.
\end{enumerate}
We establish the following proposition.
\begin{PropositionJ3}
There exists $(\Phi^*, q^*, r^*) \in [0,1]\times (0,\frac{1}{n}) \times [0,\pi^*]$ such that
\begin{equation}\label{A.24}
    q^* \Big(\Phi^*-\Phi^{**} \Big) =\frac{1}{\delta L},
\end{equation}
\begin{equation}\label{A.25}
\Phi^{-1}  (\Phi^*) = b+c+c(1-\delta)\alpha + c\delta \Big(r^* \Phi^* +(1-r^*) \Phi^{**}\Big) -\frac{c}{q^*},
\end{equation}
and
\begin{equation}\label{A.26}
    \frac{\delta \Phi^*+(1-\delta) \alpha}{\delta \Phi^{**}+(1-\delta)\alpha} \frac{r^*}{1-r^*} = \frac{\pi^*}{1-\pi^*},
\end{equation}
where
\begin{equation*}
\Phi^{**} \equiv \Phi \Big( \Phi^{-1}  (\Phi^*)-b \Big).
\end{equation*}
\end{PropositionJ3}
Mapping this into the game studied in the main text,
$\Phi^*$ is a strategic agent's probability
of accusing the principal
when he has witnessed an offense,
$q^*$ is the incremental probability of conviction when one additional agent accuses the principal, and $r^*$ is the probability that $\theta_i$ equals $1$. Equation (\ref{A.24}) is the principal's indifference condition when deciding whether to choose $\theta_i=1$ or $\theta_i=0$, Equation (\ref{A.25}) characterizes each agent's reporting cutoff when he has witnessed an offense, and Equation (\ref{A.26}) implies that, upon observing $a_i=1$, the judge attaches probability $\pi^*$ to $\theta_i=1$. Since $q(\cdot)$ is linear in the number of  accusations and $\theta_i$ is independent of $\theta_j$, the distance between the reporting cutoffs of each agent is exactly equal to $b$.

\begin{proof}
First, we calculate a lower bound on
\begin{equation*}
  \delta (  \Phi^*-\Phi^{**} )
\end{equation*}
where $(\Phi^*,r^*)$
is the solution to (\ref{A.25}) and (\ref{A.26}) when $q^*$ is fixed to be $1/n$. Let $\omega^* \equiv \Phi^{-1}  (\Phi^*)$. Equation~(\ref{A.25}) provides upper and lower bounds for $\omega^*$:
\begin{equation*}
b+c+c(1-\delta)\alpha+c\delta-cn \geq   \omega^* \geq b+c+c(1-\delta)\alpha-cn,
\end{equation*}
which shows that
\begin{equation}\label{A.27}
    \delta  (  \Phi^*-\Phi^{**} ) \geq
    \delta \min_{\omega \in [b+c+c(1-\delta)\alpha-cn, b+c+c(1-\delta)\alpha+c\delta-cn] }\Big\{
    \Phi(\omega)-\Phi(\omega-b)
    \Big\} \equiv \underline{A}
\end{equation}
In what follows, let $L > n\underline{A}^{-1}$. For every
$(\Phi^*, q^*, r^*) \in [0,1]\times [\frac{1}{n \underline{A}},\frac{1}{n}] \times [0,\pi^*]$, denite $f \equiv (f_1,f_2,f_3): [0,1]\times [\frac{1}{n \underline{A}},\frac{1}{n}] \times [0,\pi^*] \rightarrow [0,1]\times [\frac{1}{n \underline{A}},\frac{1}{n}] \times [0,\pi^*]$ by
\begin{equation}\label{A.28}
    f_1 (\Phi^*,q^*,r^*) \equiv  \Phi \Big( b+c+c(1-\delta)\alpha + c\delta \big(r^* \Phi^* +(1-r^*) \Phi \big( \Phi^{-1}  (\Phi^*)-b \big)\big) -\frac{c}{q^*} \Big),
\end{equation}
\begin{equation}\label{A.29}
    f_2(\Phi^*,q^*,r^*) \equiv \min \Big\{\frac{1}{n}, \frac{1}{\delta L} \frac{1}{\Phi^*-\Phi \Big( \Phi^{-1}  (\Phi^*)-b \Big)} \Big\},
\end{equation}
and
\begin{equation}\label{A.30}
    f_3(\Phi^*,q^*,r^*) \equiv \frac{\displaystyle  \pi^* (\delta \Phi \big( \Phi^{-1}  (\Phi^*)-b \big) +(1-\delta)\alpha)}{\displaystyle
   \delta \Big( \pi^* \Phi \big( \Phi^{-1}  (\Phi^*)-b \big) +(1-\pi^*) \Phi^*  \Big)
    +(1-\delta) \alpha}.
\end{equation}
Brouwer's fixed point theorem that $f$ has a fixed point. We show that $q^*< 1/n$ for every fixed point of $f$, which implies that any fixed point solves (\ref{A.24}), (\ref{A.25}) and (\ref{A.26}). Suppose toward a contradiction that $(\Phi^*, q^*, r^*)$ is a fixed point of $f$ with $q^*=1/n$. Then,
\begin{equation}\label{A.31}
\delta L \Big( \Phi^*-\Phi \big( \Phi^{-1}  (\Phi^*)-b \big) \Big) \leq n.
\end{equation}
However, from (\ref{A.27}) and the fact that $L >n \underline{A}^{-1}$, the LHS of (\ref{A.31}) is strictly greater than $n$, which yields the desired contradiction.
\end{proof}

\section*{K. $\quad$ Extensions}
We study two sets of extensions. Section K.1 concerns alternative specifications of agents' payoffs. Section K.2 concerns other strategies for agents' behavioral types.

\subsection*{K.1 $\quad$ Alternative Specifications of Agents' Payoffs}
\paragraph{Social Preferences:} Recall that $\overline{\theta} \equiv \max \{\theta_1,\theta_2,...,\theta_n\}$. Suppose that agent $i$'s payoff is $0$ when the principal is convicted and
\begin{equation}\label{H.1}
\omega_i - b
    \Big((1-\gamma) \theta_i +\gamma \overline{\theta} \Big)
    -ca_i
\end{equation}
when the principal is acquitted. Intuitively, an agent's  payoff depends not only on whether the principal has committed an offense observed by the agent himself, but also on whether the principal is guilty of committing any offense. The parameter $\gamma \in [0,1]$ measures the intensity of agents' social preferences. The baseline model corresponds to $\gamma=0$.

The agents decide whether to accuse the principal based on their beliefs about $\overline{\theta}$ after observing their own $\theta_i$. As a result, an agent's strategy in any given equilibrium is still characterized by two cutoffs: $\omega^*$ when $\theta_i=1$ and $\omega^{**}$ when $\theta_i=0$. Whether the principal's incentives to commit offenses are strategic complements or substitutes is still determined by the sign of $q(0,0)+q(1,1)-q(0,1)-q(1,0)$. When $L$ is large, two accusations are needed to convict the principal with positive probability and the principal's decisions to commit offenses are strategic substitutes. This induces negative correlation in agents' private information.

Agents' reporting cutoffs are given by
\begin{equation}\label{H.2}
    \omega^* =b+c-\frac{c}{q_m \Psi^{**}}
\end{equation}
and
\begin{equation}\label{H.3}
    \omega^{**}=c +\frac{b \Psi^* (1-\beta) \gamma}{
    \beta \Psi^{**} +(1-\beta) \Psi^*}-\frac{c}{\displaystyle q_m \Big(
    \beta \Psi^{**} +(1-\beta) \Psi^*
    \Big)}
\end{equation}
where $q_m \in (0,1)$ is the probability of conviction when there are two accusations, $\Psi^* \equiv \delta \Phi(\omega^*) +(1-\delta) \alpha$,
 $\Psi^{**} \equiv \delta \Phi(\omega^{**}) +(1-\delta) \alpha$ and $\beta$ is the probability of $\theta_j=0$ conditional on $\theta_i=0$.
 Compared to the baseline model,  (\ref{H.3}) contains the new term
 \begin{equation}\label{H.4}
    \frac{b \Psi^* (1-\beta) \gamma}{
    \beta \Psi^{**} +(1-\beta) \Psi^*},
 \end{equation}
 which measures the impact of social preferences on an agent's equilibrium strategy. We have
\begin{equation}\label{H.5}
0 \leq    \frac{b \Psi^* (1-\beta)\gamma}{
    \beta \Psi^{**} +(1-\beta) \Psi^*}
    \leq \gamma b.
\end{equation}
Proceeding as in Lemma 3.2 of the main text, we can show that
$\omega^*-\omega^{**} \in [0,b]$.
From~(\ref{H.2}) and (\ref{H.3}), we have
\begin{equation}\label{H.6}
    \frac{|\omega^*-c-b|}{\displaystyle \Big|\omega^{**}-c- \frac{b \Psi^* (1-\beta) \gamma}{
    \beta \Psi^{**} +(1-\beta) \Psi^*} \Big|} =\frac{\beta \Psi^{**} +(1-\beta) \Psi^*}{\Psi^{**}}=\beta +(1-\beta) \frac{\Psi^*}{\Psi^{**}}
    =\frac{(l^*+2) \mathcal{I}_m}{l^*+2\mathcal{I}_m},
\end{equation}
where $\mathcal{I}_m \equiv \Psi^*/\Psi^{**}$ measures the aggregate informativeness of agents' accusations.

As $L \rightarrow +\infty$, one can show that both $\omega^*$ and $\omega^{**}$ go to $-\infty$ proceeding as in the proof of Theorem 1. Since the difference between the denominator and the numerator of the LHS of (\ref{H.6})
is at most $b$,  the value of (\ref{H.6}) converges to $1$. This implies that $\mathcal{I}_m$  converges to $1$. Consequently, agents' accusations become  arbitrarily uninformative and the equilibrium probability of offense converges to $\pi^*$.

\paragraph{Ex Post Evidence and Punishment of False Accusers:} Suppose that when an innocent principal is convicted (i.e., $\theta_1=...=\theta_n=0$ and $s=1$), some ex post evidence arrives with probability $p^*$ that reveals his innocence and results in a penalty $l$ for every agent who has accused the principal. Our earlier analysis and results continue to apply, because this new setting is formally equivalent to increasing in the baseline setting $b$. To see this, agent $i$'s indifference condition when $\theta_i=0$ is now given by
\begin{equation}\label{H.7}
    q_m Q_0 \omega_i =-c(1-q_mQ_0)-q_mQ_0 p^* l
\end{equation}
and the lower cutoff is given by
\begin{equation}\label{H.8}
 \omega_m^{**} \equiv -p^* l -c \frac{1-q_m Q_{0}}{q_m Q_{0}}=-p^* l +c-\frac{c}{q_m Q_0}.
\end{equation}
This expression is the same as in the main text except that $b$ is replaced with  $\widetilde{b} \equiv b+p^*l$.

\paragraph{Intrinsic Motive for Truth Telling:} Suppose that each agent receives an extra benefit $d$, strictly less than $c$, from accusing the principal whenever he has observed an offense, regardless of whether the principal gets convicted. Agents' reporting cutoffs are now given by
\begin{equation*}
     \omega^* = b+c-\frac{c-d}{q \Psi^{**}}
\end{equation*}
and
\begin{equation*}
    \omega^{**} = c-\frac{c}{q (\beta \Psi^{**}+(1-\beta) \Psi^*)}
\end{equation*}
where $q \in (0,1)$ is the probability of conviction conditional on both agents accusing the principal, and $\beta$ is the probability that an agent who has not witnessed any offense assigns to the other agent not having any witnessed any offense, either. Letting $\pi$ denote the equilibrium probability that an offense takes place, we have
\begin{equation*}
\beta = \frac{1-\pi}{1-\pi/2}.
\end{equation*}
Let $l^* \equiv \pi^* /(1-\pi^*)$ and
    $\mathcal{I} \equiv \Psi^*/\Psi^{**}$.
We have $\beta = \frac{2 \mathcal{I}}{l^*+2\mathcal{I}}$.
The principal's incentive constraint is given by
\begin{equation*}
\frac{1}{L}=q \Psi^{**} (\Psi^*-\Psi^{**}).
\end{equation*}
Therefore, as $L \rightarrow +\infty$, both $\omega^*$ and $\omega^{**}$ go to $-\infty$. The expressions for agent's reporting cutoffs yield
\begin{equation*}
    \frac{c+b-\omega^*}{c-d}=\frac{1}{q \Psi^{**}}
\quad \textrm{and} \quad
    \frac{c-\omega^{**}}{c}=\frac{1}{q (\beta \Psi^{**}+(1-\beta) \Psi^*)}.
\end{equation*}
Therefore,
\begin{equation*}
    \frac{(c+b-\omega^*)/(c-d)}{(c-\omega^{**})/c}
    = \frac{(l^*+2) \mathcal{I}}{l^*+2\mathcal{I}}.
\end{equation*}
In what follows, we show that as $L \rightarrow +\infty$,
the LHS of the previous equation is bounded between $1$ and $\frac{c}{c-d}$.
The lower bound $1$ is straightforward to derive. To derive the upper bound $\frac{c}{c-d}$, notice that since $\omega^* > \omega^{**}$ and $\Psi^* > \Psi^{**}$, we have
\begin{equation*}
    \frac{c+b-\omega^*}{c-d} > \frac{c-\omega^{**}}{c} > \frac{c-\omega^*}{c}.
\end{equation*}
We consider two cases.
First, when $\omega^* > \omega^{**} +b$, we have
\begin{equation*}
    \frac{c+b-\omega^*}{c-\omega^{**}} < \frac{c-\omega^{**}}{c-\omega^{**}}=1,
\end{equation*}
which is equivalent to
\begin{equation*}
    \frac{(c+b-\omega^*)/(c-d)}{(c-\omega^{**})/c}  < \frac{c}{c-d}.
\end{equation*}
Second, when $\omega^* \leq \omega^{**} +b$, both $\omega^*$ and $\omega^{**}$ go to $-\infty$ and $\frac{c+b-\omega^*}{c-\omega^{**}} \rightarrow 1$,
which implies that
\begin{equation*}
    \frac{(c+b-\omega^*)/(c-d)}{(c-\omega^{**})/c}  \rightarrow \frac{c}{c-d}.
\end{equation*}
Therefore, we have
\begin{equation*}
 1 \leq   \frac{(l^*+2) \mathcal{I}}{l^*+2\mathcal{I}} \leq \frac{c}{c-d}
\end{equation*}
as $L \rightarrow +\infty$, which yields an upper bound on the informativeness ratio $\mathcal{I}$. This upper bound is nontrivial if and only if
\begin{equation*}
    \frac{2+l^*}{2} > \frac{c}{c-d},
\end{equation*}
which requires $d$ to be sufficiently small.

\subsection*{K.2 $\quad$ Alternative Behavioral Types}
We examine the robustness of our findings
against alternative specifications of behavioral types' strategies. We allow the behavioral types' accusations to be informative about the principal's offenses and show that when behavioral types are rare and the principal's loss from being convicted is sufficiently large, the informativeness of accusations converges to $1$ and the principal probability of guilt converges to $\pi^*$, as in the baseline model. We focus on comparing outcomes with one agent and two agents.

\subsubsection*{K.2.1 $\quad$ Model and Result}
Consider the following modification of the baseline model. With probability $\delta \in (0,1)$, the agent is a strategic type who maximizes payoff function given by (2.4) in the main text. With probability $1-\delta$, the agent is a behavioral type whose reporting cutoff is $\overline{\omega}$ when $\theta_i=1$ and $\underline{\omega}$ when $\theta_i=0$. We assume that both  $\overline{\omega}$
and $\underline{\omega}$ are finite and that $\overline{\omega} \geq \underline{\omega}$, That is, the behavioral type's accusation can be informative about $\theta$.\footnote{Our analysis also applies when behavioral types are using arbitrary strategies contingent on $(\theta_i,\omega_i)$, as long as conditional on each realization of $\theta_i$, the probability that the behavioral type accuses the principal is interior, and is weakly higher when $\theta_i=1$ than when $\theta_i=0$.}

When there is only one agent, the agent's reporting cutoffs $\omega_s^*$ and $\omega_s^{**}$ are given by (3.1) and (3.2), respectively. The probability that the principal is convicted following one accusation is $q_s$, where the tuple $(q_s,\omega_s^*, \omega_s^{**})$ satisfies
\begin{equation}\label{H.9}
    q_s \Big(
    \delta (\Phi(\omega_s^*)-\Phi(\omega_s^{**})) +(1-\delta) (\Phi(\overline{\omega})-\Phi(\underline{\omega}))
    \Big)=1/L.
\end{equation}
One can show that when $\delta \rightarrow 1$ and $L$ is larger than some cutoff $L(\delta)$, the informativeness of the agent's accusation
\begin{equation*}
  \mathcal{I}_s \equiv  \frac{\delta \Phi(\omega_s^*)+(1-\delta)\Phi(\overline{\omega})}{\delta \Phi(\omega_s^{**})+(1-\delta)\Phi(\underline{\omega})}
\end{equation*}
diverges to $+\infty$. That is, the agent's accusation becomes arbitrarily informative in the limit.

In the two-agent case, for every $i \in \{1,2\}$,
agent $i$'s probability of accusing the principal is $\Psi^* \equiv \delta \Phi(\omega_m^*)+(1-\delta) \Phi(\overline{\omega})$ conditional on $\theta_i=1$ and $\Psi^{**} \equiv \delta \Phi(\omega_m^{**}) +(1-\delta) \Phi(\underline{\omega})$ conditional on $\theta_i=0$. A strategic agent's reporting cutoffs are given by
\begin{equation}\label{H.10}
    \omega_m^* \equiv b+c-\frac{c}{q_m \Psi^{**}} \quad \textrm{and} \quad
    \omega_m^{**} \equiv c-\frac{c}{\displaystyle q_m \Big(\beta \Psi^{**}+(1-\beta)\Psi^* \Big)}.
\end{equation}
Let $\mathcal{I}_m \equiv \Psi^*/\Psi^{**}$.
When $L$ is large enough, the conviction probabilities in every equilibrium
must satisfy
$q(0,0)=q(0,1)=q(1,0)=0$ and $q(1,1) \in (0,1)$. Therefore,
the expressions for $\beta$ and $1-\beta$ are the same as in (3.15).
The distance between the two cutoffs is given by
\begin{equation}\label{H.11}
    \omega_m^*-\omega_m^{**} = b- \frac{c}{q_m} \frac{(1-\beta) (\mathcal{I}_m-1)}{\Psi^{**} (\beta+(1-\beta)\mathcal{I}_m)}
    =b-\frac{c}{q_m \Psi^{**}} \frac{l^*}{2+l^*} \frac{\mathcal{I}_m-1}{\mathcal{I}_m}.
\end{equation}
This implies $\omega_m^*-\omega_m^{**}<b$, because  for $\omega_m^*-\omega_m^{**}$ to weakly exceed $b$, we need $\mathcal{I}_m \leq 1$, which can only be true when $\omega^*_m \leq \omega^{**}_m$, leading to a contradiction.

In contrast to the baseline model, when behavioral types' accusations are informative about the principal's innocence, the strategic type agents' coordination motives can \textit{reverse} the ordering between the two cutoffs. That is to say,
$\omega_m^*$ can be strictly smaller than $\omega_m^{**}$ in equilibrium. As a result, the argument showing that $\mathcal{I}_m \rightarrow 1$ when $\omega_m^* \rightarrow -\infty$ in Lemma 3.3 of the main text no longer applies. In principle, $\omega_m^*$ could be much smaller than $\omega_m^{**}$, and the ratio between the absolute values in (3.17) could converge to a limit strictly above $1$ as $\omega_m^*$ and $\omega_m^{**}$ diverge to $-\infty$. To circumvent this problem, we take an alternative approach based on the comparison between
$\omega_m^*$ and $\omega_s^*$ to establish the following proposition.
\begin{PropositionK}
There exists $\overline{L}: \mathbb{R}_+ \times (0,1) \rightarrow \mathbb{R}_+$ such that an equilibrium exists for each $L > \overline{L}(c,\delta)$. Compared to the single-agent benchmark,$q_m > q_s$, $\omega_m^* > \omega_s^*$ and $\omega_m^{**}> \omega_s^{**}$.\\
Moreover, as $\delta \rightarrow 1$ and $L \rightarrow +\infty$ with the relative speed of convergence satisfying $L \geq \overline{L}(c,\delta)$, we have $\omega_m^*, \omega_m^{**} \rightarrow -\infty$, $\mathcal{I}_m \rightarrow 1$ and $\pi_m \rightarrow \pi^*$.
\end{PropositionK}
The proof, in the next section, distinguishes two cases. Intuitively, in the \textit{regular case} where $\omega_m^* \geq \omega_m^{**}$, one can still apply the ratio condition (3.17) to show that as $\omega_m^* \rightarrow -\infty$, the LHS converges to $1$ which implies that $\mathcal{I}_m \rightarrow 1$.
In the \textit{irregular case} where $\omega_m^* < \omega_m^{**}$, the distance between $|\omega_m^*-b-c|$ and $|\omega_m^{**}-c|$ can be strictly larger than $b$ and can explode as $\omega_m^* \rightarrow -\infty$.
However, since $\omega_m^*> \omega_s^*$ and the informativeness in the single-agent benchmark grows without bound as $L \rightarrow +\infty$, it places an upper bound on the informativeness of accusations in the two-agent scenario. Since informativeness is entirely contributed by the behavioral types in the irregular case, the value of the aforementioned upper bound converges to $1$ as $\mathcal{I}_s \rightarrow +\infty$. Summing up the two cases together, we know that the agents' accusations are arbitrarily uninformative in the limit even when the behavioral types' accusations are informative.

\subsubsection*{K.2.2 $\quad$ Proof of Proposition K}
We start by comparing between the single-agent and two-agent cases when behavioral types' accusations can be informative about $\theta$, captured by the two exogenous reporting cutoffs $\overline{\omega}$ and $\underline{\omega}$ with $\overline{\omega} \geq \underline{\omega}$.

Suppose toward a contradiction that $\omega_m^* \leq \omega_s^*$. The expressions for these cutoffs imply that
\begin{equation*}
    q_m \Big(\delta \Phi(\omega_m^{**})+(1-\delta)\Phi(\underline{\omega})\Big) \leq q_s.
\end{equation*}
Therefore,
\begin{equation*}
    q_m \Psi^{**}
    \Big(
    \delta \Phi(\omega_s^*)+(1-\delta)\Phi(\overline{\omega})
    -\delta \Phi(\omega_s^{**})-(1-\delta) \Phi(\underline{\omega})
    \Big)
    \end{equation*}
    \begin{equation*}
    \leq q_s \Big(
    \delta \Phi(\omega_s^*)+(1-\delta)\Phi(\overline{\omega})
    -\delta \Phi(\omega_s^{**})-(1-\delta) \Phi(\underline{\omega})
    \Big)=1/L
   \end{equation*}
    \begin{equation*}
    =q_m \Psi^{**} \Big(
    \delta \Phi(\omega_m^*)+(1-\delta)\Phi(\overline{\omega})
    -\delta \Phi(\omega_m^{**})-(1-\delta) \Phi(\underline{\omega})
    \Big)
\end{equation*}
or, equivalently,
\begin{equation}\label{H.12}
\Phi(\omega_m^*)-\Phi(\omega_m^{**})\geq \Phi(\omega_s^*)-\Phi(\omega_s^{**}).
\end{equation}
Since $\omega_m^*-\omega_m^{**} <b= \omega_s^* -\omega_s^{**}$ and $\omega_m^* < \omega_s^*$, we have
\begin{equation}\label{H.13}
    \Phi(\omega_m^*)-\Phi(\omega_m^{**})< \Phi(\omega_s^*)-\Phi(\omega_s^{**}),
\end{equation}
which contradicts (\ref{H.12}). This shows that $\omega_m^* > \omega_s^*$. Since
$\omega_m^*-\omega_m^{**} <b= \omega_s^* -\omega_s^{**}$, we have $\omega_m^{**}> \omega_s^{**}$.
Moreover, $\omega_m^* > \omega_s^*$ implies that $q_m \Psi^{**} > q_s$, or $1 \geq \Psi^{**} > q_s/q_m$. This implies that $q_m>q_s$.

Next, we evaluate the informativeness of agents' accusations when there are two agents and $\delta$ and $L$ are sufficiently large. First, for every $X \in \mathbb{R}_+$, there exists $\overline{\delta} \in (0,1)$ and $L^*: (\overline{\delta},1) \rightarrow \mathbb{R}_+$ such that when $\delta > \overline{\delta}$ and $L > L^*(\delta)$, the first cutoff in the single-agent case satisfies
\begin{equation}\label{H.14}
    \frac{\delta \Phi(\omega_s^*)+(1-\delta)\Phi(\overline{\omega})}{\delta \Phi(\omega_s^{*}-b)+(1-\delta)\Phi(\underline{\omega})} >X,
\end{equation}
which implies that
\begin{equation}\label{H.15}
    \delta \Phi(\omega_s^*) > (1-\delta) \Big(X \Phi(\underline{\omega})-\Phi(\overline{\omega})\Big).
\end{equation}
Next, we establish an upper bound on the informativeness of accusations in the limit of the two-agent case. Consider the two-agent setting with parameter values $(L,c,\delta)$ such that
$L \geq \overline{L}(c,\delta)$, which guarantees that there exist proper equilibria satisfying Refinements 1 and 2.
In every equilibrium such that
$\omega_m^* \geq \omega_m^{**}$, the expressions for $\omega_m^* $ and $\omega_m^{**}$
imply that
\begin{equation}\label{H.16}
    \frac{|\omega^*_m-c-b|}{|\omega_m^{**}-c|} = \frac{(l^*+2) \mathcal{I}_m}{l^*+2\mathcal{I}_m}.
\end{equation}
The LHS converges to $1$ as $\omega_m^* \rightarrow -\infty$. The expression on the RHS then implies that $\mathcal{I}_m \rightarrow 1$.

In equilibria such that $\omega_m^*< \omega_m^{**}$ we have, since $\omega_s^* < \omega_m^*$,
\begin{equation*}
    \mathcal{I}_m
    \leq \frac{\delta \Phi(\omega_m^*) +(1-\delta) \Phi(\overline{\omega})}{\delta \Phi(\omega_m^{*})+(1-\delta) \Phi(\underline{\omega})}
    \underbrace{ \leq}_{\textrm{since } \mathcal{I}_m >1 \textrm{ and } \omega_m^* > \omega_s^*} \frac{\delta \Phi(\omega_s^*) +(1-\delta) \Phi(\overline{\omega})}{\delta \Phi(\omega_s^{*})+(1-\delta) \Phi(\underline{\omega})}
    \end{equation*}
    \begin{equation}\label{H.17}
    \leq \frac{\displaystyle (1-\delta) \Big(X \Phi(\underline{\omega})-\Phi(\overline{\omega})\Big)+(1-\delta) \Phi(\overline{\omega})}{\displaystyle (1-\delta) \Big(X \Phi(\underline{\omega})-\Phi(\overline{\omega})\Big)+    (1-\delta) \Phi(\underline{\omega})}
    =\frac{X \Phi(\underline{\omega})}{X \Phi(\underline{\omega}) - \Phi(\overline{\omega}) +\Phi(\underline{\omega})},
\end{equation}
which also converges to $1$ as $X \rightarrow +\infty$.

In summary, since $\omega_m^* \rightarrow -\infty$ and $X \rightarrow +\infty$ as $\delta \rightarrow 1$ and $L \rightarrow +\infty$, the informativeness ratio $\mathcal{I}_m$ converges to $1$ irrespective of whether $\omega_m^*\geq  \omega_m^{**}$ or $\omega_m^*< \omega_m^{**}$.

\subsection*{K.3 $\quad$ Principal's Payoffs}
We allow the punishment to the principal to depend on the number of offenses that he is believed to have committed, focusing on the case of two agents. The principal receives a penalty $L$ if $\Pr(\overline{\theta}=1|\boldsymbol{a}) \geq \pi^*$, and a penalty $L' (>L)$ if
$\Pr(\theta_1=\theta_2=1|\boldsymbol{a}) \geq \pi^{**}$. The evaluator is indifferent between both sentences at the cutoff $\pi^{**}$.  If the judge's posterior belief after observing $\boldsymbol{a}$ meets both of these requirements, the principal receives a penalty $L''$ that is at least $\max\{L,L'\}$.

In every proper equilibrium that satisfies Refinements 1 and 2, each agent's strategy takes the form of two cutoffs. For $i \in \{1,2\}$, let $\Psi_i^*$ be the probability that agent $i$ accuses the principal when $\theta_i=1$ and
$\Psi_i^{**}$ be the corresponding probability when $\theta_i=0$. Our refinements imply that $\Psi_i^* > \Psi_i^{**}$ for $i \in \{1,2\}$.
From the principal's perspective, whether $\theta_1$ and $\theta_2$ are strategic complements or strategic substitutes
depends only on the expected penalty under each $\boldsymbol{a}$. Let $P:\{0,1\}^2 \rightarrow [0,+\infty)$ be the mapping from $\boldsymbol{a}$ to expected penalties. The principal's decisions are strategic complements if
\begin{equation}\label{H.18}
    P(1,1)+P(0,0)< P(1,0)+P(0,1)
\end{equation}
and are strategic substitutes otherwise. We show the following lemma.
\begin{LemmaK}
For every $\boldsymbol{a}$ and $\boldsymbol{a}'$ such that $\boldsymbol{a} > \boldsymbol{a'}$,
\begin{equation}\label{H.19}
    \Pr\Big(\overline{\theta}=1 \Big|\boldsymbol{a} \Big) > \Pr\Big(\overline{\theta}=1 \Big|\boldsymbol{a'} \Big),
\end{equation}
and
\begin{equation}\label{H.20}
        \Pr\Big(\theta_1=\theta_2=1 \Big|\boldsymbol{a} \Big) >  \Pr\Big(\theta_1=\theta_2=1 \Big|\boldsymbol{a'} \Big).
\end{equation}
\end{LemmaK}
Lemma K implies that if $P(0,1)$ or $P(1,0)$ is strictly positive, then every proper equilibrium that satisfies Refinements 1 and 2 must also satisfy $P(1,1) \geq L$. This property comes from the fact that when the equilibrium is proper and $L$ is large enough, every agent has witnessed offense with strictly positive probability. As in the proof of Theorem 1 in Online Appendix F, this leads to a uniform lower bound on the marginal increase in the expected probability of receiving punishment when the principal commits one extra offense.
\begin{proof}
The second inequality follows from Lemma 3.2. For the first inequality, it suffices to compare the following two ratios:
\begin{equation*}
 \mathcal{I}_2 \equiv   \frac{\Pr(a_1=a_2=1|\theta_1=\theta_2=1)}{\Pr(a_1=a_2=1|\sum_{i=1}^2 \theta_i \leq 1)}
\end{equation*}
and
\begin{equation*}
  \mathcal{I}_1 \equiv      \frac{\Pr(a_1=0,a_2=1|\theta_1=\theta_2=1)}{\Pr(\sum_{i=1}^2 a_i=1|\sum_{i=1}^2 \theta_i \leq 1)}.
\end{equation*}
Let $p_0$ be the probability that $(\theta_1,\theta_2)=(0,0)$ conditional on $(\theta_1,\theta_2)\neq (1,1)$,
$p_1$ be the probability that $(\theta_1,\theta_2)=(1,0)$ conditional on $(\theta_1,\theta_2)\neq (1,1)$, and $p_2$ be the probability that $(\theta_1,\theta_2)=(0,1)$ conditional on $(\theta_1,\theta_2)\neq (1,1)$.
We have
\begin{equation*}
     \mathcal{I}_2^{-1} = p_0 \frac{\Psi_1^{**}}{\Psi_1^*} \frac{\Psi_2^{**}}{\Psi_2^*}
     + p_1 \frac{\Psi_2^{**}}{\Psi_2^*} + p_2 \frac{\Psi_1^{**}}{\Psi_1^*},
\end{equation*}
and
\begin{equation*}
         \mathcal{I}_1^{-1} = p_0 \frac{1-\Psi_1^{**}}{1-\Psi_1^*} \frac{\Psi_2^{**}}{\Psi_2^*}
     + p_1 \frac{\Psi_2^{**}}{\Psi_2^*} + p_2 \frac{1-\Psi_1^{**}}{1-\Psi_1^*}.
\end{equation*}
Since $\Psi_i^* > \Psi_i^{**}$ for every $i \in \{1,2\}$, we have $\mathcal{I}_2^{-1}<\mathcal{I}_1^{-1}$, or, equivalently,
$\mathcal{I}_2<\mathcal{I}_1$.
\end{proof}

\section*{L. $\quad$ Distribution of Payoff Shocks}
We revisit the specification of the distribution of agents' payoff shock $\omega_i$. First, we motivate our model's assumption that the support of $\omega_i$ is unbounded below. Second, we explore the robustness of our results under alternative distributions of $\omega_i$.

\paragraph{Unbounded Support:} The assumption that $\omega_i$ is unbounded from below is equivalent to the following property: under every conviction rule $q: \{0,1\}^n \rightarrow [0,1]$ that is responsive to each agent's accusation. That is,
\begin{itemize}
  \item[(*)] for every $i \in \{1,...,n\}$, there exists $a_{-i} \in \{0,1\}^{n-1}$ such that $q(1,a_{-i}) \neq q (0,a_{-i})$,
\end{itemize}
each strategic type agent accuses the principal with \textit{strictly positive probability}.

Recall that $(*)$ is required by Refinement 2.
When $L$ is large, the conviction probability $q(\cdot)$ is low for all $\boldsymbol{a}$. Therefore, the unbounded support assumption ensures that (strategic) victims accuse the principal with positive probability in all equilibria that satisfy Refinement 1 when $L$ is large enough.

Next, we show that under the alternative assumption that the support of $\omega_i$ is bounded below, all symmetric equilibria of the game violate Refinement 1. As noted earlier, these equilibria contradict the key principle that defendants should not be convicted based on the judge's prior belief:
\begin{LemmaL}
If $\omega_i \geq \underline{\omega}$ with probability $1$ for some $\underline{\omega} \in \mathbb{R}$,
there exists $\overline{L} \in \mathbb{R}_+$, such that for every $L \geq \overline{L}$,
there exists no symmetric equilibrium that satisfies Refinement 1.
\end{LemmaL}
\begin{proof}
We rule out equilibria in which the probability of offense is $0$ or interior. First, suppose toward a contradiction that there exists an equilibrium in which the probability of offense is $0$. Since a behavioral agent accuses the principal with interior probability, the posterior probability of guilt is $0$ for every report profile $\boldsymbol{a}$, and the principal is never punished. The principal thus have a strict incentive to commit offenses, which leads to a contradiction. Next, suppose toward a contradiction that there exists an equilibrium in which the probability of offense is interior. For every $L \in \mathbb{R}_+$, there exists $\overline{q}_L$ such that:
\begin{equation}\label{I.1}
    \max_{\boldsymbol{a} \in \{0,1\}^n} q(\boldsymbol{a}) \leq \overline{q}_L.
\end{equation}
Since the principal is indifferent between committing one and zero offense, $\overline{q}_L \rightarrow 0$ as $L \rightarrow +\infty$. When $\theta_i=1$, agent $i$'s reporting cutoff is bounded above by
\begin{equation}\label{I.2}
    \omega_i^* \leq b+c-\frac{c}{\overline{q}_L}.
\end{equation}
When $L$ is large enough for the RHS of (\ref{I.2}) to be less than $\underline{\omega}$, agent $i$ accuses the principal with probability $0$ regardless of $\theta_i$. This implies that the principal has a strict incentive to commit offenses and leads to a contradiction.
\end{proof}

\paragraph{Alternative Distributions:} Let $\Phi$ be the cdf of $\omega_i$. We introduce two properties of $\Phi$:
\begin{Definition}
$\Phi$ is \textit{regular} if it admits a continuous density and there exists $\overline{\omega} \in \mathbb{R}$ such that
\begin{enumerate}
  \item the support of $\Phi$ contains $(-\infty,\overline{\omega})$;
  \item the density is strictly increasing when $\omega \leq \overline{\omega}$;
  \item $\Phi(\omega+b)/\Phi(\omega)$ is non-increasing in $\omega$ for some $b \in \mathbb{R}_+$
  when $\omega < \overline{\omega}$.
\end{enumerate}
$\Phi$ has \textit{thin left tail} if there exists $b \in \mathbb{R}_+$ such that
\begin{equation}\label{I.3}
 \lim_{\omega \rightarrow -\infty}   \Phi(\omega+b)/\Phi(\omega) =+\infty.
\end{equation}
\end{Definition}
Our result in the single-agent benchmark (Proposition 1) and our result on DPP (Theorem 2 and 2')
requires that $\Phi$ be regular and have a thin left tail.
Our insights that APP hurts witness credibility and harms deterrence, namely, Theorems 1 and 1' remain valid for all distributions with support that is unbounded below. The comparative statics results (Proposition 2 and Theorem 3) require $\Phi$ to be regular. The second requirement of \textit{regularity} is used to derive the following property: agents accuse the principal with strictly higher probability in environments with more agents, and the third requirement of regularity is used to conclude that informativeness is strictly lower when there are more agents since the reporting thresholds are higher and the distance between them is smaller.

To give some concrete examples, normal distributions are regular and have a thin left tail. Exponential distributions are regular but do not have a thin left tail. Any distribution that is regular with a tail thinner than the exponential distribution  has a thin left tail. Pareto distributions have a support that is unbounded below. However, they are not regular and do not have thin left tail.

\end{spacing}


\title{Online Appendix\\Crime Entanglement, Deterrence, and Witness Credibility}
\author{Harry Pei \and Bruno Strulovici}
\date{\today}

\maketitle

\appendix
\begin{spacing}{1.5}

\section*{F. $\quad$ Properties of Conviction Probabilities}\label{secC}
We show the following proposition, which implies Lemma 3.1 and Lemma D.1 of the main text.
\begin{PropositionF}\label{PropC.1}
There exists $\overline{L}>0$ such that for any $L >\overline{L}$ and proper equilibrium that satisfies Refinements 1 and 2,
\begin{enumerate}
  \item if the judge uses decision rule (2.4), then $q(0,0)=q(1,0)=q(0,1)=0$.
  \item if the judge uses decision rule (2.5), then there exists no $\mathbf{a} \in \{0,1\}^2$ such that $q(\mathbf{a})=1$.
\end{enumerate}
\end{PropositionF}
We consider three cases separately depending on whether the following expression is strictly positive, strictly negative, or equal to $0$:
\begin{equation}\label{1}
    q(1,1)+q(0,0)-q(1,0)-q(0,1).
\end{equation}
Lemma 3.2 implies that the principal's actions are complements if (\ref{1}) is negative and vice versa.
\subsection*{F.1 $\quad$ The value of (\ref{1}) is strictly positive}
First, we show that $q(1,1)$ must be strictly less than 1 regardless the decision rule used to adjudicate guilt, which implies that $q(\mathbf{a})<1$ for any $\mathbf{a}$. We the show that under decision rule (2.4), $\max\{q (0,1), q (1,0)\}>0$ implies that $q (1,1)=1$.

Suppose toward a contradiction that $q (1,1)=1$. Refinement 1 imposes that $q (0,0)=0$. This leads to the following expressions for agent $1$'s reporting cutoffs when he has and has not witnessed offense:
\begin{equation}\label{C.1}
    \omega_1^* \equiv b-c \frac{\displaystyle (1-\Psi_2^{**}) (1-q (1,0))}{\displaystyle q (1,0)+\Psi_2^{**} (1-q (1,0)-q (0,1))},
\end{equation}
\begin{equation}\label{C.2}
        \omega_1^{**} \equiv -c \frac{\displaystyle (1-X_2) (1-q (1,0))}{\displaystyle q (1,0)+X_2 (1-q (1,0)-q (0,1))},
\end{equation}
where
\begin{equation}\label{C.3}
    X_2 \equiv \frac{1-p_1-p_2}{1-p_1} \Psi_2^{**} +\frac{p_2}{1-p_1} \Psi_2^*
\end{equation}
and $p_i$ is the probability with which $\theta_i=1$. We note that
$\omega_1^*$ is increasing in $\Psi_2^{**}$ and $q (1,0)$ and decreasing in $q (0,1)$ and that
$\omega_1^{**}$ is increasing in $X_2$ and $q (1,0)$ and decreasing
in $q (0,1)$. The distance between the two cutoffs is
\begin{equation}\label{C.4}
    \omega_1^*-\omega_1^{**}=b-(\Psi_2^*-\Psi_2^{**})C_1
\end{equation}
where
\begin{equation*}
    C_1 \equiv c (1-q (0,1))(1-q (1,0)) \times \frac{p_2}{1-p_1}
    \end{equation*}
    \begin{equation}\label{C.5}
    \times \frac{1}{q (1,0)+X_2 (1-q (1,0)-q (0,1))}
    \times
    \frac{1}{q(1,0)+\Psi_2^{**} (1-q(1,0)-q (0,1))}.
\end{equation}
By symmetrically, one can obtain the expressions for $\omega_2^*$, $\omega_2^{**}$, and their difference.
Conditional on $\theta_2=0$, choosing $\theta_1 =1$ rather than $\theta_1 = 0$ increases the principal's conviction probability by
\begin{equation}\label{C.6}
    (\Psi_1^*-\Psi_1^{**})
    \Big(
    q(1,0)+\Psi_2^{**} (1-q(1,0)-q(0,1))
    \Big)
\end{equation}
Similarly, if choosing $\theta_2=1$ rather than $\theta_2 = 0$ given that  $\theta_1=0$ increases the conviction probability by
\begin{equation}\label{C.7}
 (\Psi_2^*-\Psi_2^{**})
    \Big(
    q(0,1)+\Psi_1^{**} (1-q(1,0)-q(0,1))
    \Big).
\end{equation}
In every equilibrium, both (\ref{C.6}) and (\ref{C.7}) are bounded below by $1/L$. In what follows, we establish a lower bound for the maximum of these two expressions that is independent of $L$ and that will deliver the desired contradiction when $L$ is large enough. Throughout the proof, we assume without loss of generality that $\omega_1^* \geq \omega_2^*$. The following lemma provides a comparison between $q(1,0)$ and $q (0,1)$.
\begin{LemmaF}
In every equilibrium such that $\omega_1^* \geq \omega_2^*$, we have $q (1,0) \geq q (0,1)$.
\end{LemmaF}
Lemma F is proved in Section F.4 and taken for granted for now.
\paragraph{Lower Bound on $\omega_1^*$:}
For every $\epsilon >0$,
\begin{enumerate}
  \item If $q (1,0) \geq \epsilon$, then
    $\omega_1^{**} \geq -c\frac{1-\epsilon}{\epsilon}$.
  \item If $q (1,0) < \epsilon$, then $q (0,1) \in (0,\epsilon)$, by Lemma F. Therefore,
  \begin{eqnarray}\label{C.9}
 \omega_2^* &= & b -c \frac{\displaystyle (1-\Psi_1^{**}) (1-q (0,1))}{\displaystyle q (0,1)+\Psi_1^{**} (1-q(1,0)-q(0,1))}
      {}
\nonumber\\
& \geq &
b-c \frac{\displaystyle \Big(1-\delta \Phi(\omega_1^{*}-b)-(1-\delta)\alpha \Big) (1-q (0,1))}{\displaystyle q (0,1)+\Big(\delta\Phi(\omega_1^{*}-b)+(1-\delta)\alpha \Big) \Big(1-q (1,0)-q(0,1)\Big)}
      {}
\nonumber\\
& \geq & b -c \frac{\displaystyle \Big(1-\delta \Phi(\omega_2^{*}-b)-(1-\delta)\alpha \Big) (1-q(0,1))}{\displaystyle q (0,1)+\Big(\delta\Phi(\omega_2^*-b)+(1-\delta)\alpha \Big) \Big(1-q (1,0)-q (0,1)\Big)}
      {}
\nonumber\\
& \geq & b
 -c \frac{1-\delta\Phi(\omega_2^*-b)-(1-\delta)\alpha}{(1-\epsilon)\Big(\delta \Phi(\omega_2^*-b)+(1-\delta)\alpha \Big)}.
\end{eqnarray}
Supplementary Appendix J, which proves the existence of an equilibrium, also proves that there exists a solution to the equation
\begin{equation*}
   \omega_2^* = b-c \frac{1-\delta \Phi(\omega_2^*-b)-(1-\delta)\alpha}{\displaystyle (1-\epsilon)\Big(\delta\Phi(\omega_2^*-b)+(1-\delta) \alpha \Big)}.
\end{equation*}
The RHS of (\ref{C.9}) is bounded below for each $\epsilon$, uniformly over its argument $\Phi(\omega^*_2$. This lower, which we denote $\underline{\omega}^*(\epsilon)$, is decreasing in $\epsilon$. This yields a  lower bound for $\omega_1^*$ given by
\begin{equation}\label{C.10}
    \underline{\omega}_1^* \equiv \sup_{\epsilon \in [0,1]} \Big\{
    \min \big\{b-c\frac{1-\epsilon}{\epsilon}, \underline{\omega}^*(\epsilon)
    \big\}
    \Big\},
\end{equation}
which is finite and independent of $L$.
\end{enumerate}

\paragraph{Upper Bound on $C_1$:} We provide an upper bound for
\begin{equation}\label{C.11}
    \frac{1}{q (1,0)+\Psi_2^{**} (1-q(1,0)-q(0,1))}.
\end{equation}
For every $\epsilon > 0$, there are two cases:
\begin{enumerate}
  \item If $q(1,0) \geq \epsilon$, then (\ref{C.11}) is no more than $1/\epsilon$.
  \item If $q(1,0) < \epsilon$, then Lemma F implies that $q(0,1)< \epsilon$. Let $\underline{\omega}_2^{**}(\epsilon)$ be the smallest root of the following equation:
  \begin{equation}\label{C.12}
  \omega \equiv -c \frac{\displaystyle 1- \delta \Phi(\omega)-(1-\delta)\alpha}{\displaystyle \Big(\delta \Phi(\omega)+(1-\delta)\alpha\Big) (1-\epsilon)}.
  \end{equation}
  Since $q(1,0),q(0,1) \in [0,\epsilon]$,
   $\underline{\omega}_2^{**}(\epsilon)$
  is a lower bound for $\omega_2^{**}$. An upper bound on (\ref{C.11}) is given by
  \begin{equation}\label{C.13}
      \frac{1}{q(1,0)+\Psi_2^{**} (1-q(1,0)-q(0,1))}  \leq \frac{1}{\Phi(\underline{\omega}_2^{**}(\epsilon)) (1-2\epsilon)}.
  \end{equation}
\end{enumerate}
In summary:
\begin{equation}\label{C.14}
    C_1 \leq c Y^2
\end{equation}
where
\begin{equation*}
    Y \equiv \inf_{\epsilon \in [0,1]} \Big\{\max
    \big\{
    1/\epsilon, \frac{1}{\Phi(\underline{\omega}_2^{**}) (1-2\epsilon)}
    \big\}
    \Big\}.
\end{equation*}

\paragraph{Lower Bound on the Maximum of (\ref{C.6}) and (\ref{C.7}):} Since $\phi\geq 0$ is the derivative of $\Phi$, we have for all $\omega' > \omega''$
\begin{equation}\label{C.15}
    \Phi(\omega')-\Phi(\omega'')
    \geq (\omega'-\omega'') \min_{\omega \in [\omega',\omega'']} \phi(\omega).
\end{equation}
We consider two cases. First, suppose that $\Phi(\omega_1^*)-\Phi(\omega_1^{**}) \geq \Phi(\omega_2^*)-\Phi(\omega_2^{**})$.
Using the fact that $\Psi_i^*-\Psi_i^{**}=\delta (\Phi(\omega_i^*)-\Phi(\omega_i^{**}))$, we have
\begin{equation}\label{C.16}
    \frac{\delta}{\min_{\omega \in [\omega_1^{**},\omega_1^*]} \phi(\omega)} \Big(
    \Phi(\omega_1^*)-\Phi (\omega_1^{**})
    \Big)
    \geq \omega_1^*-\omega_1^{**} =b-C_1 (\Psi_2^*-\Psi_2^{**} )
    \geq b-C_1 (\Psi_1^*-\Psi_1^{**}).
\end{equation}
This, together with (\ref{C.14}), gives an lower bound on $\Psi_1^*-\Psi_1^{**}$. Moreover,
  \begin{eqnarray}\label{C.17}
q(1,0) +\Psi_2^{**} (1-q(1,0)) & \geq & q(1,0) +\Psi_2^{**} \big( 1-q(1,0)-q(0,1)\big)
      {}
\nonumber\\
& \geq &
\frac{c\big(1-q(1,0)\big)(1-\Psi_2^{**})}{|\underline{\omega}_1^*|},
\end{eqnarray}
where the last inequality uses (\ref{C.1}) and the fact that $\underline{\omega}_1^*\leq \omega_1^*$. This provides a lower bound for $q (1,0)$ and implies a lower bound on (\ref{C.6}).

Second, consider the case in which $\Phi(\omega_1^*)-\Phi(\omega_1^{**}) < \Phi(\omega_2^*)-\Phi(\omega_2^{**})$ and let
\begin{equation}\label{C.18}
   \beta \equiv \frac{\omega_1^*-\omega_1^{**}}{b}.
\end{equation}
Since $X_2 > \Psi_2^{**}$, we have $\beta \in (0,1)$.
Recalling that $\underline{\omega}_1^*\leq \omega_1$, we have
\begin{equation}\label{C.19}
\frac{1}{\delta}(\Psi_1^*-\Psi_1^{**})=    \Phi(\omega_1^*)-\Phi(\omega_1^{**}) \geq \beta b \phi(\underline{\omega}_1^{*}-b).
\end{equation}
Moreover, (\ref{C.4}) and (\ref{C.14}) imply that
\begin{equation}\label{C.20}
    \Psi_2^*-\Psi_2^{**}
    =(1-\beta)b/C_1 \geq \frac{(1-\beta)b Y^2}{c}
\end{equation}
Since the pdf of $\omega_i$  is increasing in $\omega$ for $\omega<0$,
(\ref{C.20}) yields a lower bound on $\omega_2^{**}$. We denote this lower bound by $\widetilde{\omega}(\beta)$. By construction, $\widetilde{\omega}(\beta)$ is decreasing in $\beta$.
\begin{enumerate}
  \item When $\beta \geq 1/2$, (\ref{C.19}) implies a lower bound for $\Phi(\omega_1^*)-\Phi(\omega_1^{**})$. Inequality~(\ref{C.17}) then yields a lower bound for $q(1,0)$ and implies a lower bound on (\ref{C.6}).\footnote{The validity of inequality (\ref{C.17}) does not depend on the sign of $\Psi_1^*-\Psi_1^{**}-\Psi_2^*+\Psi_2^{**}$.}
  \item When $\beta < 1/2$, we have $\omega_2^{**} \geq \widetilde{\omega}(1/2)$ and
  \begin{equation*}
    \Psi_2^*-\Psi_2^{**} \geq \frac{b}{2C_1}.
  \end{equation*}
  The lower bound on $\omega_2^{**}$ also delivers a lower bound on $q (0,1)+\Psi_1^{**} (1-q(1,0)-q(0,1))$, since (\ref{C.2}) implies that
  \begin{equation*}
    \widetilde{\omega}(1/2) \leq \omega_2^{**} \leq \omega_2^* = -c \frac{(1- \Psi_1^{**}) (1-q(0,1))}{q(0,1)+\Psi_1^{**} (1-q(1,0)-q(0,1))},
  \end{equation*}
  which leads to
  \begin{equation}\label{C.21}
    q(0,1)+\Psi_1^{**} (1-q(1,0)-q(0,1))
    \geq \frac{(1- \Psi_1^{**}) (1-q(0,1))}{-\widetilde{\omega}(1/2)/c}.
  \end{equation}
  Since $1-\Psi_1^{**} \geq \delta -\delta \Phi(0)$ and $1-q (0,1) \geq 1/2$, the lower bound on $q(0,1)+\Psi_1^{**} (1-q(1,0)-q(0,1))$ is strictly bounded below away from $0$. This leads to a uniform lower bound on (\ref{C.7}).
\end{enumerate}

Next, we show that under decision rule (2.4), equilibrium that satisfies (\ref{1})
 $\max\{q (0,1), q (1,0)\}>0$ must also satisfy $q (1,1)=1$.
Suppose toward a contradiction that both $q (1,0)$ and $q (1,1)$ are strictly between $0$ and $1$. Then, agent $2$'s accusation does not affect the posterior belief about $\overline{\theta}$. This implies that $a_2$ is uninformative about $\theta_2$. This is only possible if $\omega_2^*=\omega_2^{**}$, which contradicts Lemma 2.1.

\subsection*{F.2 $\quad$ The value of (\ref{1}) is strictly negative}
Next, we study the case in which $q(1,0)+q(0,1) > q(0,0) +q(1,1)$, i.e., $\theta_1$ and $\theta_2$ are strategic complements. Lemma 3.2 implies that, conditional on committing an offense against one agent, the principal has a strict incentive to commit an offense against the other agent. Therefore, in such equilibria, the principal commits either he commits both offenses or no offense.

We proceed in two steps, as in the previous case. First, we prove by contradiction that $q(1,1)<1$ regardless of which decision rule is applied, which implies that $q(\mathbf{a})<1$ for any $\mathbf{a}$ and decision rule. We then show that under decision rule (2.4), $\max\{q (0,1), q (1,0)\}>0$ only if $q (1,1)=1$.

When $q(1,0)+q(0,1) > q(0,0) +q(1,1)$, Lemma 3.2 implies that $\theta_1$ and $\theta_2$ are strategic complements. Therefore, agent $i$ assigns a higher probability to agent $j$ accusing the  principal when $\theta_i=0$ than when $\theta_i=1$. This implies that
\begin{equation}\label{C.22}
    \min \{\omega_1^*-\omega_1^{**}, \omega_2^*-\omega_2^{**}\} \geq b.
\end{equation}
By setting $\theta_1=\theta_2=1$, the principal raises the probability that he is convicted by at least
\begin{equation}\label{C.23}
    (\Psi_1^*-\Psi_1^{**})\Big(
    \Psi_2^* (1-q(0,1))+(1-\Psi_2^*) q(1,0)
    \Big)
    +
    (\Psi_2^*-\Psi_2^{**})\Big(    \Psi_1^{**} (1-q(1,0))+(1-\Psi_1^{**}) q(0,1)
    \Big)
\end{equation}
compared to the case in which he sets $\theta_1=\theta_2=0$.
Therefore, the value of (\ref{C.23})
cannot exceed $2/L$.
The remainder of this proof establishes a strictly positive  lower bound on (\ref{C.23}) that applies uniformly across all $L$. This will imply that when $L$ is large enough, equilibria that exhibit strategic complementarities between $\theta_1$ and $\theta_2$ do not exist.

First, $\max \{q(0,1),q(1,0)\} \geq 1/2$ since $q(0,1) +q(1,0) \geq 1$. Without loss of generality, we assume that $q(1,0) \geq 1/2$. Second, it is a dominant strategy for agent $i$ to abstain from accusing the principal when $\omega_i >0$, which implies that $1-\Psi_i^* \geq \delta (1-\Phi(0))$. Third, player $1$'s reporting threshold when $\theta_1=1$ is
\begin{equation}\label{C.24}
    \omega_1^* =b-c \frac{(1-Q_2^H)(1-q(1,0))}{Q_2^H (1-q(0,1))+(1-Q_2^H) q(1,0)},
\end{equation}
where $Q_2^H$ is the probability that agent $2$ accuses the principal conditional on $\theta_1=1$. The RHS of (\ref{C.24}) is strictly increasing in $Q_2^H$. Therefore,
\begin{equation*}
    \omega_1^* \geq b-c\frac{1-q(1,0)}{q(1,0)} \geq b-c.
\end{equation*}
From  (\ref{C.22}), we have
\begin{equation}\label{C.25}
\frac{1}{\delta} (\Psi_1^*-\Psi_1^{**})=    \Phi(\omega_1^*)-\Phi(\omega_1^{**}) \geq b \min_{\omega \in [-b-c,0]} \phi(\omega).
\end{equation}
This yields the desired lower bound for (\ref{C.23}):
\begin{eqnarray*}
   \underbrace{(\Psi_1^*-\Psi_1^{**})}_{\textrm{bounded by (\ref{C.25})}}
   \Big(
    \underbrace{\Psi_2^* (1-q(0,1))}_{\geq 0}+\underbrace{(1-\Psi_2^*)}_{\geq \delta (1-\Phi(0))} \underbrace{q(1,0)}_{\geq 1/2}
    \Big)
    +
    \underbrace{(\Psi_2^*-\Psi_2^{**})\Big(    \Psi_1^{**} (1-q(1,0))+(1-\Psi_1^{**}) q(0,1)}_{\geq 0}
    \Big)
    \end{eqnarray*}
    \begin{eqnarray}\label{C.26}
\geq
\frac{\delta^2b}{2}(1-\Phi(0)) \min_{\omega \in [-b-c,0]} \phi(\omega)
\end{eqnarray}
and implies that $q(1,1) <1$ when $L$ is large enough.

Next, we show that under decision rule (2.4), $q(1,1)+q(0,0)<q(1,0)+q(0,1)$ implies that
$q(1,1)=1$. Suppose toward a contradiction that $q(1,1) \in (0,1)$ and $q(1,0)+q(0,1) > q(0,0) +q(1,1)$. Then, at least one of inequalities $q(0,1)>0$ and $q(1,0) >0$ must hold. If $q(0,1)>0$, the judge's posterior belief about $\overline{\theta}$ is independent of agent $1$'s report. This gives an opportunistic principal a strict incentive to choose $\theta_1=1$ and leads to a contradiction. The case $q(1,0)>0$ leads to a similar contradiction.
\subsection*{F.3 $\quad$ The value of (\ref{1}) is $0$}
\paragraph{Part I:} We show that in every equilibrium where the value of (\ref{1}) is $0$,
each agent witnesses an offense with strictly positive probability and $q(1,1)=1$. This implies that:
\begin{itemize}
  \item[1.] $q(1,0)+q(0,1)=1$.
  \item[2.] The marginal cost of committing an offense is the same across agents, i.e.,
  \begin{equation}\label{C.27}
    (\Psi_1^*-\Psi_1^{**})q(1,0)
    =
        (\Psi_2^*-\Psi_2^{**})q(0,1).
  \end{equation}
\end{itemize}
First, suppose toward a contradiction that the principal chooses $a_1=1$ with probability $0$. Then, agent $1$'s report
does not affect the judge's posterior belief about the value of
$\theta_1 \theta_2$. Therefore, $q(1,0)=q(0,0)=0$. Since the value of (\ref{1}) is $0$, we have $q(0,1)=q(1,1) \in (0,1]$. This contradicts the conclusion of Lemma 2.1 since $q$ is not responsive to agent $1$'s report.

Next, suppose toward a contradiction that $q(1,1) \in (0,1)$ and that each agent witnesses an offense with positive probability. Then, at least one of the following conditions must hold: $q(1,0) \in (0,1)$ or $q(0,1) \in (0,1)$. The previous paragraph has ruled out equilibria in which either $q(1,0)$ or $q(0,1)$ is equal to $0$. Suppose that $q(1,0), q(0,1),q(1,1) \in (0,1)$. Then, the report profiles $\boldsymbol{a}=(1,1)$, $(1,0),(0,1)$ must lead to the same posterior probability that the principal has committed at least one offense. For $i \in \{1,2\}$, let $p_i$ be the probability that $\theta_i =1$ and $\theta_{-i} = 0$ conditional on the principal having committed at least one offense.
Since the posterior probability of guilt is the same for reporting profiles $(1,1)$ and $(1,0)$, we have
\begin{equation}\label{2}
    (1-p_1-p_2)\frac{\Psi_1^* \Psi_2^*}{\Psi_1^{**} \Psi_2^{**}}
    +p_1 \frac{\Psi_1^*}{\Psi_1^{**}} +p_2 \frac{\Psi_2^*}{\Psi_2^{**}}
    = (1-p_1-p_2)\frac{\Psi_1^* (1- \Psi_2^*)}{\Psi_1^{**} (1-\Psi_2^{**})}
    +p_1 \frac{\Psi_1^*}{\Psi_1^{**}} +p_2 \frac{1-\Psi_2^*}{1-\Psi_2^{**}}.
\end{equation}
Since
\begin{equation*}
    \frac{\Psi_1^* \Psi_2^*}{\Psi_1^{**} \Psi_2^{**}} > \frac{\Psi_1^* (1- \Psi_2^*)}{\Psi_1^{**} (1-\Psi_2^{**})}
\end{equation*}
and
\begin{equation*}
    \frac{\Psi_2^*}{\Psi_2^{**}} > \frac{1-\Psi_2^*}{1-\Psi_2^{**}},
\end{equation*}
the LHS of (\ref{2}) is strictly greater than the RHS  of (\ref{2}) unless $p_1=1$.
By assumption, the principal commits an offense against each agent with strictly positive probability, so either $1-p_1-p_2>0$
or $p_2>0$, which violates (\ref{2}).

\paragraph{Part II:} We show that $q(1,1)<1$ when $L$ is large enough.

Suppose toward a contradiction that $q(1,1)=1$. We derive a lower bound on (\ref{C.27}) that holds for all $L$. Without loss of generality, we assume that $q(1,0) \geq q(0,1)$, which implies that $q(1,0) \geq 1/2$. Agent $1$'s reporting cutoffs satisfy
\begin{equation*}
    \omega_1^* =b-c \frac{q(0,1)}{q(1,0)}
    \Big(
    1-p_{x} \Psi_2^*-(1-p_x) \Psi_2^{**}
    \Big)
\end{equation*}
and
\begin{equation*}
    \omega_1^{**} =-c \frac{q(0,1)}{q(1,0)}
    \Big(
    1-p_y \Psi_2^*-(1-p_y) \Psi_2^{**}
    \Big)
\end{equation*}
where $p_x,p_y \in [0,1]$ represent agent $1$'s beliefs about $\theta_2$ conditional on each realization of $\theta_1$. This implies that
\begin{equation}\label{C.28}
    \omega_1^*-\omega_1^{**}
    =b-c \frac{q(0,1)}{q(1,0)} (p_x-p_y) (\Psi_2^*-\Psi_2^{**}).
\end{equation}
The absolute value of
\begin{equation*}
c \frac{q(0,1)}{q(1,0)} (p_x-p_y)
\end{equation*}
is at most $c$.
To bound the LHS of (\ref{C.27}) from below, we proceed according to the following two steps.

\paragraph{Step 1: Lower bound on $\omega_1^*$} The formula for $\omega_1^*$ and the assumption that $q(1,0) \geq q(0,1)$ imply that
 \begin{equation}\label{C.29}
    \omega_1^* \geq b -c \Big(1-p_{x} \Psi_2^*-(1-p_x) \Psi_2^{**} \Big) \geq b-c \delta (1-\Phi(0)).
 \end{equation}
 We note the lower bound on the RHS by $\underline{\omega}_1^*$.

\paragraph{Step 2: Lower bound on (\ref{C.27})} Since $q(1,0) \geq q(0,1)$ and $q(1,0)+q(0,1) \geq q(1,1)=1$, we have $q(1,0) \geq 1/2$. Therefore, (\ref{C.27}) will be bounded below if we establish a strictly positive lower bound on $\min\{\Psi_1^*-\Psi_1^{**}, q(0,1)(\Psi_2^*-\Psi_2^{**})\}$.

If $p_x-p_y \leq 0$, we have $\omega_1^*-\omega_1^{**} \geq b$. The lower bound on $\omega_1^*$ then implies a strictly positive lower bound on $\Psi_1^*-\Psi_1^{**}$, as desired.
If $p_x-p_y >0$, we follow same derivation as in the last step of Section F.1.  More precisely, we consider two cases.

First, suppose that $\Psi_1^*-\Psi_1^{**} \geq \Psi_2^*-\Psi_2^{**}$. Then, we have
\begin{equation}\label{C.30}
    \frac{\Psi_1^*-\Psi_1^{**}}{\phi(\underline{\omega}_1^*-b)}
    \geq \omega_1^*-\omega_1^{**} =b-c (\Psi_2^*-\Psi_2^{**} )
    \geq b-c (\Psi_1^*-\Psi_1^{**}).
\end{equation}
This yields a strictly positive lower bound on $\Psi_1^*-\Psi_1^{**}$.

Second, suppose that $\Psi_1^*-\Psi_1^{**} < \Psi_2^*-\Psi_2^{**}$. Then,
the variable $\beta \equiv (\omega_1^*-\omega_1^{**})/b$
lies between $0$ and $1$ due to the assumption that $p_x-p_y >0$. Equality (\ref{C.28}) implies that
\begin{equation*}
     \omega_1^*-\omega_1^{**}
    =b-c \frac{q(0,1)}{q(1,0)} (p_x-p_y) (\Psi_2^*-\Psi_2^{**}) \geq b-c (\Psi_2^*-\Psi_2^{**}),
\end{equation*}
which yields
\begin{equation}\label{C.31}
   \Psi_2^* \geq  \Psi_2^*-\Psi_2^{**} \geq (1-\beta) b /c.
\end{equation}
This provides a lower bound on $\omega_2^*$ that is decreasing in $\beta$ and that we denote $\widetilde{\omega}(\beta)$. We also have
\begin{equation}\label{C.32}
    \frac{1}{\delta}(\Psi_1^*-\Psi_1^{**})=    \Phi(\omega_1^*)-\Phi(\omega_1^{**}) \geq \beta b \phi(\underline{\omega}_1^{*}-b).
\end{equation}
We consider two subcases, depending on the value of $\beta$ relative to $1/2$.
\begin{enumerate}
  \item If $\beta \geq 1/2$, then (\ref{C.32}) implies that
  \begin{equation}\label{C.33}
  \Psi_1^*-\Psi_1^{**} \geq b \delta \phi(\underline{\omega}_1^{*}-b)/2.
  \end{equation}
  \item If $\beta < 1/2$, then (\ref{C.31}) implies that
  \begin{equation}\label{C.34}
    \Psi_2^*-\Psi_2^{**} \geq b/2c.
  \end{equation}
  We have
  \begin{equation}\label{C.35}
    \omega_2^* =b-c(1-Q) \frac{q(1,0)}{q(0,1)} \geq \underline{\omega}_2(\beta)
  \end{equation}
  where $Q$ is a number between $0$ and $(1-\delta) \alpha+\delta \Phi(0)$. This yields the following lower bound on $q(0,1)$:
  \begin{equation}\label{C.36}
    q(0,1) \geq \frac{b-c(1-Q)q(1,0)}{\underline{\omega}_2(\beta)} \geq \frac{b-c}{2 \underline{\omega}_2(\beta)}.
  \end{equation}
  This expression is bounded below away from $0$ for all $\beta<1/2$. This, together with (\ref{C.34}), lead to the following lower bound on the RHS of (\ref{C.27}):
  \begin{equation}\label{C.37}
    q(0,1) (\Psi_2^*-\Psi_2^{**}) \geq \frac{(b-c)b}{4 c\underline{\omega}_2(\beta)}.
  \end{equation}
\end{enumerate}

\subsection*{F.4 $\quad$ Proof of Lemma F}
Suppose toward a contradiction that there exists an equilibrium in which the value of (\ref{1}) is strictly positive, $\omega_1^* > \omega_2^*$, and  $q (1,0) < q (0,1)$. Then, (\ref{C.1}) implies that $\Phi(\omega_2^{**}) > \Phi(\omega_1^{**})$
or, equivalently, that $\omega_2^{**} > \omega_1^{**}$. This, together with $\omega_1^* > \omega_1^{**}$
and $\omega_2^* > \omega_2^{**}$,
implies that
\begin{equation}\label{C.38}
\omega_1^{**}<\omega_2^{**}<\omega_2^*<\omega_1^*.
\end{equation}
We start by showing that $p_1,p_2 >0$. Suppose that $p_1=0$ and $p_2>0$. Then, (\ref{C.3}) implies that $X_1=\Psi_1^{**}$ and, hence, that $\omega_2^*-\omega_2^{**}=b>\omega_1^*-\omega_1^{**}$, which contradicts (\ref{C.38}).
Now suppose that $p_1>0$ and $p_2=0$. Then,
\begin{equation}\label{C.39}
p_1 \frac{\Psi_1^*}{\Psi_1^{**}} + p_2 \frac{1-\Psi_2^*}{1-\Psi_2^{**}} >
p_2 \frac{\Psi_2^*}{\Psi_2^{**}} + p_1 \frac{1-\Psi_1^*}{1-\Psi_1^{**}}.
\end{equation}
This means that the judge attaches a higher probability to
$\overline{\theta}=1$ when only agent~$1$ accuses the principal than when only agent $2$ does. This implies that $q (1,0) \geq q (0,1)$, which leads to a contradiction.

Having established that $p_1,p_2$ both lie in $(0,1)$, we conclude that
(\ref{C.6}) and (\ref{C.7}) are equal to each other. Applying (\ref{C.1}) to both agents, we have
\begin{eqnarray}\label{C.40}
    \Big|\frac{\omega_1^*-b}{\omega_2^*-b}\Big|
& = &
     \frac{1-\Psi_2^{**}}{1-\Psi_1^{**}}
    \cdot
    \frac{1-q(1,0)}{1-q(0,1)}
   \cdot \frac{q(0,1) +\Psi_1^{**} (1-q(1,0)-q(0,1)) }{q(1,0) +\Psi_2^{**} (1-q(1,0)-q(0,1))}
         {}
\nonumber\\
& = & \frac{1-\Psi_2^{**}}{1-\Psi_1^{**}}
    \cdot
    \frac{1-q(1,0)}{1-q(0,1)}
   \cdot \frac{\Psi_1^*-\Psi_1^{**}}{\Psi_2^*-\Psi_2^{**}}.
\end{eqnarray}
Since
    $\frac{1-\Psi_1^{**}}{1-\Psi_2^{**}}
    < \frac{\Psi_1^*-\Psi_1^{**}}{\Psi_1^*-\Psi_2^{**}}
    \leq
    \frac{\Psi_1^*-\Psi_1^{**}}{\Psi_2^*-\Psi_2^{**}}$,
we get
\begin{equation}\label{C.41}
    1 \geq \Big|\frac{\omega_1^*-b}{\omega_2^*-b}\Big| >  \frac{1-q(1,0)}{1-q(0,1)}.
\end{equation}
The RHS of (\ref{C.41}) is greater than $1$ since $q(1,0)<q(0,1)$. This yields the desired contradiction.

\section*{G. $\quad$ Proof of Proposition 3}
In Section G.1, we construct an open interval of $L$ over which there exist equilibria in which the principal's actions are strategic complements.
In Section G.2, we show that for those values of $L$, the conviction probability is strictly concave in the number of accusations in all proper equilibria that satisfy Refinements 1 and 2.
\subsection*{G.1 $\quad$ Existence of an Equilibrium}
We construct an interval of $L$ such that under APP (i.e., decision rule (2.4)), there exists a proper symmetric Bayes Nash Equilibrium that satisfies Refinements 1 and 2 in which $q (1,1)=1$, $q (1,0)=q (0,1)=q$, and $q (0,0)=0$ with $q > 1/2$, which implies that
\begin{equation}
    q(1,1)+q(0,0)-q(1,0)-q(0,1)
\end{equation}
is strictly negative. From Lemma~3.2, an opportunistic principal's decisions to commit offenses against the two agents are strategic complements. In equilibrium, the principal chooses either $\theta_1=\theta_2=1$ or $\theta_1=\theta_2=0$ but not $\boldsymbol{\theta}=(1,0)$ or $(0,1)$. Since each agent observes an offense with interior probability, this BNE is a proper equilibrium.

First, we derive formulas for the reporting cutoffs. When $\theta_i=1$, agent $i$ accuses the principal if
\begin{equation}\label{D.1}
 \omega_i \leq    \omega^* \equiv b -\frac{c(1-q)(1-\Psi^*)}{q+\Psi^* (1-2q)}.
\end{equation}
When $\theta_i=0$, agent $i$ accuses the principal if
\begin{equation}\label{D.2}
\omega_i \leq    \omega^{**} \equiv -\frac{c(1-q)(1-\Psi^{**})}{q+\Psi^{**} (1-2q)}.
\end{equation}
The principal's indifference condition is given by
\begin{equation}\label{D.3}
    2/L=(\Psi^*-\Psi^{**}) \Big(
    (1-2q) (\Psi^*+\Psi^{**})+2q
    \Big),
\end{equation}
where
    $\Psi^* \equiv \delta \Phi(\omega^*)+(1-\delta)$ and  $\Psi^{**} \equiv \delta \Phi(\omega^{**})+(1-\delta)$.
Moreover, the equilibrium probability that the principal commits an offense, denoted by $\pi_m$, is implicitly determined by
\begin{equation}\label{D.4}
    \frac{\Psi^*(1-\Psi^*)}{\Psi^{**} (1-\Psi^{**})}=\frac{\pi^*}{1-\pi^*} \Big/ \frac{\pi_m}{1-\pi_m},
\end{equation}
where
    $\mathcal{I} \equiv \frac{\Psi^*(1-\Psi^*)}{\Psi^{**} (1-\Psi^{**})}$
measures the aggregate informativeness of accusations. In the equilibria of interest, $\mathcal{I}$ is the right measure of informativeness because one accusation is sufficient to convict the principal, which implies that the judge indifferent between $s=0$ and $s=1$ when exactly one agent accuses the principal.

Comparing (\ref{D.1}) to (\ref{D.2}), we conclude that $\omega^*-\omega^{**}>b$. We re-express (\ref{D.1}) and (\ref{D.2}) as
\begin{equation}\label{D.5}
    \frac{\omega^*-b}{c}=\frac{\Psi^*-1}{\Psi^*+(1-\Psi^*) \frac{q}{1-q}}
\end{equation}
and
\begin{equation}\label{D.6}
    \frac{\omega^{**}}{c}=\frac{\Psi^{**}-1}{\Psi^{**}+(1-\Psi^{**}) \frac{q}{1-q}}.
\end{equation}
For every $q \in [1/2,1]$, the function
\begin{equation*}
    \frac{\Psi-1}{\Psi+(1-\Psi) \frac{q}{1-q}}
\end{equation*}
is convex function in $\Psi$. Moreover, the pdf of $\omega$ is strictly increasing for $\omega \leq 0$. As a result, the function
\begin{equation}\label{eq-Om}
        \frac{\Psi (\omega)-1}{\Psi (\omega)+(1-\Psi(\omega)) \frac{q}{1-q}}
\end{equation}
is strictly increasing and convex for $\omega \leq 0$. Moreover, it takes values in $[-1,0]$.

When $\omega^*=b$, the LHS of
(\ref{D.5}) is strictly greater than its RHS. This, together with the convexity of~\eqref{eq-Om} implies that there is a unique value $\omega^*$ of $\omega$ such that the LHS of (\ref{D.5}) is equal to~\eqref{eq-Om}. This pins down the value of $\omega^*$ in equilibrium. Similarly, (\ref{D.6}) admits a unique solution, which pins down the value of  $\omega^{**}$ in equilibrium.

The RHS of (\ref{D.5}) and (\ref{D.6}) are both increasing in $q$, which implies that $\omega^*$ and $\omega^{**}$ are also increasing in $q$. Moreover, since equation (B.6) admits a unique solution $\omega^*$, the LHS of (B.6) does not depend on $q$, and the RHS is continuously increasing in $q$, the solution $\omega^*$ is continuous in $q$. Similarly, $\omega^{**}$ is also continuously increasing in $q$.

Let $\omega^*(c,q)$ and $\omega^{**} (c,q)$ denote the values of the cutoffs in equilibrium, and $L(c,q)$ denote the value of $L$, as defined by (\ref{D.3}), when $\omega^*=\omega^*(c,q)$
and $\omega^{**}=\omega^{**}(c,q)$. For every $c>0$ and
\begin{equation}
L \in  \Big[
    \min_{q \in [1/2,1]} L(c,q)
    ,
        \max_{q \in [1/2,1]} L(c,q)
    \Big],
\end{equation}
there exists $q \in (1/2,1]$, such that when the retaliation cost is $c$ and the punishment level is $L$, there exists
an equilibrium such that $q(1,1)=1$, $q(1,0)=q(0,1)=q$ and $q(0,0)=0$.

\subsection*{G.2 $\quad$ Conviction Probabilities}
For every $(c,q)$, the thresholds $\omega^*$ and $\omega^{**}$ are computing as explained in the previous section. Let $L(c,q)$ denote the punishment level that makes the principal indifferent, as defined by~\eqref{D.3}, and let $\underline{L}(c) \equiv \max_{q \in [1/2,1]} L(c,q)$.
\begin{PropositionG}
 For every $c>0$, there exists $\varepsilon>0$ such that when $L \in [\underline{L}(c),\underline{L}(c)+\varepsilon]$, in every proper equilibrium that satisfies Refinements 1 and 2,
    $q(0,0)+q(1,1)-q(1,0)-q(0,1) < 0$.
\end{PropositionG}
Equivalently, the principal's decisions are strategic complements in every equilibrium, and committing an offense against only one agent is strictly suboptimal. Since $L(c,1) \geq  \underline{L}(c)$
and $L(c,1/2) \geq  \underline{L}(c)$.
To prove Proposition G, it is sufficient to show that in every equilibrium such that $q(0,0)+q(1,1)-q(1,0)-q(0,1) \geq 0$, $L$ is strictly above $L(c,1)$ or strictly above $L(c,1/2)$. Let $(\omega_0^*,\omega_0^{**})$ denote the unique solution to
    $\frac{\omega_0^*-b}{c}=\Psi_0^*-1 $ and $\frac{\omega_0^{**}}{c} =\Psi_0^{**}-1$,
where $\Psi_0^* \equiv \delta \Phi(\omega_0^*)+(1-\delta)$ and $\Psi_0^{**} \equiv \delta \Phi(\omega_0^{**})+(1-\delta)$. By construction, $\omega_0^*$ and $\omega_0^{**}$ are the reporting cutoffs when $q(1,0)=q(0,1)=1/2$. Therefore,
\begin{equation}\label{D.7}
   L(c,1/2) = \frac{2}{\Psi_0^*-\Psi_0^{**}}.
\end{equation}
The rest of the proof consists of two parts. In Section G.3, we rule out equilibria in which $q(0,1)=q(1,0)=0$. In Section G.4, we rule out equilibria in which $\max\{q(0,1),q(1,0)\}>0$ but $q(0,1)+q(1,0) \leq 1$.

\subsection*{G.3 $\quad$ Equilibria in which $\mathbf{q(1,0)=q(0,1)=0}$}
Supplementary Appendix I shows that all proper equilibria that satisfy Refinements 1 and 2 with  $q(0,0)=q(1,0)=q(0,1)=0$ must be symmetric.
Let $q \equiv q(1,1) \in (0,1]$ be the conviction probability when both agents accuse the principal. Let $\omega_m^*$ and $\omega_m^{**}$ denote agent's reporting cutoffs, which satisfy (3.8) and (3.9) in the main text, respectively.
Let $\Psi_1^* \equiv \delta \Psi(\omega_m^*)+(1-\delta)\alpha$ and let $\Psi_1^{**} \equiv \delta \Psi(\omega_m^{**})+(1-\delta)\alpha$.
The principal's indifference condition implies that
    $\frac{1}{L}= q\Psi_1^{**} (\Psi_1^*-\Psi_1^{**})$.
We now show that $L > L(c,1/2)$. It suffices to show that
\begin{equation}\label{D.8}
    \Psi_0^*-\Psi_0^{**}
    >
    \Psi_1^{**} (\Psi_1^*-\Psi_1^{**}).
\end{equation}
The expressions for the reporting cutoffs $\omega^*$ and $\omega^{**}$ imply that
    $\omega_1^*=b+c-\frac{c}{q \Psi_1^{**}} \leq b+c-\frac{c}{\Psi_1^*} \leq b+c (\Psi_1^*-1)$ and
    $\omega_0^* = b+c(\Psi_0^*-1)$.
Since $c(\Psi(\omega)-1)$ is strictly convex in $\omega$ when $\omega<0$, and the value of $c (\Psi(\omega)-1)$
is strictly negative when $\omega=0$, we conclude that
$\omega_1^*< \omega_0^*$. Since $\omega_0^*-\omega_0^{**} >b > \omega_1^*-\omega_1^{**}$, we conclude that
\begin{equation}\label{D.9}
    \Psi_0^*-\Psi_0^{**}
    >
   \Psi_1^*-\Psi_1^{**}.
\end{equation}
This in turn implies (\ref{D.8}).

\subsection*{G.4 $\quad$ $q(1,0)$ or $q(0,1)$ is strictly positive, and (\ref{1}) is positive}
Suppose toward a contradiction that there exists an equilibrium such that, first, $q(1,0)+q(0,1)<q(1,1)+q(0,0)$, and, second, at least one of the probabilities $q(0,1)$ and $q(1,0)$ is strictly positive.
According to Lemma 3.2, the principal's incentives to commit crimes are strategic substitutes. According to Lemma 2.1, the principal commits offense with interior probability, and therefore, he is indifferent between committing no offense and committing one offense, but he commits two offenses with zero probability.

Let $q_1 \equiv q(1,0)$ and $q_2 \equiv q(0,1)$. For $i \in \{1,2\}$,
let $p_i$ be the probability that $\theta_i=1$, and $\omega_i^*$ and $\omega_i^{**}$ denote agent $i$'s reporting cutoffs, which satisfy
\begin{equation}\label{D.10}
\omega_i^* =b-c\frac{(1-\Psi_j^{**}) (1-q_i)}{q_i +\Psi_j^{**} (1-q_1-q_2)}
\end{equation}
and
\begin{equation}\label{D.11}
\omega_i^{**} =-c\frac{(1-X_j) (1-q_i)}{q_i +X_j (1-q_1-q_2)}
\end{equation}
where $j$ denotes the agent other than $i$ and $X_i \equiv
 \frac{1-p_1-p_2}{1-p_i}   \Psi_i^{**}
+\frac{p_j}{1-p_i} \Psi_i^*$.
For $i \in \{1,2\}$, let
$\mathcal{I}_i \equiv \frac{p_i}{p_i+p_j} \frac{\Psi_i^*}{\Psi_i^{**}} +\frac{p_j}{p_i+p_j} \frac{1-\Psi_j^*}{1-\Psi_j^{**}}$.
The prior probability of guilt of the principal is $p_1+p_2$. Since the principal is convicted with positive probability after one accusation, we have
\begin{equation*}
    \max\{ \mathcal{I}_1, \mathcal{I}_2 \}
    \geq l^* \frac{1-p_1-p_2}{p_1+p_2}
    \geq  \min \{ \mathcal{I}_1, \mathcal{I}_2 \}.
\end{equation*}

\paragraph{Step 1:} We rule out equilibria in which the principal's expected cost of committing the offense observed by agent $1$, relative to committing no offense, is different from the expected cost of committing the offense observed by agent $2$.
Suppose toward a contradiction that the cost of committing the offense observed by agent $1$ is strictly higher than the cost of committing the offense observed by agent $2$. This implies that $p_1=0$ and $p_2>0$ and that
\begin{equation*}
    \mathcal{I}_2 =\frac{\Psi_2^*}{\Psi_2^{**}} > 1 > \frac{1-\Psi_2^*}{1-\Psi_2^{**}} =\mathcal{I}_1.
\end{equation*}
Therefore, $q_1=0$, $q_2>0$, and the marginal cost of committing the offense associated with agent $1$ conditional on $\theta_2=0$ is
    $L (\Psi_1^*-\Psi_1^{**}) (1-q_2) \Psi_2^{**}$.
The marginal cost of committing the offense associated with agent $2$ conditional on $\theta_1=0$ is
$L (\Psi_2^*-\Psi_2^{**}) \big( (1-q_2) \Psi_1^{**} +q_2 \big)$,
which equals $1$ in equilibrium.
Since the marginal cost of committing offense against agent $1$ is strictly higher,
\begin{equation}\label{D.12}
\frac{\Psi_1^* \Psi_2^{**}-\Psi_1^{**} \Psi_2^*}{\Psi_2^*-\Psi_2^{**}} \geq \frac{q_2}{1-q_2}.
\end{equation}
This implies that $\Psi_1^*/\Psi_1^{**} > \Psi_2^*/\Psi_2^{**}$
Since the strategy profile is a proper equilibrium, we have $\omega_2^*-\omega_2^{**}=b > \omega_1^*-\omega_1^{**}$. This can occur only if $\omega_1^{**} < \omega_2^{**}$. Since the density of $\omega$ is strictly increasing when $\omega <0$,
\begin{equation}\label{D.13}
    \Psi_1^*-\Psi_1^{**} < \Psi_2^*-\Psi_2^{**} < \Psi(0)-\Psi(-b).
\end{equation}
When $L$ is close to $\underline{L}(c)$, the equilibrium conditions imply that
\begin{equation*}
    \frac{1}{\displaystyle (\Psi_2^*-\Psi_2^{**}) \Big( (1-q_2) \Psi_1^{**} +q_2 \Big)} \leq L(c,1),
\end{equation*}
or equivalently,
\begin{equation*}
    q_2 \geq -\frac{\Psi_1^{**}}{1-\Psi_1^{**}}
    +\frac{(\Psi(0)-\Psi(-b)) (2-\Psi(0)-\Psi(-b))}{2 (1-\Psi_1^{**}) (\Psi_2^*-\Psi_2^{**})}.
\end{equation*}
Since $\Psi(0), \Psi(-b)<1/2$,
this implies that
\begin{equation}\label{D.14}
    q_2 \geq 1-\frac{1}{1-\Psi_1^{**}} + \frac{\Psi(0)-\Psi(-b)}{2 (1-\Psi_1^{**}) (\Psi_2^*-\Psi_2^{**})}.
\end{equation}
Plugging (\ref{D.14}) into (\ref{D.12}) yields
\begin{equation}\label{D.15}
    \frac{\Psi_1^* \Psi_2^{**}-\Psi_1^{**} \Psi_2^*}{\Psi_2^*-\Psi_2^{**}} \geq
    \frac{(\Psi(0)-\Psi(-b)) -2 \Psi_1^{**} (\Psi_2^*-\Psi_2^{**})}{2(\Psi_2^*-\Psi_2^{**})-(\Psi(0)-\Psi(-b))}.
\end{equation}
Letting $\Delta \equiv \Psi(0)-\Psi(-b)$ and $\Delta_i \equiv\Psi_i^*-\Psi_i^{**}$, the previous inequality can be re-expressed as
\begin{equation}\label{D.16}
    2 \Delta_1 \Delta_2 \Psi_2^{**} \geq \Delta (\Psi_2^*(1-\Psi_1^{**})-(1-\Psi_1^*)\Psi_2^{**})=\Delta \Delta_2 (1-\Psi_1^{**}) +\Delta \Delta_1 \Psi_2^{**}.
\end{equation}
However, (\ref{D.13}) and the inequality $1> \Psi_1^{**}+\Psi_2^{**}$ are incompatible with (\ref{D.16}), which yields the desired contradiction.

\paragraph{Step 2:} From Step 1, the principal incurs the same marginal cost from committing the offense associated with agent $1$ and the offense associated with agent $2$. This leads to the following indifference condition:
\begin{equation}\label{D.17}
    L =\frac{1}{\displaystyle (\Psi_1^*-\Psi_1^{**}) \Big(
    \Psi_2^{**} (1-q_1-q_2) +q_1
    \Big)}
    =\frac{1}{\displaystyle (\Psi_2^*-\Psi_2^{**}) \Big(
    \Psi_1^{**} (1-q_1-q_2) +q_2
    \Big)}.
\end{equation}
Without loss of generality, we assume $q_1 \leq q_2$.
Since $q_1+q_2 \leq 1$, we have
\begin{equation*}
L=    \frac{1}{\displaystyle (\Psi_1^*-\Psi_1^{**}) \Big(
    \Psi_2^{**} (1-q_1-q_2) +q_1
    \Big)} \geq \frac{2}{\Psi_1^*-\Psi_1^{**}}.
\end{equation*}
In what follows, we show that
\begin{equation*}
     \frac{2}{ \Psi_1^*-\Psi_1^{**}} > L (c,1)=\frac{2}{(\Psi(0)-\Psi(-b)) (2-\Psi(0)-\Psi(-b))}.
\end{equation*}
Since $\Psi(b)<1/2$ and $\Psi(0)<1/2$, the above inequality is implied by
\begin{equation}\label{D.18}
\Psi(b)-\Psi(0)
   >
   \Psi_1^*-\Psi_1^{**},
\end{equation}
which holds
because $\omega_1^*-\omega_1^{**} <b$, $\omega_1^*<0$ and the density of $\omega$ is strictly increasing when $\omega<0$.

\subsection*{G.5 $\quad$ $q(1,0)$ or $q(0,1)$ is strictly positive, and (\ref{1}) is equal to zero}
We start by ruling out equilibria in which one of the agents never witnesses any offense. Suppose toward a contradiction that the principal chooses $\theta_1=0$ with probability $1$. Then, as in Section G.4, we have $q(1,0)=0$. Since also $q(0,0)=0$, (\ref{1}) implies $q(0,1)=q(1,1)$, which contradicts Lemma 2.1.

Next, we rule out equilibria in which $q(1,1) < 1$. Suppose that $q(1,1) \in (0,1)$. Then, either $q(0,1)$ and $q(1,0)$ both lie in $(0,1)$, which is ruled out in Section F.1 of this document, or one of the probabilities $q(1,0)$ and $q(0,1)$ is equal to 0, which is ruled out by Lemma 2.1 and  Refinement~1.

Therefore, we focus without loss of generality on equilibria with the following features: (1) both offenses entail the same marginal cost, and (2) $q(1,1)=1$, $q(1,0)$ and $q(0,1)$ both lie in $(0,1)$, and $q(0,1)+q(1,0)=1$.
Since the marginal costs associated with both offenses are the same, we have
\begin{equation}\label{D.19}
    L = \frac{1}{q_1 (\Psi_1^*-\Psi_1^{**})} =\frac{1}{q_2 (\Psi_2^*-\Psi_2^{**})}.
\end{equation}

\paragraph{Step 1:} First, we consider the symmetric case in which $q_1=q_2=1/2$. We need to show that
\begin{equation*}
    \frac{2}{\Psi_1^*-\Psi_1^{**}} > L(c,1)=\frac{2}{(\Psi(0)-\Psi(-b))(2-\Psi(0)-\Psi(-b))}.
\end{equation*}
This inequality is implied by
\begin{equation}\label{D.20}
    \Psi(0)-\Psi(-b) > \Psi_1^*-\Psi_1^{**}.
\end{equation}
To show (\ref{D.20}), let $\overline{\omega}$ be defined by
    $\Psi(\overline{\omega}) \equiv \Psi(\omega^*)-\Psi(0)+\Psi(-b)$.
Inequality (\ref{D.20}) is equivalent to $\overline{\omega}< \omega^{**}$, or equivalently,
\begin{equation}\label{D.21}
    \Psi(\overline{\omega}) -1 > \frac{\overline{\omega}+b}{c}.
\end{equation}
Plugging in the expression     $c=\frac{\omega^*}{\Psi(\omega^*)-1}$ into (\ref{D.21}), we have
    $\Big(
    \Psi(\omega^*)-\Psi(0)+\Psi(-b)-1
    \Big) \frac{\omega^*}{\Psi(\omega^*)-1}  > \overline{\omega}+b$.
This is equivalent to
\begin{equation}\label{D.22}
    \omega^* +\underbrace{(\Psi(0)-\Psi(-b)) \frac{\omega^*}{1-\Psi(\omega^*)}}_{>0} > \overline{\omega}+b.
\end{equation}
Since the density of $\omega$ is strictly increasing when $\omega<0$ and $\omega^*<0$, we ahve
    $\Psi(\omega^*)-\Psi(\omega^*-b) < \Psi(0)-\Psi(-b)$.
This implies that $\omega^*>\overline{\omega}+b$, which verifies inequality (\ref{D.22}).

\paragraph{Step 2:} We consider asymmetric equilibria in which $q_1 \neq q_2$. Agent $i$'s reporting cutoffs are bounded by
    $\frac{\omega_i^*}{c} \leq \frac{q_j}{q_i} (\Psi(\omega_j^*)-1)$ and
    $\frac{\omega_i^{**}+b}{c} \geq \frac{q_j}{q_i} (\Psi(\omega_j^{**})-1)$, where $\Psi(\omega) \equiv (1-\delta) \alpha +\delta \Phi(\omega)$.
The function $\Psi(\cdot)$ is convex for $\omega$ small enough. For fixed $q_1$ and $q_2$ such that $q_1+q_2=1$, this implies that  $(\omega_1^*,\omega_2^*)$ is bounded above by the largest solution to
    $\frac{\omega_1^*}{c} = \frac{q_2}{q_1} (\Psi(\omega_2^*)-1)$ and
    $\frac{\omega_2^*}{c} = \frac{q_1}{q_2} (\Psi(\omega_1^*)-1)$.
The largest solution is well-defined since $\Psi(\cdot)$ is a strictly increasing function, and therefore, for every pair of solutions $(\omega_1^*,\omega_2^*)$ and $(\omega_1',\omega_2')$, if $\omega_1^* >\omega_1'$, then $\omega_2^* > \omega_2'$.
Similarly, $(\omega_1^{**},\omega_2^{**})$ is bounded below by the smallest solution to $\frac{\omega_1^{**}+b}{c} = \frac{q_2}{q_1} (\Psi(\omega_2^{**})-1)$ and
    $\frac{\omega_2^{**}+b}{c} = \frac{q_1}{q_2} (\Psi(\omega_1^{**})-1)$.
 The smallest solution is also well-defined since $\Psi(\cdot)$ is a strictly increasing function, and therefore, for every pair of solutions $(\omega_1^{**},\omega_2^{**})$ and $(\omega_1',\omega_2')$, if $\omega_1^{**} >\omega_1'$, then $\omega_2^{**} > \omega_2'$.

Let $\{\omega_i^*(q_1)\}_{i=1}^2$ be the largest solution to the first system of equations and let
$\{\omega_i^{**}(q_1)\}_{i=1}^2$ be the smallest solution to the second system of equations.
The minimum $L$ in this class of equilibria is bounded below by
\begin{equation}\label{D.23}
    \max_{i\in\{1,2\}} \frac{1}{\displaystyle q_i \Big( \Psi(\omega_i^*(q_1))-\Psi (\omega_i^{**}(q_1)) \Big)}.
\end{equation}
We start by showing that when $q_1<1/2$, we have $\omega_2^*(q_1) > \omega_1^*(q_1)$ and $\omega_2^{**}(q_1) > \omega_1^{**}(q_1)$. Suppose toward a contradiction that $\omega_2^*(q_1) \leq \omega_1^*(q_1)$. Let $\alpha \equiv (q_2/q_1)^2$, which is strictly greater than 1. We have
$\alpha (\Psi_2^*(q_1)-1)-(\Psi_1^*(q_1)-1)>0$,
which is equivalent to
\begin{equation*}
    \underbrace{(1-\alpha) (1-\Psi_2^*(q_1))}_{<0} +\underbrace{\Psi_2^*(q_1)-\Psi_1^*(q_1)}_{<0 \textrm{ by hypothesis}}>0.
\end{equation*}
This leads to a contradiction.

Next, we show that $\omega_1^*(q_1) < \omega_1^*(1/2) < \omega_2^{*}(q_1)$ and
$\omega_1^{**}(q_1) < \omega_1^{**}(1/2) < \omega_2^{**}(q_1)$. These inequalities hold because for every $q_1 \in (0,1/2]$,
\begin{equation}\label{D.24}
    \frac{\omega_1^*(q_1) \omega_2^*(q_1)}{c^2} = \Big(\Psi_1^*(q_1)-1\Big)\Big(\Psi_2^*(q_1)-1\Big).
\end{equation}
Since $\Psi(\omega)$ is strictly convex for $\omega<0$, we have
    $\omega \geq \Psi(\omega)-1$
\textit{if and only if} $\omega \geq \omega^* (1/2)$. This together with (\ref{D.24}) implies that
 $\omega_1^*(q_1) < \omega_1^*(1/2) < \omega_2^{*}(q_1)$. Similarly, one can show that
$\omega_1^{**}(q_1) < \omega_1^{**}(1/2) < \omega_2^{**}(q_1)$.

Last, suppose toward a contradiction that
\begin{equation}\label{D.25}
    \frac{1}{2}(\Psi^*-\Psi^{**}) < \min \{
    q_1 (\Psi_1^*-\Psi_1^{**}), q_2 (\Psi_2^*-\Psi_2^{**})
    \}.
\end{equation}
This implies the following inequality, which violates $\Psi^*$'s convexity and yields the desired contradiction:
\begin{equation*}
   \Big( \omega^*(1/2)-(\omega^{**}(1/2)+b)\Big)^2 \geq 4q_1q_2 \Big(\omega_1^*(q_1)-(\omega_1^{**}(q_1)+b) \Big)
   \Big(\omega_2^*(q_1)-(\omega_2^{**}(q_1)+b) \Big).
\end{equation*}

\section*{H. $\quad$ Results with Three or More Agents}
We show Lemma E.1 in Section H.1.
Sections H.2 and H.3 prove Propositions 4 and 5 in the main text. Finally, Section H.4 shows that in symmetric equilibria, Refinement 1 implies Refinement 2.
\subsection*{H.1 $\quad$ Proof of Lemma E.1}\label{subC.1}
First, we show that when $L$ is large enough, $q(1,1,...,1)<1$ in every symmetric BNE that satisfies Refinement 1.
Suppose toward a contradiction that for every $L' \in \mathbb{R}_+$, there exists $L \geq L'$ and a symmetric equilibrium for punishment $L$ that satisfies Refinement 1 such that $q(1,1,...,1)=1$. We establish a lower bound on the marginal increase in conviction probabilities that uniformly applies across all $L$. For every $\boldsymbol{a} \succ \boldsymbol{a'}$, we have
    $\Pr(\overline{\theta}=1 | \boldsymbol{a}) >  \Pr(\overline{\theta}=1 | \boldsymbol{a'})$.
As a result, there exist $m \in \{1,2,...,n\}$ and $q \in [0,1)$ such that the principal is convicted for sure when there are $m$ accusations or more, and is convicted with probability $q$ when there are $m-1$ accusations. Refinement 1 requires that $q=0$ when $m=1$.

From Lemma 2.1, an agent's equilibrium strategy is summarized by two cutoffs, $\omega^*$ and $\omega^{**}$, such that for every $i \in \{1,2,...,n\}$,
agent $i$ accuses the principal when $\theta_i=1$ and $\omega_i \leq \omega^*$, or
when $\theta_i=0$ and $\omega_i \leq \omega^{**}$.
Let $\Psi^* \equiv (1-\delta)\alpha +\delta \Phi(\omega^*)$ and $\Psi^{**}\equiv (1-\delta)\alpha +\delta \Phi(\omega^{**})$. We have $\Psi^* > \Psi^{**}$.

For every $m \leq n-1$, let $Q(m,\theta_{-i})$ be the probability with which agents other than $i$ submit $m$ accusations given $\theta_{-i}$.
Fixing $\theta_{-i}$, changing $\theta_i$ from $0$ to $1$, the marginal increase in conviction probability is given by:
\begin{equation}\label{F.17}
    (\Psi^*-\Psi^{**}) P(m,q,\theta_{-i}),
\end{equation}
where
\begin{equation}\label{F.18}
P(m,q,\theta_{-i})  \equiv
q  Q(m-2,\theta_{-i})+
\sum_{j=m-1}^{n-1} Q(j,\theta_{-i})-qQ(m-1,\theta_{-i})-\sum_{j=m}^{n-1} Q(j,\theta_{-i}).
\end{equation}
This yields
\begin{equation}\label{F.19}
P(m,q,\theta_{-i}) = (1-q) Q(m-1,\theta_{-i}) + qQ(m-2,\theta_{-i}).
\end{equation}
Since $\theta$ is binary and the equilibrium is symmetric, the functions
$Q(m,\theta_{-i})$ and
$P (m,q,\theta_{-i})$ depend on $\theta_{-i}$ only through the number of $1$s in the entries of $\theta_{-i}$. Let $|\theta_{-i}|$ be the number of $1$s in the vector $\theta_{-i}$.
Abusing notation, we rewrite
$Q(m,\theta_{-i})$ as $Q(m,|\theta_{-i}|)$, and
$P(m,q,\theta_{-i})$ as $P(m,q,|\theta_{-i}|)$.
For any fixed values of $m$ and $q$: one of these three statements is true:
\begin{itemize}
  \item[1.] $P(m,q,|\theta_{-i}|)$ is strictly increasing in $|\theta_{-i}|$,
  \item[2.] $P(m,q,|\theta_{-i}|)$ is strictly decreasing in $|\theta_{-i}|$,
  \item[3.] $P(m,q,|\theta_{-i}|)$ is first increasing and then decreasing in $|\theta_{-i}|$.
\end{itemize}
In equilibrium, the principal is indifferent between committing an offense against $k$ agents and not committing any offense, where $k$ satisfies:
\begin{equation}\label{F.20}
    k \in \arg\min_{\widetilde{k} \in \{1,...,n\}} \frac{1}{\widetilde{k}}\sum_{j=0}^{\widetilde{k}-1} P(m,q,|\theta_{-i}|).
\end{equation}
This property must hold because from Lemma 2.1 the principal must be indifferent between committing no offense and a positive number $k$ of offenses. For this value of $k$, the average cost of committing an offense when the principal commits $k$ offenses is equal to $1$, and there does not exist $k' \in \{1,2,...,n\}$ such that if the principal commits $k'$ offenses, his average cost of committing an offense is strictly less than $1$.

The value of $k$ depends on the monotonicity of $P(m,q,\cdot)$.
When $P(m,q,\cdot)$
is strictly increasing,
$k=1$ and the principal is indifferent between committing only one offense and committing no offense.
When $P(m,q,\cdot)$
is strictly decreasing, $k=n$ and the principal is indifferent between committing no offense and committing offense against all agents.
When $P(m,q,\cdot)$
is first increasing and then decreasing, $k$ is either $1$ or $n$, depending on the parameters.
In what follows, we consider the two values of $k$ separately.

\paragraph{Strategic Substitutes:} When $k=1$, an agent's reporting cutoff when he has witnessed an offense is
\begin{equation}\label{F.21}
  \omega^*= b-c\frac{\displaystyle  \Big\{ 1-q  Q(m-2,0)-
\sum_{j=m-1}^{n-1} Q(j,0) \Big\}
    }{P(m,q,0)}
\end{equation}
Similarly, an agent's reporting cutoff when he has not witnessed any offense is
\begin{equation}\label{F.22}
    \omega^{**}=-c\frac{\displaystyle 1 - \beta \Big\{q  Q(m-2,0)+
\sum_{j=m-1}^{n-1} Q(j,0) \Big\} -(1-\beta) \Big\{ q  Q(m-2,1) +
\sum_{j=m-1}^{n-1} Q(j,1)\Big\} }{\beta P(m,q,0) +(1-\beta) P(m,q,1)}
\end{equation}
where $\beta$ is the probability with which $\theta_1=...=\theta_n=0$ conditional on $\theta_i=0$.

In the first step, we show that $\omega^*<\omega^{**}+b$. This inequality comes from the fact that $k=1$, which implies that $P(m,q,1)>P(m,q,0)$. Moreover,
\begin{equation*}
    q  Q(m-2,1)+
\sum_{j=m-1}^{n-1} Q(j,1) > q  Q(m-2,0)+
\sum_{j=m-1}^{n-1} Q(j,0).
\end{equation*}
Therefore,
\begin{equation*}
    \frac{\omega^*-(\omega^{**}+b)}{c} = \frac{\displaystyle 1 - \beta \Big\{q  Q(m-2,0)+
\sum_{j=m-1}^{n-1} Q(j,0) \Big\} -(1-\beta) \Big\{ q  Q(m-2,1) +
\sum_{j=m-1}^{n-1} Q(j,1)\Big\} }{\beta P(m,q,0) +(1-\beta) P(m,q,1)}
\end{equation*}
\begin{equation*}
    -\frac{\displaystyle  \Big\{ 1-q  Q(m-2,0)-
\sum_{j=m-1}^{n-1} Q(j,0) \Big\}
    }{P(m,q,0)} < 0.
\end{equation*}

In the second step, we bound $\omega^*$ from below using the facts that $|\omega^*-\omega^{**}|<b$ and $q(1,1,...,1)=1$.
First, for every $m \in \{0,1,...,n-1\}$ and $q$,
    $P(m,q,0) \geq P(n-1,0,0) =(\Psi^{**})^{n-1}$.
From (\ref{F.21}), we know that
\begin{equation*}
    \frac{\omega^*-b}{c} \geq - (\Psi^{**})^{-(n-1)} \geq - \Big(\delta \Phi(\omega^*-b) +(1-\delta) \alpha \Big)^{-(n-1)}.
\end{equation*}
Since the RHS of the last formula is bounded below, there exists  $\underline{\omega}^* \in \mathbb{R}_-$, independent of $L$, such that $\omega^* \geq \underline{\omega}^*$.

In the third step, we bound the value of $\Psi^*-\Psi^{**}$ from below. Let
\begin{equation*}
    X_0 \equiv q  Q(m-2,0)+
\sum_{j=m-1}^{n-1} Q(j,0)
\textrm{ and }
        X_1 \equiv q  Q(m-2,1)+
\sum_{j=m-1}^{n-1} Q(j,1).
\end{equation*}
From (\ref{F.21}) and (\ref{F.22}),
\begin{equation}\label{F.23}
    \frac{\omega^*-\omega^{**}}{c}=\frac{b}{c}
    - (1-\beta) \frac{P(m,q,1)(1-X_0)-P(m,q,0)(1-X_1)}{\displaystyle P(m,q,0)\big(\beta P(m,q,0)+(1-\beta)P(m,q,1)\big)}.
\end{equation}
We start by bounding
\begin{equation}\label{F.24}
    \frac{P(m,q,1)(1-X_0)-P(m,q,0)(1-X_1)}{\Psi^*-\Psi^{**}}
\end{equation}
from above. Since
\begin{equation*}
    P(m,q,1)(1-X_0)-P(m,q,0)(1-X_1)=(X_1-X_0) P(m,q,0) +(1-X_0) (P(m,q,1)-P(m,q,0)),
\end{equation*}
and $1-X_0$ as well as $P(m,q,0)$ are bounded from above by $1$, we only need to bound
$\frac{X_1-X_0}{\Psi^*-\Psi^{**}}$ and $\frac{P(m,q,1)-P(m,q,0)}{\Psi^*-\Psi^{**}}$
 from above.
Notice that
\begin{equation*}
    \frac{Q(j,1)-Q(j,0)}{\Psi^*-\Psi^{**}}=
        {n-2 \choose j-1} (\Psi^{**})^{j-1} (1-\Psi^{**})^{n-1-j}
 -   {n-2 \choose j} (\Psi^{**})^{j} (1-\Psi^{**})^{n-2-j}
\end{equation*}
which is bounded from above by ${n-2 \choose j-1}$. Since $X_1-X_0$ and $P(m,q,1)-P(m,q,0)$ are both linear combinations of terms of the form of $Q(j,1)-Q(j,0)$, $\frac{X_1-X_0}{\Psi^*-\Psi^{**}}$ and $\frac{P(m,q,1)-P(m,q,0)}{\Psi^*-\Psi^{**}}$
are also bounded above. Let $C_0 \in \mathbb{R}_+$ be the upper bound on (\ref{F.24}). Since
     $P(m,q,0)\big(\beta P(m,q,0)+(1-\beta)P(m,q,1)\big)$
is bounded away from $0$, we can also bound
\begin{equation}\label{F.25}
  \frac{1}{\Psi^*-\Psi^{**}} \cdot  \frac{P(m,q,1)(1-X_0)-P(m,q,0)(1-X_1)}{\displaystyle P(m,q,0)\big(\beta P(m,q,0)+(1-\beta)P(m,q,1)\big)}.
\end{equation}
Letting $C_1\in \mathbb{R}_+$ denote this bound, we have
\begin{equation}\label{F.26}
     \frac{\omega^*-\omega^{**}}{c} \geq \frac{b}{c}
    - \delta (1-\beta)C_1 (\Phi(\omega^*)-\Phi(\omega^{**})).
\end{equation}
Letting $C_2 \equiv \delta (1-\beta)C_1$, we obtain
\begin{equation}\label{F.27}
  \frac{\omega^*-\omega^{**}}{c} +C_2   (\Phi(\omega^*)-\Phi(\omega^{**})) \geq \frac{b}{c}.
\end{equation}
We have shown in the previous step that $\omega^* > \underline{\omega}^*$. Let  $\epsilon \in \mathbb{R}_+$ denote the unique solution of the equation
\begin{equation*}
    \frac{\epsilon}{c}+C_2 \epsilon \phi(\underline{\omega}^*-\epsilon) =\frac{b}{c},
\end{equation*}
whose LHS is continuous, strictly increasing in $\epsilon$, strictly greater than $\frac{b}{c}$ for $\epsilon \rightarrow \infty$,  and strictly less than
$\frac{b}{c}$ when $\epsilon \rightarrow -\infty$. From (\ref{F.27}), we have $\omega^*=\omega^{**} \geq \epsilon$, and, hence,
\begin{equation}\label{F.28}
    \Psi^*-\Psi^{**} =\delta (\Phi(\omega^*)-\Phi(\omega^{**})) \geq \delta \epsilon \phi(\underline{\omega}^*-\epsilon).
\end{equation}
The principal's incentive constraint is
$P(m,q,0) (\Psi^*-\Psi^{**})L=1$.
Since $P(m,q,0)$ is bounded  below by
  $\Big(  \delta \Phi(\underline{\omega}^*-b) +(1-\delta)\alpha \Big)^{-(n-1)}$
and $\Psi^*-\Psi^{**}$ is bounded below by (\ref{F.28}), the LHS must go to $+\infty$ as $L \rightarrow +\infty$, which leads to a contradiction.

\paragraph{Strategic Complements:} When $k=n$, an agent's reporting cutoff when he has witnessed an offense is
\begin{equation}\label{F.29}
    \omega^*= b-c\frac{\displaystyle   \Big\{ 1-q  Q(m-2,n-1)-
\sum_{j=m-1}^{n-1} Q(j,n-1) \Big\}}{P(m,q,n-1)}
\end{equation}
Similarly, an agent's reporting cutoff when he has not witnessed any offense is
\begin{equation}\label{F.30}
    \omega^{**}=-c\frac{\displaystyle \Big\{ 1-q  Q(m-2,0)-
\sum_{j=m-1}^{n-1} Q(j,0) \Big\}}{P(m,q,0)}
\end{equation}
Since $k=n$,
    $P(m,q,n-1) > P(m,q,0)$. Moreover, since $\Psi^*-\Psi^{**} >0$,
\begin{equation}\label{F.32}
   q Q(m-2,n-1)-
\sum_{j=m-1}^{n-1} Q(j,n-1)
> q Q(m-2,0)-
\sum_{j=m-1}^{n-1} Q(j,0).
\end{equation}
These inequalities imply that
the distance between the cutoffs exceeds$b$. The inequality
\begin{equation}\label{F.33}
   \frac{ \omega^*}{c} \geq - \Big((1-\delta)\alpha\Big)^{-(n-1)}
\end{equation}
provides a lower bound on $\omega^*$, which we denote by $\underline{\omega}^*$. The principal's marginal cost of committing another offense is bounded below by
\begin{equation}\label{F.34}
    L (\Psi^*-\Psi^{**}) P(m,q,n-1) \geq L \delta b \phi(\underline{\omega}^*-b) (1-\delta)^{n-1}\alpha^{n-1}.
\end{equation}
The RHS goes to infinity as $L \rightarrow \infty$, which leads to a contradiction.

\paragraph{Principal's Equilibrium Strategy:} Since $q(1,1,...,1) \in (0,1)$ and $\omega_i^* >\omega_i^{**}$ in any symmetric equilibrium that satisfies Refinement 1, we have $\Pr(\overline{\theta}=1|\boldsymbol{a}) < \Pr(\overline{\theta}=1|(1,1,...,1))=\pi^*$ for every $\boldsymbol{a} \neq (1,1,...,1)$. Therefore, $q(\boldsymbol{a})=0$ for every $\boldsymbol{a} \neq (1,1,...,1)$.
Let $q \equiv q(1,1,...,1)$. When the principal commits an offense in addition to $m \in \{0,1,2,...,m-1\}$ offenses, the probability that he gets convicted increases by
$q(\Psi^*-\Psi^{**}) (\Psi^{**})^{n-m} (\Psi^*)^m$,
which is a strictly increasing function of $m$. Therefore, if the principal commits $m \geq 2$ offenses with positive probability, the probability $\Pr(\overline{\theta}=1)$ that he commits  at least one offense is equal $1$, which contradicts Lemma 2.1. Therefore, the principal must be indifferent between committing one offense and committing no offense, which proves the second part of Lemma E.1.

\subsection*{H.2 $\quad$ Proof of Proposition 4}\label{subC.2}
Exploiting Lemma E.1, we establish the properties common to all symmetric equilibria that satisfy Refinement 1 .
We start by deriving formulas for the agents' reporting cutoffs $(\omega_n^*,\omega_n^{**})$, the informativeness of accusations $\mathcal{I}_n$ and the equilibrium probability of offense $\pi_n$. For
 every $i \in \{1,2,...,n\}$, agent $i$'s reporting cutoff when $\theta_i=1$ is
\begin{equation}\label{F.1}
    \omega_n^* =b+c-\frac{c}{q_n Q_{1,n}}.
\end{equation}
His reporting cutoff when $\theta_i=0$ is
\begin{equation}\label{F.2}
    \omega_n^{**} =c-\frac{c}{q_n Q_{0,n}}
\end{equation}
where
\begin{equation}\label{F.3}
    Q_{1,n} \equiv \Big(
    \delta \Phi(\omega_n^{**})+(1-\delta)\alpha
    \Big)^{n-1}
    \end{equation}
    and
    \begin{equation*}
     Q_{0,n} \equiv \frac{n \mathcal{I}_n}{(n-1)l^*+n \mathcal{I}_n}\Big(
    \delta \Phi(\omega_n^{**})+(1-\delta) \alpha
    \Big)^{n-1}
    \end{equation*}
        \begin{equation}\label{F.4}
    +
    \frac{(n-1)l^*}{(n-1)l^*+n \mathcal{I}_n}\Big(
    \delta \Phi(\omega_n^{**})+(1-\delta) \alpha
    \Big)^{n-2}\Big(
    \delta \Phi(\omega_n^{*})+(1-\delta) \alpha
    \Big).
\end{equation}
From Lemma E.1, the principal commits at most one offense, which yields the informativeness ratio
\begin{equation*}
    \mathcal{I}_n =\frac{\delta \Phi(\omega_n^*)+(1-\delta)\alpha}{\delta \Phi(\omega_n^{**})+(1-\delta)\alpha}.
\end{equation*}
The judge is indifferent between convicting and acquitting a principal facing $n$ accusations. This implies that
\begin{equation}\label{F.5}
    \mathcal{I}_n  =\frac{\pi^*}{1-\pi^*} \Big/ \frac{\pi_n}{1-\pi_n}.
\end{equation}
When $L$ is large enough, the principal's indifference condition for committing zero or one offense is
\begin{equation}\label{F.6}
    \frac{1}{\delta L} =
q_n \Big(\Phi(\omega_n^*)-\Phi(\omega_n^{**})\Big)
    \Big(
    \delta \Phi(\omega_n^{**})+(1-\delta) \alpha
    \Big)^{n-1}.
\end{equation}

Using these formulas, we now show that $\omega_n^*-\omega_n^{**} \in (0,b)$. Suppose toward a contradiction that $\omega_n^*-\omega_n^{**} \leq 0$. Then, comparing (\ref{F.3}) and (\ref{F.4}) yields $Q_{1,n} \geq Q_{0,n}$. Plugging this into (\ref{F.1}) and (\ref{F.2}) implies that $\omega_n^* \geq \omega_n^{**}+b$, which contradicts the presumption that $\omega_n^*-\omega_n^{**} \leq 0$ and implies $\omega_n^*-\omega_n^{**} > 0$.
Since $\omega_n^*-\omega_n^{**} > 0$, we have
$Q_{1,n} < Q_{0,n}$.
The expressions for the cutoffs then imply that $\omega_n^*-\omega_n^{**}<b$.

Next, we show that $\mathcal{I}_n \rightarrow 1$ as $\omega_n^* \rightarrow -\infty$. Equations (\ref{F.1}) and (\ref{F.2}) yield
\begin{equation}\label{F.15}
    \frac{|\omega_n^*-b-c|}{|\omega_n^{**}-c|} =\frac{Q_{0,n}}{Q_{1,n}}
    =\frac{(n-1)l^*}{(n-1)l^*+n\mathcal{I}_n} \mathcal{I}_n
    +\frac{n \mathcal{I}_n}{(n-1)l^*+n\mathcal{I}_n}.
\end{equation}
Since $\omega_n^*-\omega_n^{**} \in (0,b)$,
the LHS converges to $1$ as $\omega_n^* \rightarrow -\infty$, which implies that the RHS also converges to~$1$. This can occur only if $\mathcal{I}_n \rightarrow 1$.

In the last step, we show that
$\omega_n^* \rightarrow -\infty$
as $L \rightarrow +\infty$. Suppose toward a contradiction that there exists a finite accumulation point $\omega^* \in \mathbb{R}_-$ for $\omega_n^*$. The LHS of (\ref{F.6}) converges to $0$ when $L \rightarrow +\infty$. Therefore, at least one of the following properties must occur along some subsequence: $q_n \rightarrow 0$ or $\Phi(\omega_n^*)-\Phi(\omega_n^{**}) \rightarrow 0$.
Since $\omega_n^* \rightarrow \omega^*$,
$\Phi(\omega_n^*)-\Phi(\omega_n^{**}) \rightarrow 0$
implies that $\omega_n^*-\omega_n^{**} \rightarrow 0$.

First, suppose toward a contradiction that $q_n \rightarrow 0$ along some subsequence. From (\ref{F.1}),
$\omega_n^* \rightarrow -\infty$ along this subsequence, which leads to a contradiction.

Second, suppose toward a contradiction that $q_n$ is bounded away from $0$ along some subsequence, i.e., strictly greater than some $\underline{q} >0$. For the LHS of (\ref{F.6}) to converge to $0$,
we need $\omega_n^*-\omega_n^{**}\rightarrow 0$ along this subsequence.
Subtracting the expression of $\omega_n^*$ from that of $\omega_n^{**}$, we obtain
\begin{equation}\label{F.16}
    \frac{q_n}{c} \Big(\omega_n^*-(\omega_n^{**}+b)\Big)
    =\frac{(n-1)l^*}{(n-1)l^*+n} \Big\{
    \frac{1}{\delta \Phi(\omega_n^*)+(1-\delta)\alpha}-\frac{1}{\delta \Phi(\omega_n^{**})+(1-\delta)\alpha}
    \Big\}.
\end{equation}
The absolute value of the LHS
 is no less than $\underline{q}b/c$ in the limit as $\omega_n^*-\omega_n^{**} \rightarrow 0$. The absolute value of the RHS converges to $0$ as
 $\Phi(\omega_n^*)-\Phi(\omega_n^{**}) \rightarrow 0$, leading to a contradiction. This implies that $\omega_n^* \rightarrow -\infty$ in every equilibrium as $L \rightarrow +\infty$.

 The three parts together imply that as $L \rightarrow +\infty$, $\omega_n^* $ and $\omega_n^{**}$ go to $-\infty$, $\mathcal{I}_n\rightarrow 1$ and$\pi_n\rightarrow \pi^*$.

\subsection*{H.3 $\quad$ Proof of Proposition 5}\label{subC.3}
Equilibrium existence is established in Supplementary Appendix J. This section establishes the limiting properties of such equilibria. Since the principal is indifferent between all action profiles, his marginal cost of committing an offense equals his marginal benefit: $q L (\Psi^*-\Psi^{**})=1$
where $q \in (0,1)$ is the incremental probability of conviction after an agent accuses him, $\Psi^*$ is the probability of $a_i=1$ conditional on
$\theta_i=1$, and $\Psi^{**}$ is the probability of $a_i=1$ conditional on
$\theta_i=0$. The reporting cutoffs are $\omega^* \equiv -\frac{c (P-q)}{q} +b$  and $\omega^{**} \equiv -\frac{c (P-q)}{q}$,
where $P \equiv \Pr(s=0|a_i=0)$.
Therefore, $\omega^*-\omega^{**}=b$, which implies that $\Psi^*-\Psi^{**} \rightarrow 0$ if and only if $\omega^* \rightarrow -\infty$.
As $L \rightarrow +\infty$, the indifference condition implies that either $q \rightarrow 0$ or $\Psi^*-\Psi^{**} \rightarrow 0$ or both. The formulas for the reporting cutoffs implies that $\omega^* \rightarrow -\infty$ if and only if $q \rightarrow 0$. This implies that $\omega^* \rightarrow -\infty$ as $L \rightarrow \infty$. In the double limit $\lim_{L \rightarrow \infty} \lim_{\delta \rightarrow 1}$, the informativeness ratio
    $\mathcal{I} \equiv
    \frac{\delta \Phi(\omega_i^*) +(1-\delta)\alpha}{\delta \Phi(\omega_i^{**})+(1-\delta) \alpha}$
converges to $+\infty$.

\subsection*{H.4 $\quad$ Equilibrium Refinements}\label{subC.4}
\begin{LemmaH}
When the judge uses decision rule (2.4), every symmetric Bayes Nash Equilibrium that satisfies Refinement 1 is a proper equilibrium and satisfies Refinement 2.
\end{LemmaH}
\begin{proof}[Proof:] As with Lemma 2.1, the principal commits an offense with positive probability in every Bayes Nash Equilibrium. Furthermore, Refinement 1 implies that the principal commits an offense with probability less than $1$.
Therefore, in any symmetric Bayes Nash equilibrium, the principal chooses $\theta_i=1$ and $\theta_i=0$ with probabilities that are both strictly positive and identical across $i$.

Therefore, in these equilibria players' information sets all occur with positive probability on the equilibrium path, implying that these equilibria are proper.

Fix a symmetric Bayes Nash equilibrium and  let $P_0 \equiv \Pr(a_i=1|\theta_i=0)$ and $P_1 \equiv \Pr(a_i=1|\theta_i=1)$. We show that the equilibrium satisfies Refinement 2 by showing that $P_1>P_0$.

First, suppose toward a contradiction that $P_1=P_0$. Since the judge's decision is measurable with respect to $\boldsymbol{a}$, the probability of convicting the principal is independent of $\boldsymbol{\theta}$. This implies that the principal has a strict incentive to choose
$\boldsymbol{\theta}=(1,1,...,1)$.
Therefore, $\Pr(\overline{\theta}=1|\boldsymbol{a})=1$ for every $\boldsymbol{a} \in \{0,1\}^n$.
This implies that $q(\boldsymbol{a})=1$ for every $\boldsymbol{a} \in \{0,1\}^n$, which violates Refinement 1.

Next, suppose that $P_1<P_0$. Symmetry implies that
$\Pr(\overline{\theta}=1|\boldsymbol{a}) < \Pr(\overline{\theta}=1|\boldsymbol{a'})$ whenever $\boldsymbol{a}$ contains strictly more 1's than $\boldsymbol{a'}$. From decision rule (2.4), $ q(0,0,..,0) \geq q(\boldsymbol{a})$ for every $\boldsymbol{a} \neq (0,0,...,0)$.
Refinement 1 requires that $q(0,0...,0)=0$, which implies that $q(\boldsymbol{a})=0$ for every $\boldsymbol{a} \in \{0,1\}^n$. Under these conviction probabilities, the principal has a strict incentive to choose
$\boldsymbol{\theta}=(1,1,...,1)$, according to which $q(\boldsymbol{a})=1$ for every $\boldsymbol{a} \in \{0,1\}^n$, which leads to a contradiction.

Since $P_1>P_0$, $\Pr(\overline{\theta}=1|\boldsymbol{a}) > \Pr(\overline{\theta}=1|\boldsymbol{a'})$ whenever $\boldsymbol{a}$ contains strictly more 1's than $\boldsymbol{a'}$. Under conviction rule (2.4),
    $q(\boldsymbol{a}) \geq q(\boldsymbol{a'})$  for every  $\boldsymbol{a}>\boldsymbol{a'}$.
Suppose toward a contradiction that there exists $i \in\{1,2,...,n\}$ such that $q(0,\boldsymbol{a}_{-i})=q(1,\boldsymbol{a}_{-i})$ for every $\boldsymbol{a}_{-i} \in \{0,1\}^{n-1}$. Then, the principal has a strict incentive to choose $\theta_i=1$, which implies that
$\Pr(\overline{\theta}=1|\boldsymbol{a})=1$ for every $\boldsymbol{a} \in \{0,1\}^n$.
This leads to a contradiction, and establishes Refinement 2.\end{proof}

\end{spacing}